\documentclass[acmsmall]{acmart}

\usepackage[acronym,shortcuts,nonumberlist,hyperfirst = false]{glossaries}
\glsdisablehyper
\newglossarystyle{longacro}{%
	\setglossarystyle{long}
		{\end{longtable*}}%
}
\newacronym{802.11 AFW}{802.11 AFW}{LabVIEW Communications 802.11 Application Framework}
\newacronym{ADC}{ADC}{Analog to Digital Conversion}
\newacronym{CCE}{CCE}{Control Channel Element}
\newacronym{CFO}{CFO}{carrier frequency offset}
\newacronym{CQI}{CQI}{Channel Quality Indicator}
\newacronym{CRS}{CRS}{Cell-Specific Reference Signal}
\newacronym{D2D}{D2D}{Device-to-Device}
\newacronym{DAC}{DAC}{Digital to Analog Conversion}
\newacronym{DMRS}{DMRS}{Demodulation Reference Signal}
\newacronym{FFT}{FFT}{Fast Fourier Transform}
\newacronym{FPGA}{FPGA}{Field-Programmable Gate Array}
\newacronym{IFFT}{IFFT}{Inverse Fast Fourier Transform}
\newacronym{LTE AFW}{LTE AFW}{LabVIEW Communications LTE Application Framework}
\newacronym{LTE}{LTE}{Long-Term Evolution}
\newacronym{OFDMA}{OFDMA}{Orthogonal Frequency-Division Multiple Access}
\newacronym{OFDM}{OFDM}{Orthogonal Frequency-Division Multiplexing}
\newacronym{PDCCH}{PD\-CCH}{Physical Downlink Control Channel}
\newacronym{PDSCH}{PD\-SCH}{Physical Downlink Shared Channel}
\newacronym{PRB}{PRB}{Physical Resource Block}
\newacronym{PSS}{PSS}{Primary Synchronization Signal}
\newacronym{PUSCH}{PU\-SCH}{Physical Uplink Shared Channel}
\newacronym{RBG}{RBG}{Resource Block Group}
\newacronym{REG}{REG}{Resource Element Group}
\newacronym{RE}{RE}{Resource Element}
\newacronym{SDR}{SDR}{Software-Defined Radio}
\newacronym{SRS}{SRS}{Sounding Reference Signal}
\newacronym{TDMA}{TDMA}{Time-Division Multiple Access}
\newacronym{UERS}{UERS}{UE-Specific Reference Signal}
\newacronym{UE}{UE}{User Equipment}
\newacronym{USRP}{USRP}{Universal Software Radio Peripheral}

\newacronym{3GPP}{3GPP}{3rd Generation Partnership Project}
\newacronym{PRACH}{PRACH}{Physical Random Access Channel}
\newacronym{PUCCH}{PUCCH}{Physical Uplink Control Channel}

\newacronym{MAC}{MAC}{Medium Access Control}
\newacronym{TTI}{TTI}{Transmission Time Interval}

\newacronym{DMA}{DMA}{direct memory access}
\newacronym{FIFO}{FIFO}{First In, First Out}
\newacronym{MCS}{MCS}{modulation and coding schemes}
\newacronym{RNTI}{RNTI}{Radio Network Temporary Identifier}

\newacronym{FDMA}{FDMA}{Frequency-Division Multiple Access}
\newacronym{SC-FDMA}{SC-FDMA}{Single-Carrier Frequency-Division Multiple Access}

\newacronym{DCI}{DCI}{Downlink Control Information}
\newacronym{QPSK}{QPSK}{Quadrature Phase-Shift Keying}
\newacronym{BPSK}{BPSK}{Binary Phase-Shift Keying}
\newacronym{PSK}{PSK}{Phase-Shift Keying}
\newacronym{FSK}{FSK}{Frequency-Shift Keying}

\newacronym{H2T}{H2T}{Host-to-Target}
\newacronym{T2H}{T2H}{Target-to-Host}
\newacronym{TS}{TS}{Target-Scope}

\newacronym{MIMO}{MIMO}{Multiple-Input Multiple-Output}
\newacronym{SISO}{SISO}{Single-Input Single-Output}
\newacronym{CFI}{CFI}{Control Format Indicator}

\newacronym{UDP}{UDP}{User Datagram Protocol}
\newacronym{TCP}{TCP}{Transmission Control Protocol}

\newacronym{TDM}{TDM}{Time-Division Multiplexing}
\newacronym{FDM}{FDM}{Frequency-Division Multiplexing}

\newacronym{TDD}{TDD}{Time-Division Duplexing}

\newacronym{PSSS}{PSSS}{Primary Sidelink Synchronization Signal}
\newacronym{SSSS}{SSSS}{Secondary Sidelink Synchronization Signal}

\newacronym{PSSCH}{PSSCH}{Physical Sidelink Shared Channel}
\newacronym{PSCCH}{PSCCH}{Physical Sidelink Control Channel}
\newacronym{PSDCH}{PSDCH}{Physical Sidelink Discovery Channel}
\newacronym{PSBCH}{PSBCH}{Physical Sidelink Broadcast Channel}

\newacronym{BLER}{BLER}{Block Error Rate}
\newacronym{CDL}{CDL}{Clock-Driven Logic}
\newacronym{CP}{CP}{Cyclic Prefix}
\newacronym{RF}{RF}{Radio Frequency}
\newacronym{LLR}{LLR}{Log-Likelihood Ratio}
\newacronym{CRC}{CRC}{Cyclic Redundancy Check}

\newacronym{DTP}{DTP}{Data Transmission Protocol}
\newacronym{SINR}{SINR}{Signal-to-Interference-Plus-Noise Ratio}

\newacronym{AGC}{AGC}{Automatic Gain Control}
\newacronym{AMC}{AMC}{Adaptive Modulation and Coding}
\newacronym{QoS}{QoS}{Quality of Service}
\newacronym{CSI}{CSI}{channel state information}
\newacronym{VANET}{VANET}{Vehicular Ad Hoc Network}

\newacronym{WARP}{WARP}{Wireless Open Access Research Platform}
\newacronym{CREW}{CREW}{Cognitive Radio Experimentation World}

\newacronym{DCF}{DCF}{Distributed Coordination Function}
\newacronym{ISM}{ISM}{Industrial, Scientific, and Medical}

\newacronym{UMTS}{UMTS}{Universal Mobile Telecommunications System}
\newacronym{GSM}{GSM}{Global System for Mobile Communications}
\newacronym{UTRAN}{UTRAN}{UMTS Terrestrial Radio Access Network}
\newacronym{E-UTRAN}{E-UTRAN}{Evolved UTRAN}
\newacronym{RAN}{RAN}{Radio Access Network}
\newacronym{EPC}{EPC}{Evolved Packet Core}
\newacronym{SAE}{SAE}{System Architecture Evolution}
\newacronym{EPS}{EPS}{Evolved Packet System}
\newacronym{WCDMA}{WCDMA}{Wideband Code-Division Multiple Access}
\newacronym{ITU}{ITU}{International Telecommunication Union}
\newacronym{LAA}{LAA}{Licensed Assisted Access}
\newacronym{VoLTE}{VoLTE}{Voice over LTE}
\newacronym{IMS}{IMS}{IP Multimedia Subsystem}
\newacronym{PDN}{PDN}{Private Data Network}

\newacronym{HSS}{HSS}{Home Subscriber Server}
\newacronym{P-GW}{P-GW}{Packet Data Network Gateway}
\newacronym{S-GW}{S-GW}{Serving Gateway}
\newacronym{MME}{MME}{Mobility Management Entity}
\newacronym{eNodeB}{eNodeB}{evolved Node B; also: \emph{eNB}}
\newacronym{OSI}{OSI}{Open Systems Interconnection}
\newacronym{RRC}{RRC}{Radio Resource Control}
\newacronym{IP}{IP}{Internet Protocol}
\newacronym{RLC}{RLC}{Radio Link Control}
\newacronym{PDCP}{PDCP}{Packet Data Convergence Protocol}

\newacronym{SL-SCH}{SL-SCH}{Sidelink Shared Channel}
\newacronym{SL-DCH}{SL-DCH}{Sidelink Discovery Channel}

\newacronym{DL-SCH}{DL-SCH}{Downlink Shared Channel}
\newacronym{UL-SCH}{UL-SCH}{Uplink Shared Channel}
\newacronym{RACH}{RACH}{Random Access Channel}
\newacronym{QAM}{QAM}{Quadrature Amplitude Modulation}
\newacronym{UCI}{UCI}{Uplink Control Information}
\newacronym{C-RNTI}{C-RNTI}{Cell-RNTI}
\newacronym{SSS}{SSS}{Secondary Synchronization Signal}
\newacronym{ARQ}{ARQ}{Automatic Repeat Request}
\newacronym{WLAN}{WLAN}{Wireless Local Area Network}
\newacronym{ProSe}{ProSe}{Proximity-Based Services}
\newacronym{PAPR}{PAPR}{Peak-to-Average Power Ratio}

\newacronym{IoT}{IoT}{Internet of Things}
\newacronym{NR}{NR}{New Radio}

\newacronym{5G}{5G}{fifth-generation}
\newacronym{3G}{3G}{third-generation}
\newacronym{4G}{4G}{fourth-generation}

\newacronym{AFW}{AFW}{application framework}
\newacronym{NI}{NI}{National Instruments}
\newacronym{PCI-E}{PCI-E}{Peripheral Component Interconnect Express}
\newacronym{PXI}{PXI}{PCI eXtensions for Instrumentation}

\newacronym{NIC}{NIC}{network interface card}
\newacronym{RT}{RT}{real time}
\newacronym{ICP}{ICP}{interprocess communication protocol}

\newacronym{PARADIS}{PARADIS}{Passive RAdiometric Device Identification System}
\newacronym{COTS}{COTS}{commercial off-the-shelf}

\newacronym{TDoA}{TDoA}{time difference of arrival}

\newacronym{AP}{AP}{access point}
\newacronym{SSID}{SSID}{service set identifier}

\newacronym{AES}{AES}{advanced encryption standard}

\newacronym{TSF}{TSF}{Time Synchronization Function}

\newacronym{TOF}{TOF}{time of flight}
\newacronym{SFO}{SFO}{sampling frequency offset}
\newacronym{FDD}{FDD}{frame detection delay}

\newacronym{PUFs}{PUFs}{physical uncloneable functions}

\newacronym{PoC}{PoC}{proof-of-concept}

\newacronym{STF}{STF}{short training field}
\newacronym{LTF}{LTF}{long training field}
\newacronym{LUT}{LUT}{lookup-table}

\newacronym{MAE}{MAE}{mean absolute error}

\newacronym{FCS}{FCS}{frame check sequence}

\newacronym{nlos}{NLOS}{non-line-of-sight}
\newacronym{SAP}{SAP}{service access point}

\newcommand{\Alg}{\texttt{RF-Veil}}

\newcommand{\AlgSt}{\texttt{RF-Veil-Standalone}}
\newcommand{\Tool}{\texttt{RF-Scope}}

\usepackage{enumitem}
\usepackage{etex}
\usepackage{subcaption}
\usepackage{algorithm}
\usepackage{amsfonts}
\usepackage{amsthm}
\usepackage{amsmath}
\usepackage{array}
\usepackage{epsfig}
\usepackage{float}
\usepackage[T1]{fontenc}
\usepackage{graphicx}
\usepackage{indentfirst}
\usepackage{multicol}
\usepackage{mdframed}
\usepackage{pgfplots}
\usepackage{tikz}
\usepackage{rotating}
\usepackage{booktabs}

\usepackage[thinlines]{easytable}

\usepackage{array}
\newcolumntype{L}[1]{>{\raggedright\let\newline\\\arraybackslash\hspace{0pt}}m{#1}}
\newcolumntype{C}[1]{>{\centering\let\newline\\\arraybackslash\hspace{0pt}}m{#1}}
\newcolumntype{R}[1]{>{\raggedleft\let\newline\\\arraybackslash\hspace{0pt}}m{#1}}

\usepackage{multirow}
\usepackage{color}
\usepackage{blkarray} 
\usepackage{tikz}
\usepgfplotslibrary{fillbetween}
\usetikzlibrary{chains}
\usetikzlibrary{positioning, patterns}
\pgfplotsset{compat=newest,
	/pgfplots/ybar legend/.style={
		/pgfplots/legend image code/.code={
			\draw[##1,/tikz/.cd,bar width=1pt, yshift=-0.45em, bar shift=1pt]
			plot coordinates {(0cm,0.8em)};
		},
	},
}
\pgfplotsset{
	grid style = {
		dash pattern = on 0.025mm off 0.5mm,
		line cap = round,
		black,
		line width = 1pt
	}
}

\usepgfplotslibrary{fillbetween, statistics}

\DeclareMathOperator*{\argmin}{\arg\!\min} 
\DeclareMathSizes{10pt}{8pt}{6pt}{5pt} 	

\usepackage{todonotes}

\newmdtheoremenv{prop}{Property}

\newcommand{\sref}[1]{Section~\ref{#1}}

\newcommand{\fref}[1]{Fig.~\ref{#1}}

\usepackage{collcell}
\usepackage{hhline}
\usepackage{pgf}
\usepackage{multirow}
\usepackage{colortbl}

\def\colorModel{hsb} 

\newcommand\ColCell[1]{
	\pgfmathparse{#1<0.3?1:0}  
	\ifnum\pgfmathresult=0\relax\color{white}\fi
	\pgfmathsetmacro\compA{0}      
	\pgfmathsetmacro\compB{#1} 
	\pgfmathsetmacro\compC{1}      
	\edef\x{\noexpand\centering\noexpand\cellcolor[\colorModel]{\compA,\compB,\compC}}\x #1
} 
\newcolumntype{E}{>{\collectcell\ColCell}m{4mm}<{\endcollectcell}}  

\makeatletter
\def\@headfootfont{\sffamily \fontsize{8}{8}\selectfont}
\makeatother
\makeatletter
\let\@authorsaddresses\@empty
\makeatother

\begin{document}

\title{Stay Connected, Leave no Trace: Enhancing Security and Privacy in WiFi via Obfuscating Radiometric Fingerprints}


\author{Luis F. Abanto-Leon}
\authornote{Both authors contributed equally to this research.}
\email{labanto@seemoo.tu-darmstadt.de}
\affiliation{%
  \institution{TU Darmstadt, Secure Mobile Networking Lab (SEEMOO)}
  \country{Germany}
}

\author{Andreas B{\"a}uml}
\email{lbaeuml@seemoo.tu-darmstadt.de}
\authornotemark[1]
\affiliation{%
  \institution{TU Darmstadt, Secure Mobile Networking Lab (SEEMOO)}
  \country{Germany}
}

\author{Gek Hong (Allyson) Sim}
\email{asim@seemoo.tu-darmstadt.de}
\affiliation{%
  \institution{TU Darmstadt, Secure Mobile Networking Lab (SEEMOO)}
  \country{Germany}
}

\author{Matthias Hollick}
\email{mhollick@seemoo.tu-darmstadt.de}
\affiliation{%
  \institution{TU Darmstadt, Secure Mobile Networking Lab (SEEMOO)}
  \country{Germany}
}

\author{Arash Asadi}
\affiliation{%
  \institution{TU Darmstadt, Wireless Communication and Sensing Lab (WiSe) \& SEEMOO}
  \country{Germany}
  }


\begin{abstract}
	
	The intrinsic hardware imperfection of WiFi chipsets manifests itself in the transmitted signal, leading to a unique radiometric fingerprint. This fingerprint can be used as an additional means of authentication to enhance security. In fact, recent works propose practical fingerprinting solutions that can be readily implemented in commercial-off-the-shelf devices. In this paper, we prove analytically and experimentally that these solutions are highly vulnerable to impersonation attacks. We also demonstrate that such a unique device-based signature can be abused to violate privacy by tracking the user device, and, as of today, users do not have any means to prevent such privacy attacks other than turning off the device.
	
	
	We propose \Alg{}, {\it\textbf{ a radiometric fingerprinting solution that not only is robust against impersonation attacks but also protects user privacy by obfuscating the radiometric fingerprint of the transmitter for non-legitimate receivers}}. Specifically, we introduce a {\it \textbf{randomized pattern of phase errors}} to the transmitted signal such that only the intended receiver can extract the original fingerprint of the transmitter. In a series of experiments and analyses, we expose the vulnerability of adopting naive randomization to statistical attacks and introduce countermeasures. Finally, we show the efficacy of \Alg{} experimentally in protecting user privacy and enhancing security. More importantly, our proposed solution allows communicating with other devices, which do not employ \Alg{}.

	
	
	
\end{abstract}


\makeatletter
\def\@headfootfont{\sffamily \fontsize{8}{8}\selectfont}
\makeatother

\maketitle

\section{Introduction}
\label{s:intro}

\setlength{\abovedisplayshortskip}{2pt}
\setlength{\belowdisplayshortskip}{2pt}
\setlength{\abovedisplayskip}{2pt}
\setlength{\belowdisplayskip}{2pt}
\setlength{\jot}{1pt}

The omnipresence of WiFi devices in our daily lives demands {\it strong and quantifiable security and privacy} mechanisms to protect us from attackers. WiFi security mechanisms traditionally reside above the physical layer. This can be augmented by using physical layer characteristics (e.g., channel fading, interference, hardware impairments), which further enhance the security of WiFi. In fact, physical layer security gained momentum after a chain of acute vulnerabilities rendered these {\it high-layer} security mechanisms unsecure. This includes the disastrous RC4 vulnerability in WEP~\cite{fluhrer:2001} as well as the more recent attacks on WPA2 (e.g., KRACK~\cite{vanhoef:2017} and Kr00k~\cite{cermak:2020}). We have also witnessed a variety of masquerading attacks in which the adversary mounts a machine-in-the-middle (MitM) attack by creating a rogue access point (AP), mimicking the identity (i.e., SSID) of a legitimate AP. It has been shown that physical layer security, in particular, radiometric (radio frequency) fingerprinting can thwart such attacks~\cite{liu:2019, brik:2008, li2019location}. 

\begin{figure*}[t!]
	\begin{subfigure}{0.55\textwidth}
		\vspace{-14mm}
		\centering
		\input{figs/fig_different_phones.tex}
		\vspace{-9mm}
		\caption{}
		\label{fig_differentPhones}
	\end{subfigure}
	\hspace{2mm}
	\begin{subfigure}{0.4\textwidth}
		\centering
		\newcommand\items{5}   
		\arrayrulecolor{gray} 
		\fontsize{6}{6}\selectfont
		\begin{tabular}{m{0.75mm} m{5mm} *{\items}{E}}
			\toprule
			\multicolumn{1}{c}{} &\multicolumn{1}{c}{} &\multicolumn{\items}{c}{{\bf Predicted}} \\ 
			\cmidrule(r){3-7}		
			\multicolumn{1}{c}{} 	&\multicolumn{1}{c}{} & 
			\multicolumn{1}{c}{Phone} & 
			\multicolumn{1}{c}{Phone} & 
			\multicolumn{1}{c}{Phone} & 
			\multicolumn{1}{c}{Phone} &
			\multicolumn{1}{c}{Phone} 
			\\ 
			\multicolumn{1}{c}{} 	& \multicolumn{1}{c}{} & 
			\multicolumn{1}{c}{1} & 
			\multicolumn{1}{c}{2} & 
			\multicolumn{1}{c}{3} & 
			\multicolumn{1}{c}{4} &
			\multicolumn{1}{c}{5} 
			\\ \hline\hline \\[-0.5em]
			\multirow{9}{*}{\rotatebox{90}{{\bf Actual}}} 
			& \centering{Phone 1}	& 0.985 	& 0.0 	& 0.0 	& 0.0 & 0.0   \\		
			& \centering{Phone 2}  	& 0.0 & 0.985 & 0.085 & 0.0 & 0.0   \\ 
			& \centering{Phone 3}  	& 0.0 & 0.115 & 0.965 & 0.0 & 0.0   \\  
			& \centering{Phone 4}  	& 0.0 & 0.0 & 0.025 & 1.0 & 0.0   \\ 
			& \centering{Phone 5}  	& 0.010 & 0.0 & 0.0 & 0.0 & 0.995   \\ 
		\end{tabular}
		\vspace{-2mm}
		\caption{}
		\label{fig_differentPhones_table}
	\end{subfigure}
	\vspace{-3mm}
	\caption{WiFi radiometric fingerprints of 5 identical phone (Samsung Galaxy S6). \fref{fig_differentPhones} shows that the  \textit{fingerprints differ from one another even though the chipsets belong to the same series and manufacturer}, thus allowing to distinguish among multifarious devices. \fref{fig_differentPhones_table} shows that \textit{the devices can be distinguished with $ 96.5\% $ accuracy using a simple mean absolute error (MAE)-based classifier} (MAE threshold $= 4.5^{\circ} $).} 
	\vspace{-3mm}	 
\end{figure*}



{\it Radiometric fingerprinting techniques rely on measuring and extracting device-specific imperfections of the transmitter RF circuitry embedded in the emitted signal, which manifest in form of negligible but distinguishable errors, e.g., in phase (e.g., \cite{liu:2019}) or frequency (e.g., \cite{hua:2018}).} These imperfections are so individualized that even chipsets from the same manufacturer have different fingerprints~\cite{liu:2019, brik:2008}. 
In \fref{fig_differentPhones}, we demonstrate that the radiometric fingerprint\footnote{These fingerprints are extracted from non-linear phase errors derived from device-specific hardware imperfections~\cite{liu:2019}} of five identical phones with the same WiFi chipset are visually distinguishable. Thus, it is not surprising that these devices can be easily differentiated from one another with high success ratio (i.e., $ 96.5 \% $). 
Such degree of accuracy, on the one hand, reveals the {\it potential of radiometric fingerprinting for achieving accurate authentication, thus enhancing security}. On the other hand, it raises {\it major privacy concerns since adversaries can locate/track devices} using these unique fingerprints. Our work is motivated by the potential of radiometric fingerprinting in coping with security and privacy challenges.


{\bf Challenge I: Privacy.} 
Any unique identifier which can be easily measured/accessed by an adversary poses a significant privacy threat. Indeed, this is the motivation behind MAC address randomization in WiFi or temporary identifiers in cellular networks to prevent potential adversaries from tracking users. Radiometric fingerprints expose users to the same privacy vulnerability, and as of today, users do not have any means to prevent such privacy attacks other than turning off the device. While randomizing the physical layer characteristics of the signal is a plausible solution to enhance privacy, such procedure may degrade the communication link and disrupt or prevent legitimate radiometric fingerprinting, which brings us to the next challenges.

{\bf Challenge II: Security.} 
Radiometric fingerprints are typically considered a secure anchor for device authentication. Still, they are {\it collectible by anyone in the vicinity} of the transmitter who is capable of "overhearing" the packets, e.g., 50-100 meters for WiFi. This {\it exposes the fingerprinting methods to impersonation attacks.} Initial proposals argued that the cost of mimicking the fingerprints is too high~\cite{robyns2017noncooperative}. To date, a wide range of software-defined radios (SDRs) costing from a few hundred (up to a few thousand) euros can collect and forge the fingerprints of other devices, e.g., through modifying the phases of emitted signals, as shown in Section \ref{s3-2}. This issue is further exacerbated by the emergence of WiFi firmware patching tools~\cite{nexmon:project}, which enables commercial WiFi chipsets to shape their signals and impersonate other devices.


{\bf Challenge III: Allowing for legitimate radiometric fingerprinting.} 
There are several solutions to hide ones' fingerprint: {\it (i)} {\it Jamming}, which defeats the primary purpose of WiFi, i.e., communication; {\it (ii) } {\it Constructive interference}. The seminal work of Oh {\it et al.} on location privacy~\cite{oh2012phantom} and recent literature on privacy against WiFi sensing~\cite{yao2018aegis, qiao2016phycloak} use coordinated transmissions or a secondary signal repeater to obfuscate the physical layer information, which are not scalable and can be costly due to reliance on secondary devices; {\it (iii)}~{\it Fingerprint randomization at the transmitter} has the advantage of scalability, but it can disrupt the communication link by distorting the channel estimation at the receiver. In \cite{cominelli2020:experimental-study-csi-location-privacy}, the authors randomize the transmitted signal to obfuscate device-free localization but their approach introduces marginal impact on the quality of the communication. Furthermore, we must ensure that the randomization is reversible to allow legitimate fingerprinting.
\subsection{Our approach} 

In this paper, we propose \Alg{}, a scalable approach that enhances the user privacy by obfuscating the radiometric fingerprints of the device from adversaries while allowing the use of \gls{CSI}-based fingerprinting at legitimate receivers to strengthen the security of the network.

In essence, {\it\Alg{} adds a crafted randomized phase noise to the signal at the transmitter such that the radiometric fingerprints are obfuscated, but the quality of communication remains intact. Furthermore, we facilitate fingerprint extraction through a low-overhead synchronized random noise generation process between legitimate transmitters and receivers.}
The properties of \Alg{} are:


{\bf Privacy-preserving}. 
The latest radiometric fingerprinting solutions extract device-specific phase errors from the \gls{CSI} \cite{liu:2019, hua:2018}. 
\Alg{} introduces {\it deliberate phase noise to the subcarriers in the OFDM symbols} on a per-frame basis such that the adversary can no longer estimate the actual radiometric fingerprint by analyzing the \gls{CSI}, thus preventing the device identification/tracking via radiometric fingerprint.

{\bf Secure against impersonation.} 
We strive to maintain the possibility of legitimate fingerprinting without exposing the user to impersonation attacks. To this aim, we first devise a technique (synchronized phase noise generation), which enables only the legitimate receivers to denoise the transmitted signals and extract the original fingerprint. Secondly, we apply the obfuscation on a per-frame basis to eliminate the possibility of impersonation or reply attack via over-the-air packet sniffing. The effectiveness of this method is proven both theoretically and experimentally, even in presence of sophisticated adversaries with the capability of realizing statistical attacks.


{\bf Dual mode.} 
\Alg{} is designed to allow the legitimate use of wireless fingerprinting techniques (e.g., for authentication as in~\cite{liu:2019}) in presence of our obfuscation method. 
Furthermore, a reduced form of \Alg{} can be used to obfuscate the fingerprint of the device in order to only protect the device's privacy when fingerprinting is not used as an additional security feature, i.e., reversing the phase noise is not required.  {\it We refer to this second operational mode as \AlgSt{}}. In this mode, we can hide the fingerprint of the transmitter by executing the obfuscation blocks without any handshake or coordination with other receivers. As a result, we can ensure privacy protection in a much broader scenario, e.g., communicating with non-\Alg{}-enabled devices, in absence of any active connections, or in connection establishment phase.

{\bf Low-overhead and scalable.} 
\Alg{} has low overhead from both computational and signaling/control message perspective. 
Our simple yet effective obfuscation technique enabled extraction of \gls{CSI}-based radiometric fingerprints at the legitimate receiver without any additional complex signal processing. Furthermore, \Alg{} is highly scalable since it is implemented directly at the transmitter and does not rely on any secondary device~\cite{yao2018aegis, qiao2016phycloak}. Therefore, any WiFi device can obfuscate its fingerprint easily and independently. 

\vspace{-2mm}
\subsection{Our contributions} 


To the best of our knowledge, {\it this is the first work exposing privacy and security vulnerabilities of radiometric fingerprints as well as devising practical methods to resolve them.} Note that prior works such as~\cite{rahbari2015secrecy, rahbari2014friendly} propose techniques for preventing the exposure of unencrypted fields (e.g., headers and payload information) to counter, for instance, reactive jamming attacks that can adapt to the rate of transmission. While the approach prevents data analysis and acquisition of transmission cues, it does not protect the radiometric fingerprint of the device. The following summarizes our main contributions:
{\it (i)} Showing vulnerabilities of recent \gls{CSI}-based radiometric fingerprinting solutions~\cite{liu:2019, xie2018precise,brik:2008} to impersonation attacks (\sref{s_adversary_model});
{\it (ii)} Proposing a method for injecting artificial noise to the fingerprint without impacting communication quality (\sref{s4}). 
{\it (iii)} Designing \Tool{}, a benchmarking tool to assess the effectiveness of radiometric fingerprint obfuscation against statistical attacks (\sref{Vulnerability_analysis}). Specifically, \Tool{} is a maximum-likelihood-based estimator of the \gls{CSI}, which we prove to be near-optimal through derivation of Cramer-Rao bounds (Appendix~\ref{s:crb});
{\it (iv)}  Devising \Alg{}, a fingerprint obfuscation framework that circumvents the privacy issues without impacting the communication quality (\sref{s_algorithm}). 
{\it (v)} We prove the efficacy of our proposals, both analytically and experimentally.

\section{Fingerprinting Primer} \label{s_fingerprintin_primer}
Radio signal analysis to identify devices and distinguish between friends and foes dates back to the time of the Vietnam war. In the same line, \textit{radiometric fingerprinting} has gained momentum in recent years with the surge of attacks that leverage hardware impairments to breach privacy and security in wireless networks. Recently, \gls{CSI}-based radiometric fingerprinting gained popularity due to the availability of \gls{CSI} extraction tools~\cite{gringoli:2019, halperin2011tool, xie2018precise}. These tools allow  per-frame \gls{CSI} collection from commercial WiFi chipsets (e.g., Intel, Qualcomm, Broadcom), making \gls{CSI}-based fingerprinting practical and feasible for all devices. In the following, a short overview of \gls{CSI} estimation and \gls{CSI}-based fingerprinting is provided.


\subsection{Channel estimation in WiFi} \label{s3-1}
\begin{figure} [t!]
	\vspace{-1mm}
	\centering
	\includegraphics[width=0.8\textwidth]{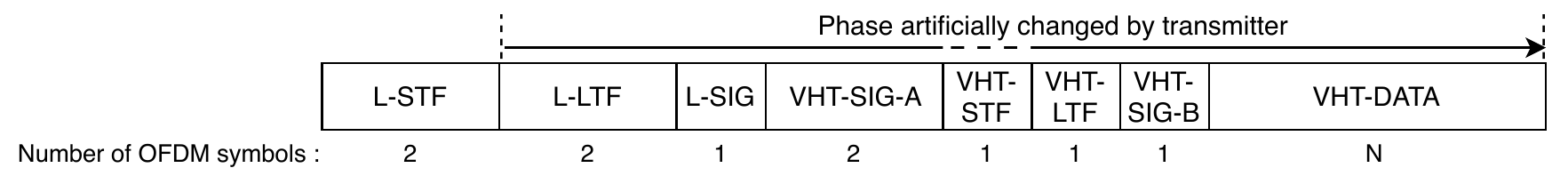}
	\vspace{-2mm}	
	\caption{IEEE 802.11ac PHY frame format. 
	} 
	\label{fig_phy-frame}
	\vspace{-4mm}
\end{figure}

As a prelude to \gls{CSI}-based fingerprinting, we describe channel estimation in WiFi throughout this section. \fref{fig_phy-frame} shows the IEEE 802.11ac PHY frame structure, wherein we recognize four distinct fields: \gls{STF}, \gls{LTF}, SIG, and DATA (cf. 17.3 in \cite{80211}). The receiver uses the STF field for signal detection, automatic gain control, time synchronization, and coarse \gls{CFO}. The LTF field is employed for fine CFO estimation and channel estimation. Channel estimation is performed by sending BPSK pilots over the LTF subcarriers of two consecutive OFDM symbols. The SIG and DATA fields convey the MCS level and the payload, respectively. The OFDM symbols in the SIG and DATA fields are equalized using the channel estimated by the preceding LTF field. The prefix "L-" denotes the legacy fields, which are included for compatibility with IEEE 802.11a.

Let $ K $ denote the number of subcarriers and $ \mathbf{s} = \left[ s_1, \cdots, s_K \right]^T \in \mathbb{C}^{K \times 1} $ the BPSK pilot symbols (defined in Equation 19-23 of \cite{80211}). Also, let $ \mathbf{F} \in \mathbb{C}^{K \times K} $ and $ \mathbf{F}^H \in \mathbb{C}^{K \times K} $ denote the discrete Fourier transform (DFT) matrix and the inverse DFT (IDFT) matrix, respectively. Moreover, $ \mathbf{F} \mathbf{F}^H = \mathbf{I} $, with $ \left( \cdot \right)^H $ representing the Hermitian transpose. The discrete-time OFDM symbol is given by $ \mathbf{x} = \mathbf{F}^H \mathbf{s} $ \cite{abanto2020:swarm-based-low-complexity-papr-reduction}. In order to improve the signal robustness against multi-path interference, the periodic OFDM symbol $ \tilde{\mathbf{x}} $ is produced by appending the cyclic prefix (CP) to $ \mathbf{x} $. The CP consists of the last $ L $ samples of $ \mathbf{x} $, thus $ \tilde{\mathbf{x}} = \left[ \mathbf{x}^T_{\left[ K-L+1:K \right]},  \mathbf{x}^T  \right]^T \in \mathbb{C}^{(K+L) \times 1} $. Appending the CP to $ \mathbf{x} $ transforms the linear convolution between $ \tilde{\mathbf{x}} $ and the channel $ \mathbf{c} = \left[ c_1, \cdots, c_J \right]^T \in \mathbb{C}^{J \times 1} $  (with $ J < L $ paths) into a circular convolution. This has the advantage of simplifying OFDM demodulation and equalization at the receiver. Upon transmitting $ \tilde{\mathbf{x}} $ over the channel $ \mathbf{c} $, the receiver obtains $ \tilde{\mathbf{r}} = \tilde{\mathbf{x}} \ast \mathbf{c} $, where $ \ast $ denotes the convolution operator. To remove the impact of inter-block interference between adjacent LTF frames (caused by multi-path propagation), we discard the first $ L $ elements of $ \tilde{\mathbf{r}} $, thus yielding $ \mathbf{r} = \tilde{\mathbf{r}}_{\left[ L+1:L+K \right] } \in \mathbb{C}^{K \times 1} $. The received signal $\mathbf{r} $ can be expressed as:
\begin{equation} \label{e3-1}
	\underbrace{\begin{pmatrix}
		r_1 \\
		\vdots \\
		r_K
		\end{pmatrix}}_{\mathbf{r}}
	=
	\underbrace{\begin{pmatrix}\vspace{-2.5mm}
		c_1      & 0       & \cdots & 0       & c_J      & \cdots & c_2 \\ \vspace{-1mm}
		\vdots   & \vdots  &        & \vdots  & \vdots   &        & \vdots \\ \vspace{-1mm}
		c_J      & c_{J-1} &        & 0       & 0        &        & 0 \\ \vspace{-2mm}
		0        & c_J     &        & 0       & 0        &        & 0 \\\vspace{-2mm}
		\vdots   & \vdots  &        & \vdots  & \vdots   &        & \vdots \\ \vspace{-2mm}
		0        & 0       & \cdots & c_J     & c_{J-1}  & \cdots & c_1 \\\vspace{-4mm}
		\end{pmatrix}}_{\mathbf{C}}
	\underbrace{\begin{pmatrix}
		x_1 \\
		\vdots \\
		x_K
		\end{pmatrix}}_{\mathbf{x}}
	+
	\underbrace{\begin{pmatrix}
		w_1 \\
		\vdots \\
		w_K
		\end{pmatrix},}_{\mathbf{w}}
\end{equation}
where $ \mathbf{w} \sim \mathcal{CN} \left( 0, \sigma^2 \mathbf{I} \right) $ denotes circularly-symmetric complex Gaussian noise. The receiver demodulates the received signal $ \mathbf{r} $ by multiplying it with $ \mathbf{F} $ to obtain $ \mathbf{y} = \mathbf{F} \mathbf{r} = \mathbf{F} \mathbf{C} \mathbf{x} + \mathbf{F} \mathbf{w} $. The convolution matrix $ \mathbf{C} \in \mathbb{C}^{K \times K} $ is circulant, which is a consequence of adding the CP to the transmitted signal. Circulant matrices can be expressed via eigen-decomposition as $ \mathbf{C} = \mathbf{F}^H \mathbf{H} \mathbf{F} $, where $ \mathbf{H} = \mathrm{diag} \left( \left[ h_1, \cdots, h_K \right] \right) $ represents a diagonal matrix containing the eigenvalues of $ \mathbf{C} $ \cite{gray2006:toeplitz-circulant-matrices-review}. As a result, the demodulated signal $ \mathbf{y} $ collapses to $ \mathbf{y} = \mathbf{F} \big( \mathbf{F}^H \mathbf{H} \mathbf{F} \big) \big( \mathbf{F}^H \mathbf{s} \big) + \mathbf{F}^H \mathbf{w} = \mathbf{H} \mathbf{s} + \mathbf{w} $. More specifically, 
\begin{equation} \nonumber
		\underbrace{\begin{pmatrix}
		y_1 \\
		\vdots \\
		y_K
		\end{pmatrix}}_{\mathbf{y}}
		=
		\underbrace{\begin{pmatrix}
			h_1      & \cdots 	& 0       \\
			\vdots   & \ddots     & \vdots  \\
			0        & \cdots       	& h_K    
		\end{pmatrix}}_{\mathbf{H}}
		\underbrace{\begin{pmatrix}
			s_1 \\
			\vdots \\
			s_K
		\end{pmatrix}}_{\mathbf{s}}
		+
		\underbrace{\begin{pmatrix}
			w_1 \\
			\vdots \\
			w_K
		\end{pmatrix}.}_{\mathbf{w}}
\end{equation}

Thus, for any subcarrier $ k \in \mathcal{K} = \left\lbrace 1, \cdots, K \right\rbrace $, the received symbol is expressed as
\begin{equation} \label{e3-2}
y_k = h_k s_k + w_k = \left| h_k \right| e^{j \phi_k} s_k + w_k,
\end{equation}
which shows that the channel affects each pilot symbol $ s_k $ by a complex-valued factor $ h_k = \left| h_k \right| e^{j \phi_k} $ and additive noise $ w_k $. Since the pilot symbols $ \mathbf{s} $ are known by the receiver, the \gls{CSI} vector $ \mathbf{h} = \left[ h_1, \cdots, h_K \right]^T $ can be obtained upon equalizing each received symbol $ y_k $ with the compensation factor $ \frac{s^*_k}{\left| s_k \right|^2 } $. Thus, the estimated channel in subcarrier $ k $ is given by:
\begin{equation} \label{e3-3}
\tilde{h}_k = h_k s_k \frac{s^*_k}{\left| s_k \right|^2} + w_k \frac{s^*_k}{\left| s_k \right|^2} = \left| h_k \right| e^{j \phi_k} + w_k.
\end{equation}

\subsection{\gls{CSI}-based fingerprinting} \label{s3-2}


%
%
{\it \gls{CSI}-based radiometric fingerprinting techniques consist of analyzing the \gls{CSI} to extract features that are unique to the transmitting device.}
Specifically, Zhuo \textit{et al.} \cite{zhuo:2017} found that WiFi chipsets exhibit non-linear phase errors that change across subcarriers and are analogous to a sinusoidal function, as shown in \fref{fig_differentPhones}. These phase errors are caused by I/Q imbalance as a result of hardware imperfections. It was shown in \cite{zhuo:2017} that these errors are latent signatures that can be extracted upon removing the linear phase errors from the \gls{CSI}. Building on this finding, Liu \textit{et al.} \cite{liu:2019} harness these non-linear errors as \gls{CSI}-based \emph{radiometric fingerprints} for identification, thus preventing impersonation by unauthorized WiFi devices. Following the same notation in (\ref{e3-2}) and (\ref{e3-3}), we denote the \gls{CSI} phases by $ \boldsymbol{\Phi} = \left[ \phi_1, \cdots, \phi_K \right]^T $, which can be further decomposed into
\begin{equation} \label{e3-4}
\boldsymbol{\Phi} ~ = ~ \underbrace{\boldsymbol{\varphi} ~ + ~ \boldsymbol{\omega} ~ + ~ \boldsymbol{\theta} ~ + ~ \boldsymbol{\psi}}_\text{linear errors} ~~~ + \underbrace{\boldsymbol{\epsilon}}_{\substack{\text{non-linear error (fingerprint)}}},
\end{equation}
where $\boldsymbol{\varphi} \in \mathbb{R}^{K \times 1} $ represents the phase of the signal at the transmitter while $ \boldsymbol{\omega} \in \mathbb{R}^{K \times 1} $, $ \boldsymbol{\theta} \in \mathbb{R}^{K \times 1}$ and $\boldsymbol{\psi} \in \mathbb{R}^{K \times 1} $ denote the phase errors due to sampling frequency offset, frame detection delay and time of flight, respectively. By using the mirror subcarriers, the linear part of the phase errors can be canceled \cite{liu:2019}. Hence, the non-linear phase errors $ \boldsymbol{\epsilon} \in \mathbb{R}^{K \times 1} $ are obtained by the following equation
\begin{equation} \label{e3-5}
\boldsymbol{\epsilon} =  \boldsymbol{\Phi} - \left( 2 \pi \lambda \cdot \mathbf{v} + \mathbf{1} Q^\star \right),
\end{equation}
where $ \mathbf{v} = \big[ -K/2, \cdots, -1, 1, \cdots, K/2 \big]^T $ 
and $ \lambda $ is a constant used for nullifying the linear phase rotation in a specific frame whereas $ Q^\star $ is used for phase error normalization \cite{liu:2019}.


The authors show that the non-linear phase errors exhibit both time and location invariance and change significantly even across devices of the same manufacturer. As a result, these non-linear phase errors can be used as highly distinctive radiometric fingerprints for device identification by leveraging the above described approach in \cite{liu:2019}. Even though the difference is very small, the phones are distinguishable from one another, as illustrated in \fref{fig_differentPhones_table}. As a result, the authors conclude that these fingerprints can be used as countermeasures against impersonation attacks. However, we show in the next section that impersonation is indeed possible. 


\section{Adversary Model and Attack Scenario} \label{s_adversary_model}
In this section, we introduce the adversary model and devise two attack scenarios, which aim at breaching privacy and security.

\subsection{Adversary model} 

\begin{figure*}[t!]
\begin{minipage}{0.52\textwidth}
	\centering
	\includegraphics[width=\textwidth]{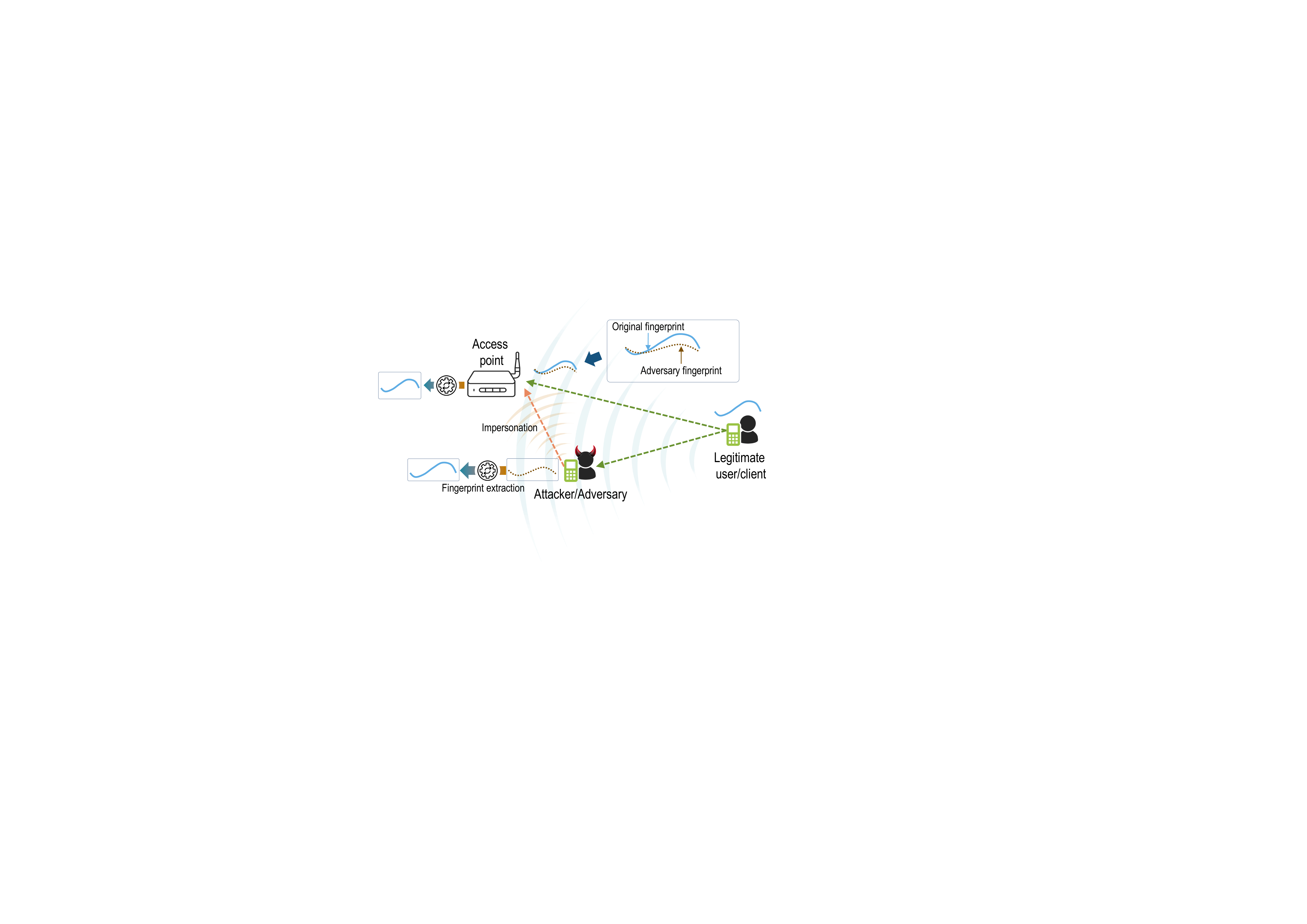}
	\vspace{-5mm}
	\caption{Adversary model. \textit{We consider an impersonation scenario, where the adversary has captured the fingerprint of the victim (legitimate user). The adversary forges the fingerprint of the victim by introducing additional phase rotations to its own fingerprint.}}
	\label{fig_adversaryModel}
\end{minipage}
\hspace{2mm}
\begin{minipage}{0.44\textwidth}
\vspace{-1mm}

\begin{tikzpicture}[scale=0.55]
\centering
\usetikzlibrary{patterns}
\scriptsize
\definecolor{orange}{RGB}{255,127,0}

\definecolor{p1-gt}{HTML}{396AB1}
\definecolor{p2-gt}{HTML}{DA7C30}
\definecolor{p3-gt}{HTML}{006d2c}
\definecolor{p4-gt}{HTML}{a50f15}
\definecolor{p5-gt}{HTML}{535154}

\draw[->] (0,0.4) -- (10.5*0.8,0.4);
\node at (3.5,-0.35) {time[minutes]};
\draw	(0,0.4) node[anchor=north] {0}
		(1*0.8,0.4) node[anchor=north] {1}
		(2*0.8,0.4) node[anchor=north] {2}
		(3*0.8,0.4) node[anchor=north] {3}
		(4*0.8,0.4) node[anchor=north] {4}
		(5*0.8,0.4) node[anchor=north] {5}
		(6*0.8,0.4) node[anchor=north] {6}
		(7*0.8,0.4) node[anchor=north] {7}
		(8*0.8,0.4) node[anchor=north] {8}
		(9*0.8,0.4) node[anchor=north] {9}
		(10*0.8,0.4) node[anchor=north] {10};


\draw (-0.25,1) node[anchor=east] {Phone 1}
(-0.25,2) node[anchor=east] {Phone 2}
(-0.25,3) node[anchor=east] {Phone 3}
(-0.25,4) node[anchor=east] {Phone 4}
(-0.25,5) node[anchor=east] {Phone 5};

\draw[densely dotted, p5-gt!50] (0,0.4) -- (0,5.5);
\draw[densely dotted, p5-gt!50] (1*0.8,0.4) -- (1*0.8,5.5);
\draw[densely dotted, p5-gt!50] (2*0.8,0.4) -- (2*0.8,5.5);
\draw[densely dotted, p5-gt!50] (3*0.8,0.4) -- (3*0.8,5.5);
\draw[densely dotted, p5-gt!50] (4*0.8,0.4) -- (4*0.8,5.5);
\draw[densely dotted, p5-gt!50] (5*0.8,0.4) -- (5*0.8,5.5);
\draw[densely dotted, p5-gt!50] (6*0.8,0.4) -- (6*0.8,5.5);
\draw[densely dotted, p5-gt!50] (7*0.8,0.4) -- (7*0.8,5.5);
\draw[densely dotted, p5-gt!50] (8*0.8,0.4) -- (8*0.8,5.5);
\draw[densely dotted, p5-gt!50] (9*0.8,0.4) -- (9*0.8,5.5);
\draw[densely dotted, p5-gt!50] (10*0.8,0.4) -- (10*0.8,5.5);

\draw[p5-gt, fill=p5-gt] (0,5.05) rectangle (2.02*0.8,5.4);

\draw[p4-gt, fill=p4-gt] (2.1*0.8,4.05) rectangle (4*0.8,4.4);

\draw[p3-gt, fill=p3-gt] (4.1*0.8,3.05) rectangle (5*0.8,3.4);

\draw[p2-gt, fill=p2-gt] (5.1*0.8,2.05) rectangle (10*0.8,2.4);

\draw[p1-gt, fill=p1-gt] (7.02*0.8,1.05) rectangle (8*0.8,1.4);

\draw[p5-gt, pattern=north west lines, thin, pattern color=p5-gt] (1.92*0.8,4.6) rectangle (2.027*0.8,4.95);
\draw[p5-gt, pattern=north west lines, pattern color=p5-gt] (0,4.6) rectangle (1.528*0.8,4.95);

\draw[p4-gt, pattern=north west lines, pattern color=p4-gt] (2.19*0.8,3.6) rectangle (4.03*0.8,3.95);

\draw[p3-gt, pattern=north west lines, pattern color=p3-gt] (4.29*0.8,2.6) rectangle (4.929*0.8,2.95);

\draw[p2-gt, pattern=north west lines, pattern color=p2-gt] (5.19*0.8,1.6) rectangle (10*0.8,1.95);

\draw[p1-gt, pattern=north west lines, pattern color=p1-gt] (7.029*0.8,0.6) rectangle (8.22*0.8,0.95);
\draw[p1-gt, pattern=north west lines, pattern color=p1-gt] (5.73*0.8,0.6) rectangle (5.74*0.8,0.95);



\draw[black!90, fill=white] (6.8*0.8,3.6) rectangle (9.7*0.81, 4.8);

\fill[draw=black!90, fill=black!90] (7*0.8,4.25) rectangle (7.5*0.8,4.65);
\draw[black!90, pattern=north west lines, pattern color=black!90] (7*0.8,3.75) rectangle (7.5*0.8,4.15);

\node at (8.5*0.8,4.5) {Actual};
\node at (8.7*0.8,4) {Predicted};


\end{tikzpicture}

	\caption{Adversary prediction on the presence time of $5$ different phones turned on and off at different time intervals. {\it An adversary can accurately determine the presence of a specific user in a network by tracking the radiometric fingerprints.}}
	\label{fig_privacy_attack}
\end{minipage}
\vspace{-3mm}
\end{figure*}

We consider a scenario where the legitimate device (i.e., client) communicates with an AP and vice versa, see \fref{fig_adversaryModel}. 
We further consider an adversary in transmission range of the legitimate communication with the following capabilities: 
{\it (i)} sniffing packets sent by the legitimate device; {\it (ii)} extracting the fingerprint of the legitimate device from the \gls{CSI}; 
{\it (iii)} knowing its own fingerprint and the ability to change it arbitrarily.
Hence, the adversary can breach the user privacy upon extracting the fingerprint from the sniffed packets. 
Even if the client employs MAC address randomization to remain anonymous, the adversary can identify and track the client via the radiometric fingerprint (Attack scenario I). 
Further, having the ability to change its fingerprint arbitrarily, the adversary can subsequently modify its own fingerprint to impersonate the client, thus compromising the security of the system (Attack scenario II). 
Note that we do not consider an adversary launching a denial-of-service attack by jamming the WiFi signals as this will disrupt the communication in WiFi channels as a whole. 



\subsection{Attack scenario I: Violating user privacy by tracking the radiometric fingerprints}

This attack focuses on tracking the presence of specific devices in the network using their radiometric fingerprints. In this scenario, the privacy-invading adversary silently sniffs the encrypted traffic over a WiFi network, extracts the fingerprints from the \gls{CSI}, and creates a database recording time and duration in which a device was present in the vicinity. In order to show the efficacy of this attack experimentally, we setup an adversary and $5$ phones that are entering and leaving the network at different times over the course of 10 minutes. We depict the results of this experiment in \fref{fig_privacy_attack}, in which the ground truth is presented in solid bars, whereas the hatched bars indicate the adversary's prediction. {\it We observe that the adversary is able to determine the presence time of the different phones with fairly high accuracy.} Note that MAC layer anonymization techniques cannot stop our adversary from tracking the presence of users across networks since such techniques do not conceal the inherent physical cues of the device, i.e., radiometric fingerprint. In an era where smartphones, smartwatches, and other WiFi-enabled wearables are omnipresent, these simple attacks expose us to significant privacy risks at workplace and at home.

%

\subsection{Attack scenario II: Compromising security via impersonation attacks}


\begin{figure*}[t!]
	\begin{minipage}{0.601\textwidth}
		\centering
		\includegraphics[width=\textwidth]{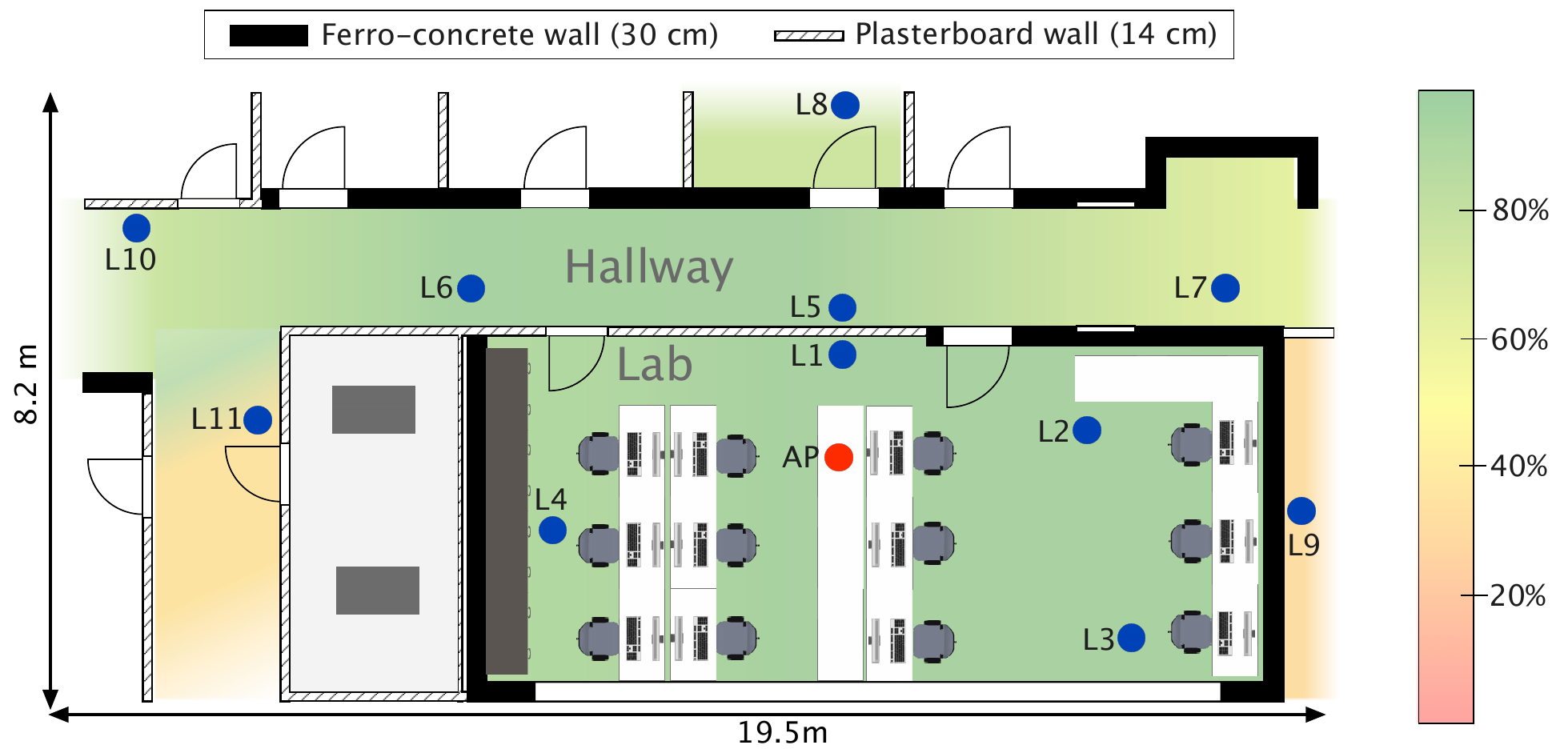}
		\fontsize{5.7}{5.7}\selectfont
		\setlength\tabcolsep{1.8pt} 
		\arrayrulecolor{black}
		\begin{tabular}{|p{8mm}|c|c|c|c|c|c|c|c|c|c|c|}
			\hline
			\textbf{Location}
			&L1 &L2 &L3 &L4 &L5 &L6 &L7 &L8 &L9 &L10 &L11 \\ \hline
			\textbf{2.4 GHz}	&$99.8\%$ &$99.6\%$ &$98.3\%$ &$88.7\%$ &$98.3\%$ &$99.4\%$ &$63.9\%$ &$76.5\%$ &$31.9\%$ &$78.3\%$ &$34.9\%$			\\ \hline
			\textbf{5 GHz}	&$95.8\%$ &$99.7\%$ &$99.5\%$ &$97.6\%$ &$99.6\%$ &$72.1\%$ &$74.9\%$ &$79.4\%$ &$0\%$ &$0\%$ &$0\%$ \\ \hline
		\end{tabular}
		\caption{The success rate of impersonation attacks is shown as a heatmap throughout the experimental site when using $2.4$ GHz. \textit{Impersonation is possible even in challenging scenarios, i.e., through a 30-centimeter thick ferro-concrete wall.} The table shows the success rate at $2.4$ and $5$ GHz bands. {\it The success rate at $5$ GHz is lower due to higher propagation and penetration loss compared to $2.4$ GHz.}} 
		\label{fig_impersonation_attack} 	
	\end{minipage}
	\hspace{2mm}
	\begin{minipage}{0.37\textwidth}
	\vspace{-5mm}
		\begin{tikzpicture}
		\definecolor{victim}{HTML}{4292c6}
		\definecolor{adversary}{HTML}{4d4d4d}
		\definecolor{forged}{HTML}{a50f15}
		
		\begin{axis}
		[
		xlabel = {Subcarrier index ($ k $)},
		ylabel = {Fingerprint [deg]},
		xmin = -29,
		xmax = 29, 
		ymin = -60, 
		ymax = 30,
		width = 5.2cm,
		height = 4.8cm,
		ymajorgrids,
    	grid style={line width=0.3pt, draw=gray!70},
		ticklabel style = {font=\fontsize{7}{7}\selectfont},
		xtick = {-28, -23, -18, -13, -8, -3, 3, 8, 13, 18, 23, 28},
		xticklabels={1, 6, 11, 16, 21, 26, 31, 36, 41, 46, 51, 56},
		ytick = {-45, -30, -15, 0, 15, 30},
		ylabel style = {text width = 5cm, align = center, font=\fontsize{7}{7}\selectfont},
		xlabel style = {align = center, font=\fontsize{7}{7}\selectfont},
		legend columns = 1,
		legend pos = north east,
		legend style = {at = {(0.4,0.015)}, anchor = south west, font = \scriptsize, fill = none,  font=\fontsize{6}{6}\selectfont, /tikz/every even column/.append style={column sep=0.03cm}, legend style={row sep=-1pt}, inner sep = 0.5pt, align = left, draw=none},
		legend cell align={left},
		]

		\addplot [color=victim, very thick]  coordinates{
		(-28, -30.587179873652076)
		(-27, -34.87433150777646)
		(-26, -37.2957568738877)
		(-25, -39.77584790777659)
		(-24, -38.44421424097001)
		(-23, -38.933642189289216)
		(-22, -41.64547993297777)
		(-21, -40.83525170788236)
		(-20, -38.240618449939475)
		(-19, -39.06906992546752)
		(-18, -36.27077847243319)
		(-17, -36.299774807946164)
		(-16, -34.32546751605907)
		(-15, -32.44540054159832)
		(-14, -31.077624226316836)
		(-13, -29.080078455830556)
		(-12, -27.012099093077268)
		(-11, -23.0113311812685)
		(-10, -21.156364868952352)
		(-9, -19.589179418676384)
		(-8, -18.383081531917878)
		(-7, -15.058370031663019)
		(-6, -14.006360634371799)
		(-5, -10.548832329790162)
		(-4, -8.640532130953174)
		(-3, -7.172557865837094)
		(-2, -5.179429732004489)
		(-1, -2.263359933706721)
		(1, 2.263359933706721)
		(2, 4.119659605586747)
		(3, 6.278267837125653)
		(4, 7.288054452902437)
		(5, 8.330268065795938)
		(6, 10.861956473523193)
		(7, 10.84914404494362)
		(8, 13.427814527322457)
		(9, 14.058669277538632)
		(10, 16.630767795609778)
		(11, 17.675356245137603)
		(12, 17.414526793149793)
		(13, 18.255458716046434)
		(14, 17.918520239272553)
		(15, 17.073683169197817)
		(16, 16.449408082227333)
		(17, 17.31346596475241)
		(18, 17.107975460251936)
		(19, 13.67873919721356)
		(20, 13.842750747968571)
		(21, 9.68507990858796)
		(22, 8.050284804151188)
		(23, 2.9720230097320215)
		(24, -1.0996425606211138)
		(25, -5.766413541194344)
		(26, -11.97375249277619)
		(27, -19.831299455286402)
		(28, -30.587179873652076)
		};
		\addlegendentry{Victim \\fingerprint}
		
		\addplot [only marks, color=forged, mark=x, mark size = 2, line width = 0.6pt] coordinates{
		(-28, -30.96491205702591)
		(-27, -34.654352879264366)
		(-26, -37.130373913862634)
		(-25, -40.0147304299201)
		(-24, -38.429631610721124)
		(-23, -39.301473094382)
		(-22, -41.75516702874225)
		(-21, -40.69139304900578)
		(-20, -38.36891527199736)
		(-19, -38.82534810769676)
		(-18, -36.49774857482267)
		(-17, -36.18417603483163)
		(-16, -33.81710057637735)
		(-15, -32.29717268547586)
		(-14, -31.187895871776167)
		(-13, -28.86020131895919)
		(-12, -26.573072602222545)
		(-11, -23.44032962564473)
		(-10, -21.14935706376476)
		(-9, -19.998991171966217)
		(-8, -18.54807709410699)
		(-7, -15.71935730499756)
		(-6, -14.297640580002597)
		(-5, -10.52184531714076)
		(-4, -8.403021498136354)
		(-3, -7.097354253000309)
		(-2, -5.039349428667424)
		(-1, -2.087300733781847)
		(1, 2.087300733781847)
		(2, 4.081852989778596)
		(3, 6.104637298379357)
		(4, 7.3168525453469355)
		(5, 8.423560778278565)
		(6, 10.53034581751104)
		(7, 10.778094472693716)
		(8, 13.575852054630694)
		(9, 13.83017544563649)
		(10, 16.830985514973527)
		(11, 17.859727889726795)
		(12, 17.111699359951903)
		(13, 18.111102648639157)
		(14, 17.36871848370088)
		(15, 17.450433991753467)
		(16, 16.660984111992786)
		(17, 17.477276337026705)
		(18, 17.411431570857548)
		(19, 13.625070119454044)
		(20, 13.36793261236501)
		(21, 9.560277085695414)
		(22, 8.070300922245167)
		(23, 3.02525777930004)
		(24, -1.411910502388837)
		(25, -6.1812586568467625)
		(26, -12.434378860573766)
		(27, -19.564326124754423)
		(28, -30.96491205702591)
		};
		\addlegendentry{Forged \\fingerprint}
		
		\addplot [color=adversary, densely dashed, ultra thick] coordinates {
		(-28, -8.724418771432365)
		(-27, -8.439286256811124)
		(-26, -8.116099542033389)
		(-25, -7.772643983032523)
		(-24, -7.331329145124692)
		(-23, -6.973521444308991)
		(-22, -6.790646513516643)
		(-21, -6.479013392640296)
		(-20, -6.066475206729605)
		(-19, -5.878176757163387)
		(-18, -5.4143983231325965)
		(-17, -5.218561154804257)
		(-16, -4.857141439883513)
		(-15, -4.537171861163109)
		(-14, -4.203998551481996)
		(-13, -3.86137999055051)
		(-12, -3.627730731082427)
		(-11, -3.1088717398404455)
		(-10, -2.8487774597910156)
		(-9, -2.699800589323169)
		(-8, -2.365395864631592)
		(-7, -1.9243182686951992)
		(-6, -1.7961587793485536)
		(-5, -1.3926296384326773)
		(-4, -1.1537892664281029)
		(-3, -0.8240942177484938)
		(-2, -0.5556632631864318)
		(-1, -0.2939915754304769)
		(1, 0.2939915754304783)
		(2, 0.6927055708549323)
		(3, 0.941603918585178)
		(4, 1.0874468546927962)
		(5, 1.265874936917047)
		(6, 1.6757186639730735)
		(7, 1.7243728738706974)
		(8, 2.0557332019785974)
		(9, 2.158468501954902)
		(10, 2.509592848059068)
		(11, 2.7024780248498943)
		(12, 2.7773810370124195)
		(13, 2.9777947104998237)
		(14, 3.0046427220171577)
		(15, 3.022646844198024)
		(16, 3.0486864095591377)
		(17, 3.160952256894526)
		(18, 3.068967409033754)
		(19, 2.8672935509300723)
		(20, 2.654231199481268)
		(21, 2.3231452261371075)
		(22, 1.904370101858792)
		(23, 1.2489479816416824)
		(24, 0.4153986655742872)
		(25, -0.7824199456351725)
		(26, -2.4716795865639316)
		(27, -5.049136437445395)
		(28, -8.724418771432365)
		};
		\addlegendentry{Adversary \\fingerprint}
		
		\end{axis}
		\end{tikzpicture}
	\centering
	\caption{Impersonation attack on \gls{CSI}-based radiometric fingerprinting.  
	\textit{Impersonation is feasible when the adversary is capable of introducing additional crafted phase rotations per subcarrier to match the fingerprint of the victim.}}
	\label{fig_fakeFingerprint}
\end{minipage}
\vspace{-4mm}
\end{figure*}

Radiometric fingerprinting can enhance the security of networks by enabling means of additional authentication based on physical layer properties of devices \cite{liu:2019, brik:2008, li2019location}.
However, we found that {\it an adversary can easily impersonate other devices} exploiting the fingerprinting scheme proposed by Liu \textit{et al.} \cite{liu:2019}.
We mount such an \textit{impersonation attack} using an SDR as follows: we first compute the fingerprint of the SDR by connecting it to another receiver (e.g., another SDR, signal analyzer). 
This needs to be done only once. 
Next, we measure the fingerprint of the target device, which only requires the adversary to sniff one (encrypted or unencrypted) packet sent by the target device. 
Knowing the proprietary fingerprint and that of a target, we can compute the phase offset on each subcarrier. 
These phase offsets are added to the LTF subcarriers and all subcarriers in the succeeding OFDM symbols (cf. \fref{fig_phy-frame}) at the SDR. 
Consequently, the fingerprint extracted by the receiver matches that of the target device. 
The SDR provides flexible processing capabilities that allow us to introduce phase rotations to the transmission chain easily. 
In \fref{fig_fakeFingerprint}, we demonstrate how accurately the SDR (i.e., adversary) can replicate the fingerprint of another device (i.e., victim).

We now analyze the efficacy of the \textit{impersonation attack} in a real-world scenario (i.e., an office building).
We set up an AP which employs the \gls{CSI}-based fingerprinting mechanism in~\cite{liu:2019} for authentication. 
To show the severity of the attack, we mount the attack in different locations in the vicinity of the victims, as depicted in  \fref{fig_impersonation_attack}. 
The adversary is transmitting 1000 packets in each location from which the access point calculates the fingerprints and compares them against a reference fingerprint of the legitimate user.
We conduct this experiment for each of the WiFi bands (i.e., $2.4$ and $5$ GHz) and show the results in the table in \fref{fig_impersonation_attack}.
In the $2.4$ GHz band, we observe that the adversary can successfully impersonate the victim in all locations. 
While the attack is very successful in line-of-sight scenarios (i.e., inside the lab), we observe that it still yields very high success rates in \gls{nlos} scenarios, e.g., the hallway or even in the office across the hallway. 
We kept all doors closed throughout the experiments. 
We also observe that the impersonation attack is possible even in highly challenging scenarios, i.e., behind a 30-centimeter thick ferro-concrete wall. However, the success rate is lower due to high signal attenuation. 
Due to higher propagation and penetration loss at $5$ GHz band, the success rate in the \gls{nlos} locations (i.e., L5 to L11) is lower. In particular, in locations L9 to L11, no signal was received by the AP.
However, locations L5 to L8 yield similar results in $5$ GHz and $2.4$ GHz bands.
We conclude that, as long as the adversary is in range of the access point, they can successfully effectuate an impersonation attack regardless of the frequency band used by the access point.



\subsection{Takeaway} 
In this section, we emphasize the need for a secure and privacy-preserving fingerprinting solution. 
Existing fingerprinting solutions based on \gls{CSI} are capable of {\it distinguishing between different devices, even of the same model, allowing adversaries to track the presence of users in a network}.
Further, we show that {\it an adversary can successfully impersonate the victim's device even through thick composite steel–concrete walls, which are among the most disruptive construction materials for wireless signals}.
Hence, there are two main takeaway messages from this section: {\it $(i)$ device fingerprints can be used to invade privacy of users and, as of the writing of this paper, there is no protection for users; and $(ii)$ active deployments of \gls{CSI}-based fingerprinting schemes can be attacked.}

\section{How to Inject Artificial Noise to Fingerprints without Impacting Communication}\label{s4}

Here we discuss our method for injecting artificial noise (i.e., randomized phase rotation) to the radiometric fingerprints. 
We further {\it prove why it does not impact the quality of communication}.

\begin{figure*}[t!]
	\centering
	\includegraphics[width=0.9\columnwidth]{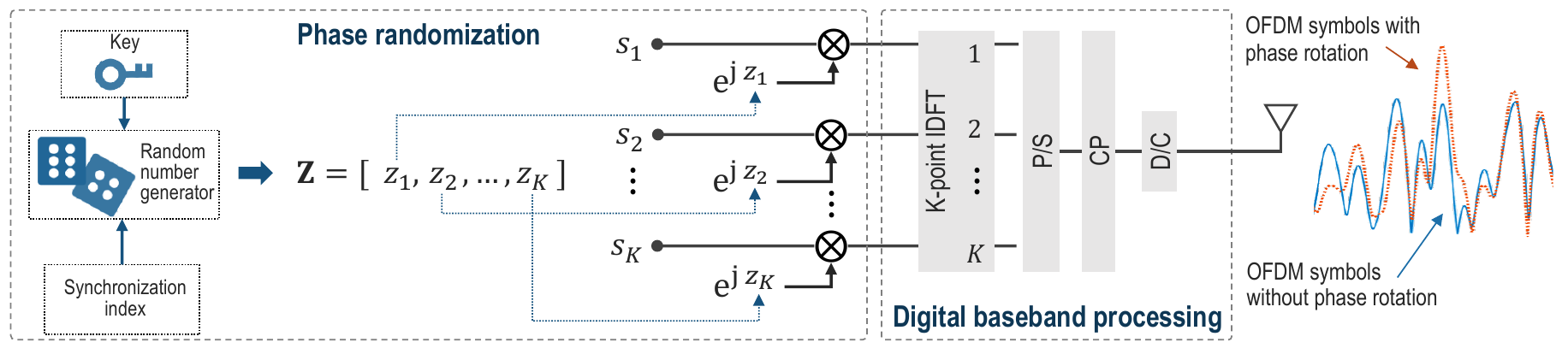}	
	\caption{Diagram illustrating the phase randomization method at the transmitter. \textit{Additional random phase rotation at every subcarrier protects the fingerprint of the transmitter and therefore prevents impersonation.}}
	\label{fig_rand_overview}
	\vspace*{-4mm}
\end{figure*}
\begin{figure*}[t!]
\begin{minipage}{0.48\textwidth}
	\centering
	\vspace{-8mm}
\definecolor{original}{HTML}{4d4d4d}
\definecolor{uniform}{HTML}{a50f15}
\definecolor{gaussian}{HTML}{4292c6}

\begin{tikzpicture}
\begin{axis}
[
xlabel = {Subcarrier index ($ k $)},
ylabel = {Fingerprint [deg]},
xmin = 0,
xmax = 57, 
ymin = -350, 
ymax = 300,
width = 7cm,
height = 4cm,
ymajorgrids,
  	grid style={line width=0.3pt, draw=gray!70},
ticklabel style = {font=\fontsize{7}{7}\selectfont},
xtick = {1, 6, 11, 16, 21, 26, 31, 36, 41, 46, 51, 56},
ytick = {-200, 0, 200},
ylabel style = {text width = 5cm, align = center, font=\fontsize{7}{7}\selectfont},
xlabel style = {align = center, font=\fontsize{7}{7}\selectfont},
legend columns = 2,
legend pos = north east,
legend style = {at = {(0.01,0.0005)}, anchor = south west, font = \scriptsize, fill = none,  font=\fontsize{6}{6}\selectfont, /tikz/every even column/.append style={column sep=0.2mm}, legend style={row sep=-2pt}, inner sep = 0.5pt, align = left, draw=none},
legend cell align={left},
]

\addplot[color = original, mark = none, mark options = {scale = 1, fill = original, solid}, line width = 2pt, densely dotted] table {figs_data/OriginalFingerprint.txt};
\addlegendentry[]{Original fingerprint} 

\addplot[color = uniform, mark = *, mark options = {scale = 0.6, fill = uniform, solid}, line width = 0.3pt, smooth] table {figs_data/Fig6-RecoveredFingerprintUniformMean0Var1WithoutRecovery.txt}; 
\addlegendentry[]{Obfuscated | Uniform} 

\addplot[color = gaussian, mark = triangle*, mark options = {scale = 0.8, fill = gaussian, solid}, line width = 0.3pt, smooth] table {figs_data/Fig6-RecoveredFingerprintGaussianMean0Var1WithoutRecovery.txt}; 
\addlegendentry[]{Obfuscated | Gaussian} 

\end{axis}
\end{tikzpicture}
	\vspace{-13mm}
	\caption{Fingerprints before and after obfuscation. \textit{Upon including random phase rotations in the subcarriers, the recovered fingerprint differs from the original.}}
	\label{fig_Rand_fingerprint}
\end{minipage}
\hspace{2mm}
\begin{minipage}{0.48\textwidth}
	\centering
	\vspace{-8mm}
\definecolor{original}{HTML}{4d4d4d}
\definecolor{uniform}{HTML}{a50f15}
\definecolor{gaussian}{HTML}{4292c6}

		\begin{tikzpicture}
		
		\begin{axis}
		[
		xlabel = {Time [seconds]},
		ylabel = {Throughput [Mbit/s]},
		ymin = 31.25,
		ymax = 33.25, 
		xmin = 0,
		xmax = 60,
		width = 7cm,
		height = 4cm,
		ymajorgrids,
		grid style={line width=0.3pt, draw=gray!70},
		ticklabel style = {font=\fontsize{7}{7}\selectfont},
		ytick = {31.5, 32, 32.5, 33},
		ylabel style = {text width = 5cm, align = center, font=\fontsize{7}{7}\selectfont},
		xlabel style = {align = center, font=\fontsize{7}{7}\selectfont},
		legend columns = 1,
		legend pos = north east,
		legend style = {at = {(0.01,0.001)}, anchor = south west, font = \scriptsize, fill = none,  font=\fontsize{6}{6}\selectfont, /tikz/every even column/.append style={column sep=-2mm}, legend style={row sep=-2pt}, inner sep = 0.5pt, align = left, draw=none},
		legend cell align={left},
		]
		
		\addplot[color = original, mark = none, mark options = {scale = 1, fill = original, solid}, line width = 1pt, densely dotted, smooth, mark repeat=3, mark phase=1] table {figs_data/Fig6-ThroughputReference.txt};
		\addlegendentry[]{Throughput | Original} 
		
		\addplot[color = uniform, mark = *, mark options = {scale = 0.6, fill = uniform, solid}, line width = 0.3pt, smooth, mark repeat=3, mark phase=1] table {figs_data/Fig6-ThroughputUniformVar1.txt}; 
		\addlegendentry[]{Throughput | Uniform} 
		
		\addplot[color = gaussian, mark = triangle*, mark options = {scale = 0.8, fill = gaussian, solid}, line width = 0.3pt, smooth, mark repeat=3, mark phase=1] table {figs_data/Fig6-ThroughputGaussianVar1.txt}; 
		\addlegendentry[]{Throughput | Gaussian}

		\end{axis}

		\end{tikzpicture}
	\vspace{-13mm}
	\caption{Throughput before and after obfuscation. \textit{The throughput is not affected by the phase rotations, as these can be reverted at the receiver.} }
	\label{fig_Rand_fingerprint_throughput}
\end{minipage}
\vspace{-3mm}
\end{figure*}

Randomizing the radiometric fingerprints is the first logical step towards maintaining user privacy. However, if not carefully designed, the randomization can potentially break/degrade the communication link. As described in Section~\ref{s3-1}, the WiFi receiver relies on the LTF field for channel estimation. To ensure that {\it obfuscation via randomization does not disrupt communication, we maintain the introduced phase rotations on each subcarrier constant for the duration of the whole frame}. As a result, the estimated \gls{CSI} from the preambles remains valid for the succeeding VHT-DATA frame, as shown in \fref{fig_phy-frame}, thus allowing successful decoding of information. In the following, we describe the process. \fref{fig_rand_overview} shows the transmitter chain of our proposed fingerprint obfuscation method, in which we include deliberate phase randomization across all OFDM subcarriers. Having the \textit{pre-shared key} and \textit{randomization index}, the receiver can decode the message and extract the fingerprint without impacting the communication.


%


Let $ z_k $ denote the phase rotation in subcarrier $ k $ intentionally included by the transmitter. From~\eqref{e3-2}, the signal received in the $ k $-th subcarrier is given by $ y_k = h_k s_k e^{j z_k} + w_k $. Using \eqref{e3-3}, the \gls{CSI} at the receiver is expressed as
\begin{equation} \label{e4-3}
	\tilde{h}_k = h_k e^{j z_k} + w_k = \left| h_k \right| e^{j (\phi_k + z_k)} + w_k.
\end{equation}

Compared to (\ref{e3-3}), the factor $ e^{j z_k} $ in (\ref{e4-3}) obfuscates the legitimate \gls{CSI} by shifting its phase information. As a result, the phase $ z_k $ will appear in the radiometric fingerprint extracted by an adversary, thus safeguarding the device original fingerprint. The effects of this phase randomization mechanism can only be reverted by a trusted receiver that is aware of $ z_k $.
In particular, we assume that the phase $ z_k $ is a realization of a random variable $ Z_k $, that can be generated locally at the receiver since the pre-shared key to the random generator is known. 
As a result, the receiver generates $ z_k $ and multiplies the perturbed $ \tilde{h}_k $ in (\ref{e4-3}) by $ e^{-j z_k} $ yielding $ e^{-j z_k} \left( \left| h_k \right| e^{j (\phi_k + z_k)} + w_k \right) = \left| h_k \right| e^{j \phi_k} + w_k $, which is equivalent to (\ref{e3-3}), and therefore showing that the \gls{CSI} remains unaffected as the phase randomization can be removed. In addition, we denote the capacity of the channel in (\ref{e3-3}) by $ C' = \log_2 \left( 1 +  \left| \left| h_k \right| e^{j \phi_k} \right|^2 / \sigma^2 \right) = \log_2 \left( 1 + \left| h_k \right|^2 / \sigma^2 \right) $. Similarly, the channel capacity of (\ref{e4-3}) is denoted by $ C'' = \log_2 \left( 1 + \left| \left| h_k \right| e^{j (\phi_k + z_k)} \right|^2 / \sigma^2 \right) = \log_2 \left( 1 + \left| h_k \right|^2 / \sigma^2 \right) $, thus revealing the equivalence $ C' \equiv C'' $. This shows that the channel capacity before and after randomization does not change. Therefore, for a given MCS level, the throughput is not altered by phase randomization as long as the phase $ z_k $ is generated correctly at each receiver. We generalize this idea for every subcarrier $ k \in \mathcal{K} $. 
In \fref{fig_Rand_fingerprint}, we illustrate the original fingerprint of a device as well as the obfuscated versions, in which the random phase rotations are obtained from uniform and Gaussian distributions, i.e., $ Z_k \sim \mathcal{U} \big( \mu_k, \xi^2_k \big) $ and $ Z_k \sim \mathcal{N} \big( \mu_k, \xi^2_k \big) $ with $ \mu_k = 0 $ deg and $ \xi^2_k = 60 $ deg\textsuperscript{2} (deg $ \equiv ~ ^\circ $). Since the additional randomized phase $ z_k $ differs for each subcarrier, the adversary cannot leverage the linear phase error difference among subcarriers to identify the users. 
\emph{In particular, if the same phase rotation $ z_k $ is used for all $ K $ subcarriers, the original fingerprint can be easily extracted via the method proposed in \cite{liu:2019} as such method exploits the phase difference among adjacent subcarriers, which in this case would be constant and easy to remove.} Moreover, we corroborate experimentally that the throughput is not affected by our proposed obfuscation method. In particular, for the obfuscated signals depicted in \fref{fig_Rand_fingerprint}, we show the throughput in \fref{fig_Rand_fingerprint_throughput}.
 
In the next section, we discuss the robustness of this approach against statistical attacks.



\section{\Tool{}: A Benchmarking Tool for Assessing Vulnerability to Statistical Attacks}\label{Vulnerability_analysis}


Statistical attacks are common in cryptography where the adversary exploits statistical weaknesses of the underlying random number generators or hashing algorithms to discover the  secrets, e.g., birthday attacks \cite{bellare2004hash}. In the course of our experiments, we discovered that an adversary can mount similar attacks on phase randomization to restore the original fingerprint. {\it Viewing this as an estimation problem, we devise \Tool{}, which is a maximum likelihood-based approach design to restore the legitimate (unimpaired) \gls{CSI} from a set of captured \gls{CSI} measurements with obfuscated fingerprints}. Thus, if the legitimate \gls{CSI} is restored accurately, the radiometric fingerprint can be extracted by the method described in Section \ref{s3-2} and used for malicious purposes. {\it In essence, we designed \Tool{} as a tool to evaluate the efficiency of RF-fingerprint obfuscation against statistical attacks.} Specifically, we designed an experiment in which the adversary captures 10000 \gls{CSI} samples (within $\sim$ 10 seconds) and uses \Tool{} to estimate the legitimate \gls{CSI}. This experiment showed that {\it an adversary can denoise the fingerprint even without the knowledge of the probability density function used for phase randomization}. We will elaborate on \Tool{} and the experimental results in Section~\ref{ss_bison}. We prove that this vulnerability stems from the zero-mean nature of the selected distributions, see Section~\ref{ss_bison_analysis}.

\begin{figure*}[!t]
	\centering
	\input{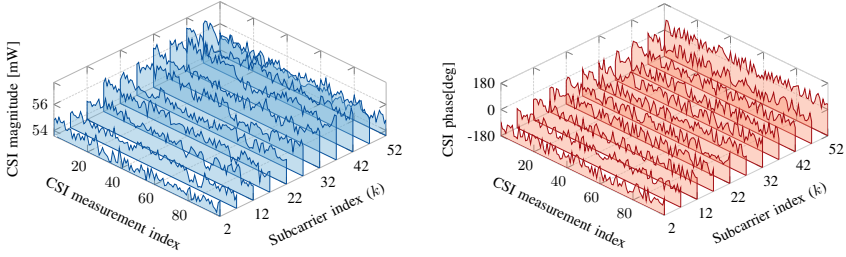}
	\vspace{-5mm}
	\caption{Collected \gls{CSI} measurements with additional synchronization phase rotations obtained from a zero-mean unit-variance Gaussian probability density function.}
	\label{fig_rand_phase_mag}
	\vspace*{-6mm}
\end{figure*}

\fref{fig_rand_phase_mag} shows the magnitude and phases of \gls{CSI} measurements. We assume that the channel impulse response is invariant for a short interval $ \tau $ compliant with the channel coherence time $ T_c $. Thus, small-scale oscillations in the \gls{CSI} magnitude are attributed to noise. On the other hand, the \gls{CSI} phase changes abruptly between contiguous measurements due to phase randomization.

\subsection{A maximum-likelihood-based estimator for evaluating statistical attacks}
\label{ss_bison}

\Tool{} minimizes the overall approximation error between the unknown \gls{CSI} and the collected measurements. The premise is that adversaries do not have information on the probability density function used for \gls{CSI} phase randomization. Let $ \mathbf{M} = \left[ \mathbf{m}_1, \cdots, \mathbf{m}_N \right] \in \mathbb{C}^{K \times N}$ denote a matrix that collects $ N $ measurements in all $ K $ subcarriers, where vector $ \mathbf{m}_n \in \mathbb{C}^{K \times 1} $ represents the \gls{CSI} (contaminated with phase randomization and noise) in the $ n $-th captured LTF frame. Also, let $ \mathbf{u} =  \left[ \left| h_1 \right| e^{j \phi_1}, \cdots, \left| h_K \right| e^{j \phi_K} \right]^T \in \mathbb{C}^{K \times 1} $ denote the unknown unrandomized \gls{CSI} vector. Further, $ \boldsymbol{\Delta} = \left[ \mathbf{m}_1 - \mathbf{u}, \cdots, \mathbf{m}_N - \mathbf{u} \right] = \mathbf{M} - \left( \mathbf{1}^T \otimes \mathbf{u} \right) $ represents the error matrix between the unknown \gls{CSI} $ \mathbf{u} $ and the measurements $ \mathbf{M} $, where $ \otimes $ is the Kronecker product. We define the following problem:
\begin{equation} \label{e4-4}
		\setlength{\abovedisplayskip}{1.0pt}
		\setlength{\belowdisplayskip}{1.0pt}
		\mathcal{B}: \mathbf{u}^\star = \argmin_{ \mathbf{u} \in \mathbb{C}^{K \times 1} } 
		\underbrace{\left\| \mathbf{M} - \mathbf{1}^T \otimes \mathbf{u} \right\|^2_\mathrm{F},}_J	
\end{equation}
where $ \left\| \cdot \right\|^2_\mathrm{F} $ denotes the Frobenius norm. To solve problem $ \mathcal{B} $, we have used several Kronecker product properties specified in \emph{Appendix~A}. Recalling that $ \left\| \boldsymbol{\Delta} \right\|^2_\mathrm{F} = \mathrm{Tr} \left( \boldsymbol{\Delta}^T \boldsymbol{\Delta} \right) $, the objective function can be recast as $ J = \mathrm{Tr} \Big( \big( \mathbf{M} - \mathbf{1}^T \otimes \mathbf{u} \big)^T \big( \mathbf{M} - \mathbf{1}^T \otimes \mathbf{u} \big) \Big) $. By employing \emph{Property 1} and \emph{Property 2}, the objective collapses to $ J = \mathrm{Tr} \big( \mathbf{M}^T \mathbf{M} - \left( \mathbf{1} \otimes \mathbf{u}^T \right) \mathbf{M} - \mathbf{M}^T \left( \mathbf{1}^T \otimes \mathbf{u} \right) + \mathbf{L} \otimes \left( \mathbf{u}^T \mathbf{u} \right) \big) $, where $ \mathbf{L} = \mathbf{1} \mathbf{1}^T $. To find a critical point $ \mathbf{u}^\star $ that minimizes $ J $, we compute the gradient of $ J $ with respect to $ \mathbf{u} $ and equate it to zero, i.e., $ \nabla_\mathbf{u} J = 0 $. To this purpose, we resort to the use of differentials. Thus, $ d J = \mathrm{Tr} \Big( - \big( \mathbf{1} \otimes d \mathbf{u}^T \big) \mathbf{M} - \mathbf{M}^T \big( \mathbf{1}^T \otimes d \mathbf{u} \big) \nonumber + \mathbf{L} \otimes \big( d \mathbf{u}^T \mathbf{u} \big) + \mathbf{L} \otimes \big( \mathbf{u}^T d \mathbf{u} \big) \Big) $, where $ d $ denotes the differential operator and $ d \mathbf{M} = 0 $, $ d \mathbf{L} = 0 $. Using \emph{Property 2}, \emph{Property 3} and \emph{Property 4}, the differential of $ J $ is expressed as $ dJ = \mathrm{Tr} \Big( - \big( \mathbf{M} \mathbf{1} \big) \otimes d \mathbf{u}^T - \mathbf{1}^T \otimes \big( \mathbf{M}^T d \mathbf{u} \big) + \mathbf{L} \otimes \big( \big( d \mathbf{u}^T \big) \mathbf{u} \big) + \mathbf{L} \otimes \big( \mathbf{u}^T d \mathbf{u} \big) \Big) $. Now, by means of \emph{Property 4} and \emph{Property 5} we obtain  $ dJ = 2 \mathrm{Tr} \Big( \big(  - \big( \mathbf{M} \mathbf{1} \big)^T + N \mathbf{u}^T \big) d \mathbf{u} \Big) $. The Frobenius inner product of two matrices $ \mathbf{A} $ and $ \mathbf{B} $ is defined as $ \left\langle \mathbf{A}, \mathbf{B} \right\rangle_\mathrm{F} \equiv \mathrm{Tr} \big( \mathbf{A}^T \mathbf{B} \big) $. Therefore, $ dJ = 2 \left\langle - \mathbf{M} \mathbf{1} + N \mathbf{u}, d \mathbf{u} \right\rangle_\mathrm{F} $, from where we obtain $ \nabla_\mathbf{u} J = 2 \big( - \mathbf{M} \mathbf{1} + N \mathbf{u} \big) $. Upon equating $ \nabla_\mathbf{u} J $ to zero, we obtain $ \mathbf{u}^\star = \frac{1}{N} \mathbf{M} \mathbf{1} = \frac{1}{N} \sum^N_{n = 1} \mathbf{m}_n $. The denoised \gls{CSI} phase $ \boldsymbol{\Phi} $ for all subcarriers is computed as 
\begin{equation} \label{e4-5}
	\boldsymbol{\Phi}^\star = \arctan \left( \mathfrak{Im} \left\{ \frac{1}{N} \sum^N_{n = 1} \mathbf{m}_n \right\} \oslash \mathfrak{Re} \left\{ \frac{1}{N} \sum^N_{n = 1} \mathbf{m}_n \right\} \right).
\end{equation}

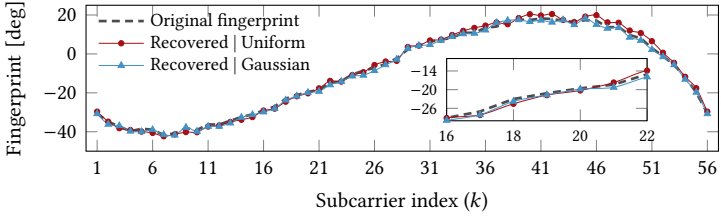
\begin{figure*}[!t]
	\vspace*{-10mm}
\definecolor{original}{HTML}{4d4d4d}
\definecolor{uniform}{HTML}{a50f15}
\definecolor{gaussian}{HTML}{4292c6}

		\begin{tikzpicture}
		\begin{axis}
		[
		xlabel = {Subcarrier index ($ k $)},
		ylabel = {Fingerprint [deg]},
		xmin = 0,
		xmax = 57, 
		ymin = -50, 
		ymax = 25,
		width = 10cm,
		height = 3.5cm,
		ticklabel style = {font=\fontsize{8}{8}\selectfont},
		xtick = {1, 6, ..., 56},
		ytick = {-40, -20, 0, 20},
		ylabel style = {text width = 5cm, align = center, font=\fontsize{8}{8}\selectfont},
		xlabel style = {align = center, font=\fontsize{8}{8}\selectfont},
		legend columns = 1,
		legend pos = north east,
		legend style = {at = {(0.015,0.48)}, anchor = south west, font = \scriptsize, fill = none,  font=\fontsize{7}{7}\selectfont, /tikz/every even column/.append style={column sep=-2mm}, legend style={row sep=-2pt}, inner sep = 0.5pt, align = left, draw=none},
		legend cell align={left},
		]

		\addplot[color = original, mark = none*, mark options = {scale = 1, fill = original, solid}, line width = 1pt, densely dashed] table {figs_data/OriginalFingerprint.txt};
		\addlegendentry[]{Original fingerprint} 
		
		\addplot[color = uniform, mark = *, mark options = {scale = 0.5, fill = uniform, solid}, line width = 0.3pt, smooth] table {figs_data/Fig8-RecoveredFingerprintUniformMean0Var1WithRecoveryBISON.txt}; 
		\addlegendentry[]{Recovered | Uniform} 
		
		\addplot[color = gaussian, mark = triangle*, mark options = {scale = 0.7, fill = gaussian, solid}, line width = 0.3pt, smooth] table {figs_data/Fig8-RecoveredFingerprintGaussianMean0Var1WithRecoveryBISON.txt}; 
		\addlegendentry[]{Recovered | Gaussian} 
		
		\end{axis}
		
		\begin{axis}
			[
				xmin = 16,
				xmax = 22, 
				ymin = -30, 
				ymax = -10,
				width = 4.25cm,
				height = 2.4cm,
				ytick = {-14, -20, -26},
				tick label style={font=\fontsize{6}{6}\selectfont,},
				shift={(48mm,4mm)},axis background/.style={fill=white}
			]
			
			\addplot[color = original, mark = none*, mark options = {scale = 0.9, fill = original, solid}, line width = 1pt, densely dashed] table {figs_data/OriginalFingerprint.txt};
			
			\addplot[color = uniform, mark = *, mark options = {scale = 0.5, fill = uniform, solid}, line width = 0.3pt, smooth] table {figs_data/Fig8-RecoveredFingerprintUniformMean0Var1WithRecoveryBISON.txt};
			
			\addplot[color = gaussian, mark = triangle*, mark options = {scale = 0.7, fill = gaussian, solid}, line width = 0.3pt, smooth] table {figs_data/Fig8-RecoveredFingerprintGaussianMean0Var1WithRecoveryBISON.txt}; 
			
		\end{axis}
		
		\end{tikzpicture}
	\vspace*{-10mm}
	\caption{Restored fingerprint upon \gls{CSI} denoising. {\it We observe that an adversary capable of mounting a statistical attack can obtain the original fingerprint even after randomization.}}
	\vspace*{-5mm}
	\label{fig_recovered_bison}
\end{figure*}

Since denoised \gls{CSI} is available through (\ref{e4-5}), the radiometric fingerprint $ \boldsymbol{\epsilon}^\star $ can be extracted using (\ref{e3-5}). \cite{liu:2019} \fref{fig_recovered_bison} shows the restored fingerprints for uniform and Gaussian distributions with mean $ \mu_k = 0^{\circ} $ and variance $ \xi^2_k = 60 $ deg\textsuperscript{2}, for all subcarriers $ k \in \mathcal{K} $. In both cases, we have collected $ N = 10000 $ measurements. We observe that the obtained fingerprints exhibit a small deviation with respect to the original one. When uniform distribution is used, the mean absolute error (MAE) is $ 0.7489^{\circ} $, whereas that of Gaussian distribution is $ 1.2252^{\circ}$. Although both distributions have a variance of $ \xi^2_k = 60 $ deg\textsuperscript{2}, for the uniform case, this signifies that the range of phase rotations is bounded to $ \left[ -99.2^{\circ}; 99.2^{\circ} \right]  $. However, for the Gaussian case, the range of rotation phases spans $ \left[ -180^{\circ}; 180^{\circ} \right] $. 

In \emph{Appendix B},  we analyze the \Tool{} estimator under the Cramer-Rao bound (CRB) framework. We show that \Tool{} attains near-optimality in estimating the \gls{CSI}.

\subsection{Statistical rationale for \gls{CSI} denoising feasibility via \Tool{}} \label{ss_bison_analysis}

If an efficient estimator does not exist for an unknown variable, the maximum-likelihood estimation often yields an asymptotically efficient estimator for sufficiently large number of samples. Based on this premise, we expect the effect of randomization to be averaged out. Thus, {\it motivated by the outcome of \Tool{}, we  justify why the effect of randomization, introduced in Section \ref{s4}, can be removed}. By assuming that an adversary is capable of collecting an infinite number of measurements, we \Tool{} within the law of large numbers; which states that the average of outcomes obtained from a large number of experiments approximates the expected value.

\noindent{\textbf{\textit{Assumption:}} \textit{Let $ f_{Z_k} \big( z_k \big) $ be a symmetric zero-mean probability density function governing the random phase rotation $ Z_k $, spanning an interval with upper and lower bounds $ z^U_k = R_k $ and $ z^L_k = -R_k $, respectively.}}

Invoking the assumption above, the expected value of the corrupted \gls{CSI} information in subcarrier $ k $ according to (\ref{e4-3}) is defined as $ \mathbb{E} \left[ \tilde{h}_k \right] = \mathbb{E} \left[ h_k e^{j Z_k} \right]  + \mathbb{E} \left[ w_k \right] $, where $ \mathbb{E} \left[ h_k e^{j Z_k} \right] = h_k \mathbb{E} \left[ e^{j Z_k} \right] = h_k \int_{-R_k}^{R_k} e^{j z_k} f_{Z_k} \big( z_k \big) d z_k $.  Using integration by parts, $ \mathbb{E} \left[ e^{j Z_k} \right]  $ can be recast as, 
    \begin{equation} \label{e4-6}
      \mathbb{E} \left[ e^{j Z_k} \right] =  2 \sin \big( R_k \big) f_{Z_k} \big( R_k \big) - \int_{-R_k}^{R_k} \sin \big( z_k \big) f'_{Z_k} \big( z_k \big) d z_k 
      		+ j \int_{-R_k}^{R_k} \cos \big( z_k \big) f'_{Z_k} \big( z_k \big) d z_k = {\beta^{\mathrm{real}}_k} + j {\beta^{\mathrm{imag}}_k},
    \end{equation}
where the equivalence $ \int u dv = uv -\int v du $ is used assuming that $ u = f_{Z_k} \big( z_k \big) $, $ dv = e^{j z_k} d z_k $ and $ f_{Z_k} \big( -R_k \big) = f_{Z_k} \big( R_k \big) $ due to symmetry. In the following, we instantiate three fundamental corollaries that allow us to gain insights on the characteristics of (\ref{e4-6}).

\noindent{\textbf{\textit{Corollary 1:}} \textit{If $ g(x) $ is an even function, then its derivative $ g'(x) $ is an odd function.}}

\noindent{\textbf{\textit{Corollary 2:}} \textit{If $ g(x) $ is even and $ h(x) $ is odd, then $ q(x) = g(x) h(x) $ is odd.}}

\noindent{\textbf{\textit{Corollary 3:}} \textit{If $ g(x) $ is odd, then $ \int^a_{-a} g(x) dx = 0 $ for $ a > 0 $.}}

By means of \textit{Corollary 1}, we assert that $ f'_{Z_k} \big( z_k \big) $ is an odd function. Also, via \textit{Corollary 2}, the function $ \cos(z_k) f'_{Z_k} \big( z_k \big) $ is odd. Finally, by means of \textit{Corollary 3} the value of $ \beta^{\mathrm{imag}}_k = \int_{-R_k}^{R_k} \cos \big( z_k \big) f'_{Z_k} \big( z_k \big) = 0 $. As a result, $ \mathbb{E} \left[ h_k e^{j Z_k} \right] =  \beta^{\mathrm{real}}_k h_k $, which shows that (on average) the \gls{CSI} in every subcarrier $ k $ is affected only by a real-value attenuation factor $ \beta^{\mathrm{real}}_k $ without altering the phase. 

\noindent{\textbf{\textit{Claim:}} \textit{When we obfuscate the fingerprints through phase randomization using symmetric zero-mean distributions, \Tool{} produces an unbiased estimator for the \gls{CSI} phase.}}

Harnessing this outcome, we compute the expected value of the proposed \Tool{} estimator, i.e., $ \mathbb{E} \left[ \mathbf{u}^\star \right]  = \frac{1}{N} \sum^N_{n = 1} \mathbb{E} \left[  \mathbf{m}_n \right]  = \frac{1}{N} \sum^N_{n = 1} \mathbb{E} \left[  \mathrm{diag} \left( e^{ j \mathbf{z}_n } \right) \mathbf{h} + \mathbf{w}_n \right] $, where $ \mathbf{z}_n = \left[ Z_{n,1}, \cdots, Z_{n,K} \right]^T $ and $ \mathbf{w}_n = \left[ w_{n,1}, \cdots, w_{n,K} \right]^T $. Thus, $ \mathbb{E} \left[  \mathbf{u}^\star \right] $ reduces to
\begin{align} \label{e4-9}
	\mathbb{E} \left\lbrace \mathbf{u}^\star \right\rbrace
	=
	\begin{pmatrix}
		\beta^\mathrm{real}_1   	& \cdots   & 0 \\
		\vdots         			    & \ddots   & \vdots       \\
			0              			&  \cdots  & \beta^\mathrm{real}_K \\
	\end{pmatrix}
	\begin{pmatrix}
		\left| h_1 \right| e^{j \phi_1} \\
		\vdots \\
		\left| h_K \right| e^{j \phi_K}
	\end{pmatrix}.
\end{align}

From (\ref{e4-9}), we note that when the randomization scheme in Section \ref{s4} is used for \gls{CSI} obfuscation, its effect can be removed via \Tool{}. Essentially, the restored \gls{CSI} magnitudes $ \left| h_k \right| $ are scaled by $ \beta^{\mathrm{real}}_k $ but the phases $ \phi_k $ remain unaffected. As a result, an adversary can extract the radiometric fingerprint $ \boldsymbol{\epsilon} $ (defined in (\ref{e3-4})) from the restored \gls{CSI} phase $ \boldsymbol{\Phi} $. In order to prevent this outcome that infringes secrecy, a specific type of probability density function is required that prevents \gls{CSI} denoising from collected measurements. This aspect is elaborated thoroughly in Section~\ref{ss_obfus_tx}.

\subsection{Takeaway}
Any system relying on randomization for improving security/privacy should prove robust against statistical attacks. Here we propose \Tool{} to {\it assess the vulnerability of fingerprint randomization against these attacks}. This tool will be later used to demonstrate the robustness of our proposed fingerprint obfuscation method (i.e., \Alg{}) against statistical attacks. Furthermore, we analyze the statistical rationale behind the aforementioned vulnerability. {\it This analysis is then leveraged to devise suitable countermeasures in the next section}.


\section{RF-Veil: A Privacy- and Security-preserving Solution for Radiometric Fingerprinting} 
\label{s_algorithm}


\begin{figure*}[!t]
	\centering
	\includegraphics[width=0.99\textwidth]{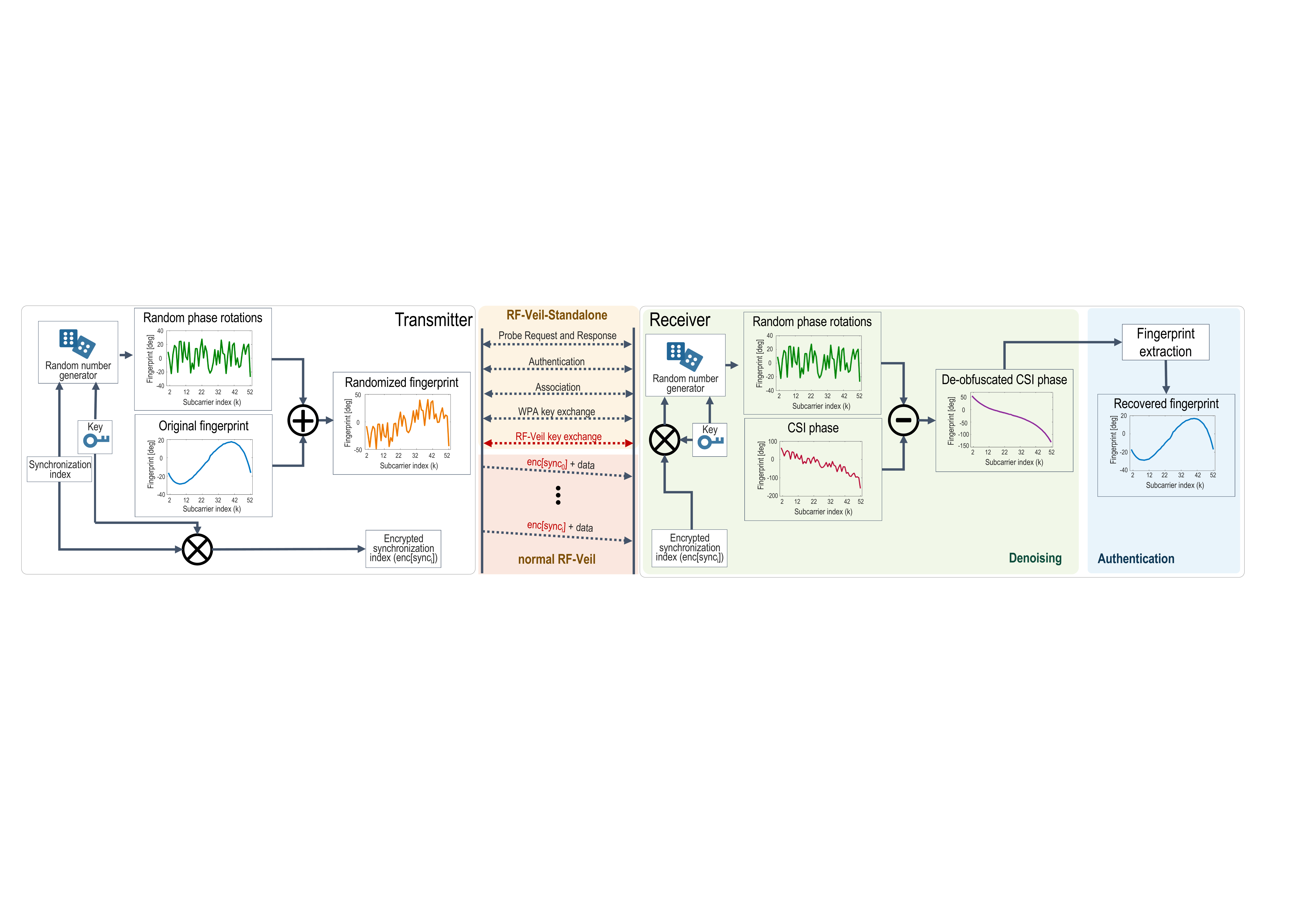}
	\caption{Schematic overview of an \Alg{} transmitter and receiver. The flow-diagram in the center depicts the WiFi connection establishment and data exchange procedure (changes due to \Alg{} marked in red color). The ACK messages are not shown in the figure for readability. They are not modified in \Alg{}. }
	\label{fig_algStructure}
	\vspace{-5mm}
\end{figure*}


In this section, we introduce our proposed technique \Alg{}, which injects crafted artificial noise to fingerprints in order to improve the robustness of WiFi transmissions against statistical attacks aiming at fingerprint acquisition. 
In \fref{fig_algStructure}, we illustrate the building blocks of \Alg{}. Note that, in \AlgSt{} mode, we only need a subset of the blocks at the transmitter since the receiver does not perform any radiometric fingerprinting. To avoid repetition, we highlight the \AlgSt{}-specific blocks and the algorithm workflow in this mode in \sref{ss:standalone}. 

{\bf A short overview of \Alg{}.} As shown in Fig.~\ref{fig_algStructure}, {\it the transmitter uses a random number generator to generate a pattern that obfuscates its radiometric fingerprint on a per-frame basis. The random number generator follows a specific distribution that is robust against statistical attacks. The receiver requires the seed to the random generator in order to generate the same pattern, which is used for \gls{CSI} denoising and fingerprint extraction.} The details about standard compliancy, random sequence generation, and key exchange are elaborated below.


\subsection{Association}
\label{ss_assoc}

In WiFi, every new device first associates to the AP upon arrival to the network. 
This includes exchanging the probe, authentication, and association request and response messages. 
It is within this stage that the AP and the device establish a secure connection. In \Alg{}, we require the access point and the client to exchange one more key, which is used as one of the inputs to the random number generator, as shown in \fref{fig_rand_overview}. We choose to use a pre-shared key due to ease of implementation.
However, one can leverage alternative secret key extraction methods that rely on channel response~\cite{liu:2013}. 
As a result, the receiver and transmitter do not require a security handshake in advance but use physical layer information to generate the secret keys. 

At this stage, the AP can extract the real fingerprint of the client after obtaining the shared key. We elaborate further on this in Section~\ref{ss_deobfus_rx}. 

\subsection{Obfuscation at the transmitter}
\label{ss_obfus_tx}
The main task of the transmitter consists in obfuscation, as depicted on the left-hand side in \fref{fig_algStructure}. For every frame, a random sequence is generated using the pre-shared key and the synchronization index. 

{\bf Pre-shared key.} In our implementation, we used a 128-bit key, which is refreshed every time the device re-associates with the AP. As a privacy protection measure, we obfuscate the fingerprint even before the association with an AP takes place. Hence, any frame transmitted from the devices (e.g., beacon, discovery) has an obfuscated fingerprint. In this case, it is advised to generate a new key periodically in order to protect against statistical attacks (see Section~\ref{s4} ). We leave the frequency of key renewal as a design choice. Since renewing the pre-shared key does not impose considerable overhead, we suggest to lean towards higher security.

{\bf Synchronization index}. 
Attaching a synchronization index to each frame has two purposes: {\it (i)} synchronize the random generator between the receiver and the transmitter and {\it (ii)} protect the receiver from replay attacks. 
The synchronization is important because the pre-shared key only ensures that the random generators at both ends produce the same string of random numbers. 
However, if a frame is lost, then the receiver may try to de-obfuscate the frame with the wrong pattern. 
To prevent this, we attach an index for each frame, so that the receiver can use this index in combination with the pre-shared key to generate a synchronized and secure randomization pattern. 
We intentionally refrained from using the existing 12-bit MAC frame sequence number due to its vulnerability to replay attacks. 
Even at low data rates, the 12-bit sequence number resets within seconds, whereas our 32-bit sequence number takes 24 days to reset at the rate of 1000 frames per second. 
We expect the WiFi connection to be re-initiated within such an interval. 
Even though the exposure of this synchronization index does not expose legitimate users to security threats, it can still be abused for tracking. 
Therefore, we encrypt this index with the pre-shared key via XOR operations. 
We further discuss this approach in \sref{s_discus}.

Once the obfuscation pattern is generated for all subcarriers, the symbols of the regular WiFi transmitter are rotated accordingly. The frame sequence number is then updated for the next frame and stored in a \gls{LUT}. Finally, the symbols with phase rotations can be sent out over the air. However, one question still remains:  {\it how do we ensure robustness against statistical attacks?}

{\bf Robustness against statistical attacks.} In Section~\ref{Vulnerability_analysis}, we showed experimentally and analytically that obfuscation with symmetric zero-mean distributions is susceptible to statistical attacks. Recalling the analysis therein, a robust distribution against such attacks should have the following properties. 

\begin{center}
$
\boxed{
	\quad \left( \mathrm{\mathbf{P_1}} \right): f_{Z} \big( z \big) \geq 0 \quad \left( \mathrm{\mathbf{P_2}} \right): \int_{-\infty}^{\infty} f_{Z} \big( z \big) d z = 1 \quad \left( \mathrm{\mathbf{P_3}} \right): f_{Z} \big( z \big) \neq f_{Z} \big( -z \big) \quad \left( \mathrm{\mathbf{P_4}} \right): \mathbb{E} \left[ f_{Z} \big( z \big) \right] \neq 0 \quad
}
$
\end{center}

Essentially, $ \left( \mathrm{\mathbf{P_1}} \right) $ and $ \left( \mathrm{\mathbf{P_2}} \right) $ are inherent properties of all probability density functions, i.e., they are non-negative, and the total area under the graph $ f_{Z} \big( z \big) $ is equal to unity. On the one hand, $ \left( \mathrm{\mathbf{P_3}} \right) $ requires the probability density function to be non-symmetric while $ \left( \mathrm{\mathbf{P_4}} \right) $ states that it must not be centered around zero. These properties ensure that the effect of the random phase rotations will prevail even if a statistical attack is perpetrated. In Section \ref{s_eval}, we corroborate experimentally that probability density functions complying with $ \left( \mathrm{\mathbf{P_1}} \right) $, $ \left( \mathrm{\mathbf{P_2}} \right) $, $ \left( \mathrm{\mathbf{P_3}} \right) $ and $ \left( \mathrm{\mathbf{P_4}} \right) $ can conceal the radiometric fingerprint effectively.

In the following, we justify the necessity for $ \left( \mathrm{\mathbf{P_3}} \right) $ and $ \left( \mathrm{\mathbf{P_4}} \right) $.  From (\ref{e4-6}), we note that $ \mathbb{E} \left[ e^{j Z_k} \right] = \beta^{\mathrm{real}}_k + j \beta^{\mathrm{imag}}_k $ must be complex-valued in order to prevent the phase randomization effect from being removed. This is attained when the term $ \beta^{\mathrm{imag}}_k = \int_{-R_k}^{R_k} \cos \big( z_k \big) f'_{Z_k} \big( z_k \big) d z_k \neq 0 $, which produces a non-zero phase shift that is absorbed by the \gls{CSI} phase thus concealing the fingerprint. In order for this to hold, $ \cos \big( z_k \big) f'_{Z_k} \big( z_k \big) $ must not be an odd function according to \textit{Corollary 3}. Since $ \cos \big( z_k \big) $ is an even function, this also signifies that $ f'_{Z_k} \big( z_k \big) $ must not be an odd function according to \textit{Corollary 2}. Via \textit{Corollary 1}, this requirement is satisfied when $ f_{Z_k} \big( z_k \big) $ is not an even function. Therefore, it is revealed that we can design arbitrary probability density functions $ f_{Z_k} \big( z_k \big) $ that are not even with non-zero mean, thus yielding the desired effect that prevents phase randomization removal. 

{\it A simple yet effective manner to meet the above criteria is using a shifted even probability density function (e.g., shifted Gaussian or uniform distribution). We will experimentally prove that in \sref{ss:algo_performance}}. 

\subsection{De-obfuscation and authentication at the receiver}
\label{ss_deobfus_rx}

The right-hand side of \fref{fig_algStructure} shows the two main tasks of the receiver: de-obfuscation and authentication.


{\bf De-obfuscation}. Having the synchronization index and pre-shared key, the receiver can re-generate the obfuscation pattern (i.e., randomized phase rotations) of the transmitted frame. This allows the receiver to extract the original fingerprint. This is done easily by subtracting the obfuscation pattern from the phase of the received signal. 

{\bf Authentication}. The receiver verifies the restored fingerprint against the original fingerprint of the transmitter to authenticate the received frame. In addition, the receiver verifies that the synchronization index is larger than that in the last received frame. A frame whose synchronization index is less than or equal to the last frame is probably sent from an adversary attempting a replay attack. {\it We highlight that with \Alg{}, WiFi devices can always obfuscate their fingerprint.}  We mentioned in Section~\ref{s4} that \Alg{} is designed such that obfuscation does not impact the communication performance. Hence, user privacy is always ensured through fingerprint concealment. 

\subsection{\AlgSt{} mode}
\label{ss:standalone}
In this mode, we allow the transmitting device to hide its fingerprint by executing the obfuscation blocks without any handshake or coordination with other receivers. Specifically, the device generates locally a synchronization index and the key, which are used as inputs for fingerprint obfuscation, as depicted in the transmitter side of \fref{fig_algStructure}. As a result, we can ensure privacy protection in a much broader scenario, e.g., communicating with non-\Alg{}-enabled devices, in absence of any active connections, or in connection establishment phase.

\subsection{SDR implementation}
\label{s:setup}

\begin{figure}
	\centering
	\includegraphics[width=0.7\textwidth]{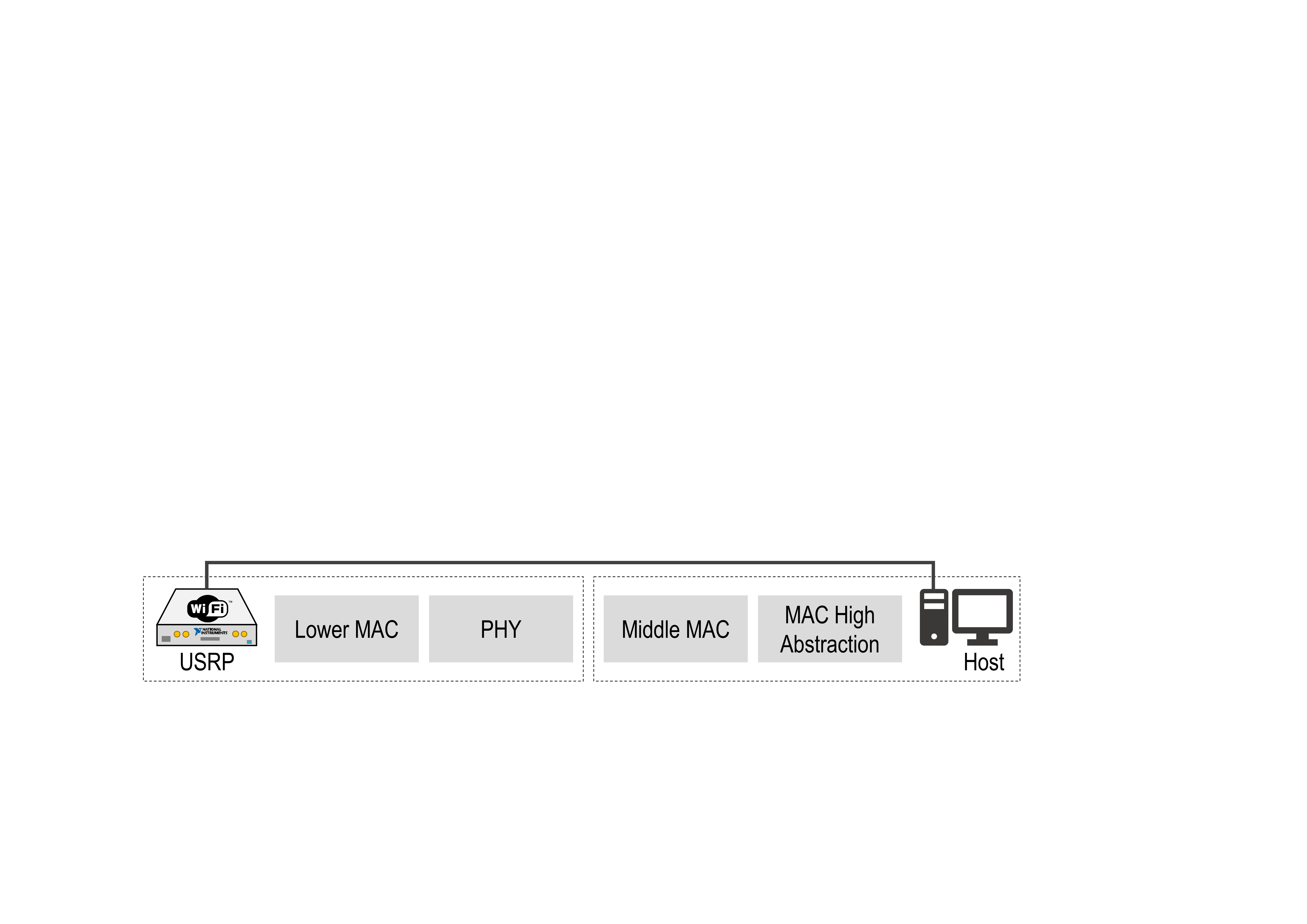}
	\vspace{-3mm}	
	\caption{Schematic overview of the hardware setup.}
	\label{fig_setup}
	\vspace{-4mm}
\end{figure}

We have implemented \Alg{} using the USRP 2954R SDR platform. A simplified overview of the hardware used in our setup is depicted in \fref{fig_setup}.
Each USRP is connected via PCI-e interface to a host machine running NI-Linux RT (kernel version 4.1.13-rt15-nilrt).
We build  \Alg{} using NI 802.11 application framework (AFW) \footnote{http://www.ni.com/pdf/manuals/376779f.pdf}, which provides the physical layer and lower MAC layer functions in the FPGA, while the rest of the MAC procedures run at the host (Linux RT in our setup). 
We provide a detailed overview of the existing implementation in Appendix \ref{appendix_application_framework}.
Due to space constraints, we do not delve into the SDR implementation details. Our implementation and data is available online\footnote{https://github.com/seemoo-lab/RF-Veil}. 
The following briefly describes the setup. 



\textbf{Fingerprint extraction at the receiver. }
The physical layer implementation of 802.11 AFW already includes \gls{CSI} estimation in the FPGA. For our implementation, we have transferred the \gls{CSI} from the FPGA to the host via a \gls{T2H} FIFO on a per-frame basis. This enables fast prototyping while maintaining real-time operation of the testbed. Having the \gls{CSI}, we implemented the radiometric fingerprinting using non-linear phase errors, as described in Section~\ref{s3-2}.


\textbf{Fingerprint modification at the transmitter. } 
These are required modifications at both the FPGA and the host. At the host, we compute the obfuscation pattern, which is sent to the FPGA on a per-frame basis. We made use of the interprocess communication protocol by NI to send packets containing the additional phase rotations. Then, we modified the transmitter chain at the FPGA to read the obfuscation pattern and  multiply each outgoing symbol with the corresponding phase rotations. 
This increases the latency of the transmission chain by 5 clock cycles (12.5 ns). 


\textbf{Secure fingerprinting. }
We implement \Alg{} on top of the \textit{Fingerprint Extraction} and \textit{Fingerprint Modification} modules on the host. 
We extend the packet headers so as to also carry the 32-bit synchronization index chosen at the transmitter. 
When a new packet is being prepared for transmission, the MAC header is used to obtain the key and synchronization index from the \gls{LUT}.
Then, the obfuscation pattern is generated using the key and synchronization index.
This pattern serves as input for the \textit{Fingerprint Modification} module, which then pushes the values to the PHY.


At the receiver side, the \gls{CSI} is written into the \textit{T2H Channel Estimation} FIFO at the PHY. 
The frame reception continues on the FPGA while the implementation of \Alg{} runs on the output of the FIFO at the host.
After the information for random pattern generation is obtained, and the randomization is reverted, the fingerprint is calculated by the \textit{Fingerprint Extraction} module. 
The obtained fingerprint is then passed to the \textit{Matcher} to be compared with the original fingerprint for authentication.

\subsection{Takeaway} 

\textit{In this section, we elaborated on the workings behind \Alg{} and its standalone-mode. 
We devised the idea of synchronized obfuscation with special probability density functions to counter the statistical attacks introduced in \sref{Vulnerability_analysis}, as well as the tracking and impersonation attacks introduced in \sref{s_adversary_model}.
The prototype implementation of \Alg{} on a USRP SDR platform enables us to experimentally evaluate the performance of our approach. 
The takeaway message is that \Alg{} introduces low overhead to the existing WiFi message flow while providing enhanced privacy for users and a secure way of physical layer device identification.}

\section{Evaluation} \label{s_eval}

In this section, we first evaluate the efficacy of the impersonation attack introduced in \sref{s_adversary_model}.
We then leverage \Tool{} to provide a broader assessment of the performance of naive randomization (i.e., obfuscation via zero-mean distributions) and \Alg{} against statistical attacks.



\subsection{Performance of naive randomization}

\begin{figure*}[!t]
	\centering
	\input{figs/error_distribution_80211ac}
	\label{fig:zero_mean_mae}
\end{figure*}

In Section~\ref{Vulnerability_analysis}, we demonstrated the vulnerability of obfuscation, with zero-mean distributions, to statistical attacks experimentally and analytically. In particular, we showed that an adversary can easily restore the original fingerprint from 10000 frames. However, we have neither studied the impact of number of samples, nor considered the effect of the distributions variance on the accuracy of the restored fingerprint by the adversary. To this aim, in \fref{fig:zero_mean_mae}, we show the mean absolute error (MAE) of the adversary's estimate of the original fingerprint when using \Tool{} in 802.11ac.
The figure demonstrates the results under four distributions, namely, uniform, Gaussian, Laplacian, and triangular. For each distribution, we compute the MAE with four variances. Here we make two key observations: {\it (i) } the adversary can {\it restore the original fingerprint with very high accuracy by just processing the \gls{CSI} of $\sim$2000 frames} (a couple of seconds\footnote{In estimating the time for collecting a given number of frames, we assume that the user transmits at $\sim$ 8Mbps. This number is referential and intendeds to provide an estimate of how fast an adversary can mount an statistical attack.}), and {\it (ii)} the \gls{CSI}-recovery error increases with the variance of randomization {\it since larger variance leads to higher entropy of the obfuscated fingerprints}. This behavior is mainly observed when the number of samples is low. As more samples are processed, the estimation error converges to nearly the same value (this is also supported by equation (\ref{e-B7}) in Appendix \ref{s:crb}). Nonetheless, an adversary can still obtain accurate estimates of the original fingerprint with negligible error even when distributions with large variances are employed. For instance, with only 1000 \gls{CSI} samples, the MAE is below 3$^{\circ} $ for distributions with a variance of 60 deg$^2$. For small variances such as $ \xi^2 = $ 5$~\text{deg}^2$, with only 500 samples (roughly 0.5 seconds), the estimation error is consistently below 1$^{\circ} $ for all distributions. We observe a similar trend with 802.11a, whose results are available in Appendix \ref{appendix_experimental_results_80211a}.

\emph{Remarks: We have shown experimentally that the effect of naive randomization can be removed if an attacker is capable of collecting a few thousand samples to mount an statistical attack. Thus, naive randomization does not protect the fingerprint of devices.}

%

\subsection{\Alg{} performance}
\label{ss:algo_performance}

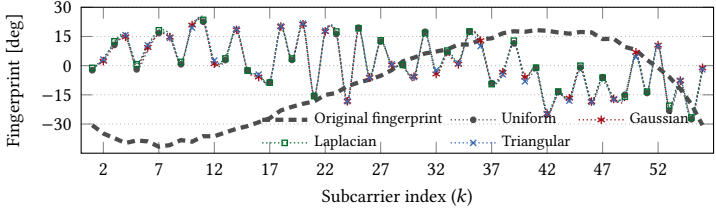
\begin{figure*}[!t]
	\centering
	\definecolor{light_blue}{HTML}{4292c6}
\definecolor{adversary}{HTML}{4d4d4d}
\definecolor{red}{HTML}{a50f15}
\definecolor{uni}{HTML}{4d4d4d}
\definecolor{gau}{HTML}{a50f15}
\definecolor{lap}{HTML}{006d2c}
\definecolor{tri}{HTML}{396AB1}
\begin{tikzpicture}
		\begin{axis}
		[
		xlabel = {Subcarrier index ($ k $)},
		ylabel = {Fingerprint [deg]},
		xmin = 0,
		xmax = 57, 
		ymin = -45, 
		ymax = 30,
		width = 10cm,
		height = 3.5cm,
		ymajorgrids,
    	grid style={line width=0.3pt, draw=gray!70},
		ticklabel style = {font=\fontsize{7}{7}\selectfont},
		xtick = {2, 7, 12, 17, 22, 27, 32, 37, 42, 47, 52},
		ylabel style = {text width = 3.5cm, align = center, font=\fontsize{7}{7}\selectfont},
		xlabel style = {align = center, font=\fontsize{7}{7}\selectfont, yshift=2pt},
		ytick = {-30, -15, 0, 15, 30},
		legend columns = 3,
		legend pos = north east,
		legend style = {at = {(0.28,-0.03)}, anchor = south west, fill = white,  font=\fontsize{6}{6}\selectfont, /tikz/every even column/.append style={column sep=0.03cm}, legend style={row sep=-1pt}, inner sep = 0.5pt, align = left, draw=none, fill=none},
		legend cell align={left},
		]

		\addplot[color = adversary, mark = none, ultra thick, mark options = {scale = 1, fill = light_blue, solid}, densely dashed] table {figs_data/OriginalFingerprint.txt};
		\addlegendentry[]{Original fingerprint} 

		\addplot[color = uni, densely dotted, mark = *, mark options = {scale = 0.5, fill = uni, solid}, line width = 0.5pt, smooth, mark repeat=2, mark phase=1] table {figs_data/Fig12-RecoveredFingerprintUniformVar1TemplateBISON.txt}; 
		\addlegendentry[]{Uniform} 
		
		\addplot[color = gau, densely dotted, mark = asterisk, mark options = {scale = 0.8, fill = gau, solid}, line width = 0.5pt, smooth, mark repeat=2, mark phase=2] table {figs_data/Fig12-RecoveredFingerprintGaussianVar1TemplateBISON.txt}; 
		\addlegendentry[]{Gaussian} 
				
		\addplot[color = lap, densely dotted, mark = square, mark options = {scale = 0.5, fill = lap, solid}, line width = 0.5pt, smooth, mark repeat=2, mark phase=1] table {figs_data/Fig12-RecoveredFingerprintLaplacianVar1TemplateBISON.txt}; 
		\addlegendentry[]{Laplacian} 
		
		\addplot[color = tri, densely dotted, mark = x, mark options = {scale = 0.8, fill = tri, solid}, line width = 0.5pt, smooth, mark repeat=2, mark phase=2] table {figs_data/Fig12-RecoveredFingerprintTriangularVar1TemplateBISON.txt}; 
		\addlegendentry[]{Triangular} 
		
		\end{axis}
\end{tikzpicture}
\vspace{-5mm}
\caption{Restored fingerprint after obfuscation with \Alg{}. {\it Note that \Alg{} prevents potential adversaries from infringing privacy and security since the original fingerprint cannot be recovered. In this experiment, the \Alg{} transmitters use the same values for shifting the means across the subcarriers. Hence, it is the expected behavior that the restored fingerprint for the different random distributions are similar. }}
\label{fig:non_zero_mean}
	\vspace{-9mm}
\end{figure*}
\begin{figure*}[!t]
	\centering
	\input{figs/performance_ideal_pdfs_80211ac}
	\vspace{-2mm}
\end{figure*}

In this experiment, we evaluate the security and privacy enhancement achieved by \Alg{}. Following the conditions for randomization patterns that are robust to statistical attacks (see Section~\ref{ss_obfus_tx}), we obfuscate the original signature of the device using the same four distributions whose mean values are now shifted according to a random pattern, unlike the previous experiment. 
The results of this experiment in legacy mode 802.11a are provided in Appendix \ref{appendix_experimental_results_80211a}.
As depicted in \fref{fig:non_zero_mean}, the recovered fingerprints using \Tool{} deviate from the original fingerprint substantially and follow the course of a random pattern. 
Essentially, when an adversary uses statistical analysis to identify devices, the extracted fingerprint will  not match with the original fingerprint. To shed light on this aspect, \fref{fig:non_zero_mean_mae} depicts the MAE of fingerprints restored by \Tool{} with increasing sample sizes and under different variances. As compared to the low MAE in \fref{fig:zero_mean_mae} (3$^{\circ}$) where zero-mean distributions are used, the MAE in \fref{fig:non_zero_mean_mae} increases approximately by 4-fold (13$^{\circ}$) regardless of the variances and sample sizes used. While all the variances lead to nearly the same error when $ N $ is large, we observe that a large variance produces more variability in the MAE, specially with a small number of samples $ N $. On the other hand, small variances produce a more condensed range of MAE values throughout all $ N $. {\it This result demonstrates two promising properties of \Alg{}: {\it (i)} the adversary's estimate does not improve even with large number of samples, and {\it (ii)} the impact of the variance on estimation is almost negligible as all the errors converge to a similar value (also supported by equation (\ref{e-B7}) in Appendix \ref{s:crb}).} Note that the random pattern in \fref{fig:non_zero_mean} is generated approximately within the same phase-error range as the original fingerprint (i.e., between $ -25^{\circ} $ and $ 40^{\circ} $). Therefore, the attained MAE is not excessively large. However, the MAE can be arbitrarily larger if we construct the pattern spanning a wider range.

\emph{Remarks: \Alg{} protects users' privacy by preventing adversaries from estimating the original fingerprint for tracking/locating the user. Furthermore, the security is also enhanced, since the adversary cannot successfully forge the original fingerprints of other devices.}

\textbf{Effect of \Alg{} on throughput. } 
In Section~\ref{s4}, we analytically showed that \Alg{} does not impact the throughput of WiFi communication. Here, we confirm our analysis with experiments.


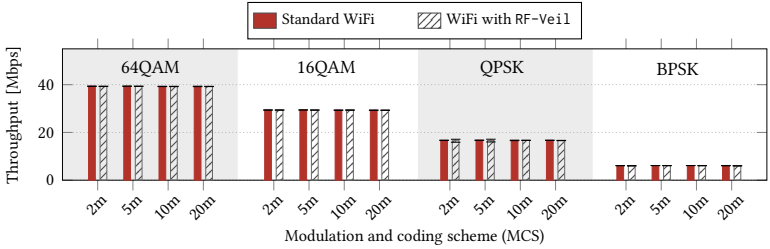
\begin{figure}[!t]
\vspace{-6mm}
\definecolor{ref}{RGB}{179,51,43}
\definecolor{uni}{RGB}{77,77,77}
\definecolor{gau}{HTML}{a50f15}
\definecolor{lap}{HTML}{006d2c}
\definecolor{tri}{HTML}{253494}
\begin{tikzpicture}[scale=0.9]
	\draw [fill=gray, opacity=0.15, draw=gray!0] (5.25,0) rectangle +(2.6,1.92);
	\draw [fill=gray, opacity=0.15, draw=gray!0] (0.0,0) rectangle +(2.6,1.92);
	\draw (1.3, 1.62) node {\scriptsize $64$QAM};	
	\draw (3.9, 1.62) node {\scriptsize $16$QAM};	
	\draw (6.5, 1.62) node {\scriptsize QPSK};
	\draw (9.1, 1.62) node {\scriptsize BPSK};
\pgfplotsset{compat=1.3}
	\begin{axis}[
		ybar,
		xlabel = {Modulation and coding scheme (MCS)},
		ylabel = {Throughput [Mbps]},
		xmin = 0,
		xmax = 20, 
		ymin = 0,
		ymax = 55,
		width = 12cm,
		height = 3.5cm, 
		bar width= 1mm,
		ymajorgrids,
    	grid style={line width=0.3pt, draw=gray!70},
		ticklabel style = {font=\fontsize{7}{7}\selectfont},
		xtick = data,
        xticklabels = {\strut $2$m, \strut $2$m, \strut $2$m, \strut $2$m, \strut $5$m, \strut $5$m, \strut $5$m, \strut $5$m, \strut $10$m, \strut $10$m, \strut $10$m, \strut $10$m, \strut $20$m, \strut $20$m, \strut $20$m, \strut $20$m},
        xticklabel style={yshift=1ex, rotate=45},
		xlabel style = {yshift=1.5ex, align = center, font=\fontsize{7}{7}\selectfont},
		scaled x ticks=false,
		ylabel style = {text width = 4cm, align = center, font=\fontsize{7}{7}\selectfont},
		legend columns = 2,
		legend pos = north east,
		legend style = {at = {(1.1,1.1)}, anchor = south west, font = \scriptsize, fill = white,  font=\fontsize{6}{6}\selectfont, /tikz/every even column/.append style={column sep=0.5cm}, legend style={row sep=1pt}, inner sep = 2pt, align = left},
		legend cell align={left},
		legend entries={Standard WiFi, WiFi with \Alg{} },
		legend to name = {legend},
		legend image code/.code={
   		\draw[#1, draw=black!70] (0cm,-0.1cm) rectangle (0.3cm,0.1cm);
    		},
		name = border,
		cycle list={ {ref}, {uni}},
        every axis plot/.append style={fill, fill opacity=1},
	]

	\addplot[draw=ref, fill=ref, pattern color=ref, error bars/.cd, y dir=both, y explicit] 
	coordinates {
	(1, 39.34137022950819) +- (0.06153663928350338,0.06153663928350338)
	(6, 29.372880131147536) +- (0.061444380402110296,0.061444380402110296)
	(11, 16.689777967213114) +- (0.07207895122914884,0.07207895122914884)
	(16, 6.097092196721314) +- (0.043516234205601044,0.043516234205601044)
	
	(2, 39.35361980327868) +- (0.055721251660833734,0.055721251660833734)
	(7, 29.40110740983606) +- (0.04011079340509439,0.04011079340509439)
	(12, 16.719603016393442) +- (0.022036621973799585,0.022036621973799585)
	(17, 6.1231891147541) +- (0.01622434338775774,0.01622434338775774)
	
	(3, 39.26077508196721) +- (0.06408027621312606,0.06408027621312606)
	(8, 29.317490754098362) +- (0.08201542679529843,0.08201542679529843)
	(13, 16.65751050819672) +- (0.0963246108593268,0.0963246108593268)
	(18, 6.121058754098363) +- (0.01702623366865173,0.01702623366865173)
	
	(4, 39.26307947540983) +- (0.08971259593774225,0.08971259593774225)
	(9, 29.295654557377052) +- (0.07890247335243655,0.07890247335243655)
	(14, 16.701494950819672) +- (0.020627128352719577,0.020627128352719577)
	(19, 6.083244852459019) +- (0.070507374840831,0.070507374840831)
	};
	
	\addplot[draw=black!60, fill=uni, pattern=north east lines,  pattern color=uni, error bars/.cd, y dir=both, y explicit]
	coordinates {
	(1, 39.29237193442622) +- (0.06873508871733021,0.06873508871733021)
	(6, 29.371282360655734) +- (0.038516249020693415,0.038516249020693415)
	(11, 16.51242544262295) +- (0.6284505258964769,0.6284505258964769)
	(16, 6.0251925245901665) +- (0.07500639556372853,0.07500639556372853)
	
	(2, 39.321152) +- (0.050161188408543916,0.050161188408543916)
	(7, 29.362228327868845) +- (0.03485923664068442,0.03485923664068442)
	(12, 16.58137901639344) +- (0.5769067229016583,0.5769067229016583)
	(17, 6.093979278688527) +- (0.018087014126281234,0.018087014126281234)
	
	(3, 39.25082990163936) +- (0.06889996083886189,0.06889996083886189)
	(8, 29.351151999999992) +- (0.06729050646801021,0.06729050646801021)
	(13, 16.68445206557377) +- (0.032501093840794956,0.032501093840794956)
	(18, 6.08217967213115) +- (0.036791149657742495,0.036791149657742495)
	
	(4, 39.251362491803285) +- (0.06071060666522452,0.06071060666522452)
	(9, 29.293075016393445) +- (0.08187461544006612,0.08187461544006612)
	(14, 16.61468275409836) +- (0.06208274492639033,0.06208274492639033)
	(19, 6.015073311475413) +- (0.09053092873103089,0.09053092873103089)
	};

	\end{axis}
	\node[above] at  (border.north) {\ref{legend}};

\end{tikzpicture}
\vspace{-4mm}
\caption{Average throughput of regular 802.11ac and \Alg{} in different distances and with different MCS over the course of 60 seconds. {\it Here, we prove experimentally that \Alg{} does not impact the throughput.}}
\label{fig_throughput_measurement}
\vspace{-5mm}
\end{figure}


We design an experiment in which we measure the throughput of two WiFi devices at different distances (up to 20m) and under distinct \gls{MCS} (up to 64 QAM) with/without \Alg{}. \fref{fig_throughput_measurement} demonstrates that \Alg{} does not impact the throughput of the system, thus confirming our analysis. This is because the fingerprint obfuscation of \Alg{} is based only on phase rotations of the I/Q symbols within a frame. In particular, such rotations do not affect the WiFi channel estimation since their effect is removed at the legitimate receivers. In the figure, we only show the result of obfuscation with uniform random distribution with $\xi^2 = 60^\circ$. However, we report that the other random distributions (Gaussian, Laplacian, and triangular) do neither impact the throughput. In this experiment, we also measured the computational overhead of \Alg{}. Our measurements show that \Alg{} has an average execution time of 49.495 microseconds, even though we implemented most parts of \Alg{} on the host (i.e., a windows machine). We expect the execution time to drop by at least an order of magnitude in real-time kernel or FPGA implementation. 

\emph{Remark: \Alg{} has low computational overhead and does not impact the communication quality. }

\section{Discussion} \label{s_discus}

In this section, we discuss some of the practical aspects of \Alg{}.

{\bf To share or not to share?}
For the secure CSI-denoising when using \Alg{}, we use a symmetric shared key as it provides the easiest way of synchronizing the random number generators at the transmitter and receiver.
Furthermore, the synchronization index can be easily encrypted by XOR-ing the index with the shared key. 
An alternative to the key exchange we use in \sref{ss_assoc} is a key extraction mechanism based on physical layer properties, such as~\cite{liu:2013}. 
This key extraction method leverages channel response information at the transmitter and receiver to generate symmetric keys. 
{\it Note that \Alg{} is compatible with both methods.} Regardless of the method, the key should be renewed at certain intervals, which brings us to the next point in our discussion.

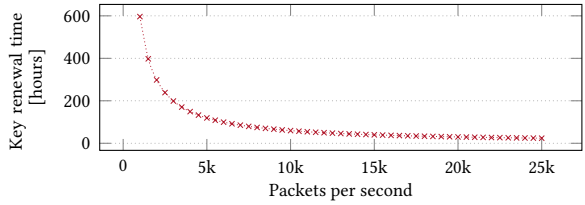
\begin{figure*}[!t]
	\centering
	\definecolor{color1}{HTML}{b2182b}
\begin{tikzpicture}[scale=1]
		\begin{axis}
		[
		xlabel = {Packets per second},
		ylabel = {Key renewal time \newline [hours]},
		width = 8cm,
		height = 3.5cm,
		ymajorgrids,
    	grid style = {line width=0.3pt, draw=gray!70},
		ticklabel style = {font=\fontsize{7}{7}\selectfont},
		xtick = {0, 5000, 10000, 15000, 20000, 25000},
		xticklabels = {0, 5k, 10k, 15k, 20k, 25k},
		ylabel style = {text width = 1.9cm, align = center, font=\fontsize{7}{7}\selectfont},
		xticklabel style={
        /pgf/number format/fixed,
        /pgf/number format/precision=5000
		},
		scaled x ticks=false,
		xlabel style = {align = center, font=\fontsize{7}{7}\selectfont, yshift=+3pt},
		]


		\addplot [mark=x,densely dotted, color=color1, mark options = {scale = 0.7, fill = color1, solid}, line width = 0.35pt, smooth, mark phase=1] coordinates {
		(1000, 596.5232355555556)
		(1500, 397.682157037037)
		(2000, 298.2616177777778)
		(2500, 238.60929422222222)
		(3000, 198.8410785185185)
		(3500, 170.43521015873017)
		(4000, 149.1308088888889)
		(4500, 132.5607190123457)
		(5000, 119.30464711111111)
		(5500, 108.45877010101009)
		(6000, 99.42053925925924)
		(6500, 91.77280547008549)
		(7000, 85.21760507936509)
		(7500, 79.53643140740742)
		(8000, 74.56540444444445)
		(8500, 70.17920418300653)
		(9000, 66.28035950617284)
		(9500, 62.791919532163746)
		(10000, 59.652323555555554)
		(10500, 56.81173671957672)
		(11000, 54.229385050505044)
		(11500, 51.871585700483095)
		(12000, 49.71026962962962)
		(12500, 47.72185884444445)
		(13000, 45.88640273504274)
		(13500, 44.18690633744856)
		(14000, 42.60880253968254)
		(14500, 41.13953348659004)
		(15000, 39.76821570370371)
		(15500, 38.485370035842294)
		(16000, 37.28270222222223)
		(16500, 36.15292336700337)
		(17000, 35.08960209150327)
		(17500, 34.08704203174603)
		(18000, 33.14017975308642)
		(18500, 32.244499219219215)
		(19000, 31.395959766081873)
		(19500, 30.590935156695153)
		(20000, 29.826161777777777)
		(20500, 29.098694417344174)
		(21000, 28.40586835978836)
		(21500, 27.74526677002584)
		(22000, 27.114692525252522)
		(22500, 26.512143802469136)
		(23000, 25.935792850241548)
		(23500, 25.383967470449175)
		(24000, 24.85513481481481)
		(24500, 24.34788716553288)
		(25000, 23.860929422222224)
		};
		
		
		\end{axis}
\end{tikzpicture}
	\vspace{-5mm}
	\caption{Time to renew the key for different sending rates.}
	\label{fig_key_renewal_rates}
	\vspace{-6mm}
\end{figure*}

{\bf How often should we renew the key?}
The monotonically increasing 32-bit synchronization index ensures that, even if the transmitter keeps the symmetric key static for a certain time, they do not repeat the pattern of phase rotations. 
If the synchronization index wraps around at its maximum of $2^{32}=4\,294\,967\,296$, the pattern of random phase rotations is repeated, and an adversary can potentially launch a replay attack. 
In order to thwart such an attack, the key has to be renewed before the transmitter starts to re-use their synchronization indices for this key.
Furthermore, it is important to point out that the transmitter initializes the synchronization index with a number between $0$ and $2\,147\,483\,648$, which ensures that a part of the key cannot be guessed from the encrypted synchronization index in the frame.
We plot the minimum time for key refreshment under different transmission rates in \fref{fig_key_renewal_rates} (for the worst case in which the transmitter chose to start at $2\,147\,483\,648$).
We observe that the key has to be refreshed every $596$ hours ($24.8$ days), assuming an average of $1000$ packets per second.
Even if we assume that the transmitter sends on average $25\,000$ packets per second, which would imply a rate of $462.4$ Mbps, it will exhaust the number of available synchronization indices in $23.8$ hours. Thus, even at very high rates ($5$ TByte of traffic per day), the key exchange is not too frequent.
Note that increasing the frequency of key exchange does not decrease the security level but increase the overhead since the keys are either transmitted encrypted by WPA2 or extracted by both transmitter and receiver using key extraction methods~\cite{liu:2013}. Subsequently, an exposure of this key would affect the privacy and security of the connection until a new key is exchanged.


{\bf What if a frame is rejected?}
A frame can be rejected for two different events: \textit{$(i)$ the calculated \gls{FCS} of the frame does not match the actual \gls{FCS} in the frame}, and \textit{$(ii)$ the fingerprinting algorithm rejects the frame (e.g., the extracted fingerprint differs from the expected one)}.
In both cases, we let the MAC layer handle the re-transmission. 
In the case of a rejected frame, an \Alg{} transmitter does not re-use the synchronization index of the frame that is to be re-transmitted; instead, it increases the count as if a new frame was transmitted. 
This is crucial to guarantee the security of the system as the encryption of the same synchronization index would lead to the same cipher-text.

{\bf What if a device is not yet connected?}
If a device is not connected to an AP, it can still obfuscate its fingerprint by using \AlgSt{} mode in order to lead a potential privacy-intruding adversary astray.
Once the device is connected to an AP and the pre-shared key has been established, it can switch into \Alg{} mode, allowing the AP to securely extract the unrandomized fingerprint. 
This same mechanism applies to the probe requests and acknowledgments in response to probe responses from APs during active scanning. Note that, when the device is not associated to an AP, it can simply use a random key.

{\bf How does an \Alg{} transmitter communicate with a non-\Alg{} receiver?}
Recalling \sref{s4}, the obfuscation of fingerprints does not degrade the channel quality as the channel estimation and equalization at the receiver can handle the arbitrary phase shifts introduced by the transmitter. Specifically, the additional phase rotations are absorbed by the \gls{CSI}, and as long as the same phase rotation pattern is used for all the subcarriers within the frame, the receiver will assume that such \gls{CSI} is legitimate. Hence, a receiver that is not aware of \Alg{} will simply revert the phase shifts together with the channel effects. In other words, a transmitter using the \AlgSt{} mode can still communicate with a legacy receiver. This receiver, however, will not be able to extract the correct fingerprint of the transmitter.

{\bf Can we implement \Alg{} on \gls{COTS} devices?} In recent years, a number of research groups have developed firmware modification/patching tools which allow manipulating MAC/PHY layer operations of the WiFi chipset. Although out of scope of this work, we believe that \Alg{} can be implemented on \gls{COTS} devices using such tools. In particular, Schulz \textit{et al.}~\cite{schulz:2018} demonstrate the feasibility of modifying IQ symbols in commercial APs equipped Broadcom chipsets using their firmware patching framework, i.e., nexmon\footnote{https://github.com/seemoo-lab/nexmon}.



\section{Related Work}
\label{s_related}

To date, we have not found any prior work on radiometric fingerprint obfuscation. 
Prior works only focused on thwarting identification techniques that used packet metadata (frame size, data rate, inter-packet time, etc.) and friendly jamming \cite{rahbari:2015}, upper-layer characteristics such as jitter of beacon timestamps~\cite{arackaparambil:2010}, rate switching mechanisms~\cite{corbett2008passive}, and under-specification of the MAC layer protocols and procedures~\cite{cache:2006, bratus:2008}.
The proposed countermeasures for these upper-layer fingerprinting techniques consist of pattern randomization~\cite{vanhoef2016mac,rahbari2015secrecy,  gruteser2005enhancing, jafarian2016multi}, similar to ours. {\it However, unlike \Alg{}, their approach eliminates the possibility of legitimate fingerprinting. Furthermore, the  solutions therein are not tested against statistical attacks}.  

In the following, we provide a broader overview of the radiometric fingerprinting solutions, which can be categorized into \textit{transient-based} and  \textit{modulation-based} approaches.

\subsection{Transient-based approaches} 
The transient refers to the part of the signal in which the amplitude rises from background noise to full power~\cite{rasmussen:2007}. Given its dependence on the hardware characteristics, a transient is a reliable feature for device identification by tracking the small but measurable differences in the turn-on transients. 
This can, for example, include the duration of turn-on transient~\cite{rasmussen:2007} or standard deviation of normalized amplitude, phase, and frequency~\cite{hall:2004}.
These approaches are cumbersome since they rely on the exact extraction of the transient portion of signals, which further depends on the channel noise. To ensure accurate and timely detection of the transient despite the channel noise, a very high sampling rate is required, which is typically achievable by high-end oscilloscopes (e.g., 4 Giga samples per second in~\cite{danev:2009}).

\subsection{Modulation-based approaches} 
Modulation- or steady-state approaches, as the name suggests, make use of errors in the modulated signal. 
The seminal work of Brik \textit{et al.}~\cite{brik:2008} proposes to collect the fingerprints from five features of the modulated signal, that is, magnitude, phase and frequency error, I/Q origin offset, and SYNC correlation. They show experimentally that their solutions, called PARADIS, can differentiate among 130 identical IEEE 802.11b devices with an accuracy above 99\% even under mobility and varying noise conditions. Similar to the transient-based approaches, their approach requires additional equipment since they rely on high-end vector analyzers for channel sampling. Motivated by their work, recent approaches~\cite{liu:2019, hua:2018} propose to use the \gls{CSI} obtained from the pilot symbols which are readily available on WiFi chipsets, such as the Intel 5300 or Atheros AR9380. Specifically, Hua \textit{et al.}~\cite{hua:2018} propose to compute the fingerprint using a combination of CFO extracted from the \gls{CSI} and \gls{TDoA} computed from capturing 5000 adjacent frames. Furthermore, they require the device to remain stationary for at least 10 seconds for authenticating a device based on the previously collected fingerprint. The most recent work on radiometric fingerprinting~\cite{liu:2019} makes use of the non-linear phase errors extracted from \gls{CSI}. Their work takes advantage of non-linear phase error extraction methods proposed by Zhuo \textit{et al.} in \cite{zhuo:2017}. In this paper, we work toward obfuscating the radiometric fingerprints caused by non-linear phase errors~\cite{liu:2019, zhuo:2017} since it neither relies on RF equipment with very high sampling rates nor requires large number of frames or stationary user behavior for fingerprinting. Nonetheless, \Alg{}'s approach can be extended to other features of the signal, which is controllable at the chipset, such as CFO and amplitude. 

\section{Conclusions}
\label{s_conclusions} 

Radiometric fingerprinting is typically considered a secure method for device identification~\cite{liu:2019, robyns2017noncooperative, xu2015device}. In this paper, we first {\it demonstrate the vulnerability of the latest \gls{CSI}-based radiometric identification schemes to impersonation attacks}, which emphasizes the need for fingerprinting solutions that are robust against adversarial attacks on user security and privacy. We also illustrate that a {\it naive fingerprint-randomization approach does not withhold adversaries capable of mounting statistical attacks (i.e., \Tool{} in this paper)}. Consequently, we devise \Alg{}, a framework that enhances user privacy against fingerprint-based tracking/localization attacks, and is robust to statistical, impersonation, and replay attacks.

\textit{To the best of our knowledge, this is the first article that addresses the vulnerabilities of radiometric fingerprints. Hence, we foresee a few avenues of research as future work. 
Leveraging the randomization patterns to create a side-channel between the receiver and transmitter is an interesting method for exchanging the synchronization index. Furthermore, extending \Alg{} to support MIMO transmissions or other signal characteristics such as \gls{CFO} is another direction to further enhance user privacy. Randomizing the STFs and its impact on the communication and radiometric fingerprints is also an interesting research avenue. Further, investigating new distributions functions for the phase rotations that not only preserve security and communication but also reduce the peak-to-average power ratio is an interest research direction, especially for achieving high energy efficiency in low-power IoT devices.}



\begin{acks}
This research is conducted in the context of the DFG-funded project SenShield (447586980). This work is in part supported by the B5G-Cell project in SFB 1053 MAKI and by the LOEWE initiative (Hesse, Germany) within the emergenCITY center. We would like to thank Clemens Felber and Dr. Walter P. Nitzold from NI, Dresden, for their valuable guidance in modification of WiFi LabVIEW AFW. 
\end{acks}

\bibliographystyle{ACM-Reference-Format}
\bibliography{biblio}


\appendix

\setlength{\abovedisplayshortskip}{3pt}
\setlength{\belowdisplayshortskip}{3pt}
\setlength{\abovedisplayskip}{3pt}
\setlength{\belowdisplayskip}{3pt}
\setlength{\jot}{0pt}

\section{Kronecker Product Properties}

\noindent{\textbf{Property 1} \textit{(Transpose of a Kronecker product)}:} 
	Let ${\bf A} \in \mathbb{C}^{m \times n}$, ${\bf B} \in \mathbb{C}^{r \times s}$, then $ \left( \mathbf{A} \otimes \mathbf{B} \right)^T = \mathbf{A}^T \otimes \mathbf{B}^T $.

\vspace{0.1cm}
\noindent{\textbf{Property 2} \textit{(Product of two Kronecker products)}:}
	Let ${\bf A} \in \mathbb{C}^{m \times n}$, $ {\bf B} \in \mathbb{C}^{r \times s} $, $ {\bf C} \in \mathbb{C}^{n \times p}$, and ${\bf D} \in \mathbb{C}^{s \times t} $, then $ {\bf AB} \otimes {\bf CD} = ({\bf A} \otimes {\bf C})({\bf B} \otimes {\bf D}) $.

\vspace{0.1cm}
\noindent{\textbf{Property 3} \textit{(Trace of a Kronecker product of matrices)}:}
	Let ${\bf A} \in \mathbb{C}^{m \times m}$, ${\bf B} \in \mathbb{C}^{n \times n}$, then $ \mathrm{Tr} \big( \mathbf{A} \otimes \mathbf{B} \big) = \mathrm{Tr} \big( \mathbf{A} \big) \big( \mathbf{B} \big) $.

\vspace{0.1cm}
\noindent{\textbf{Property 4} \textit{(Cyclic permutation of the trace)}:}
	Let ${\bf A} \in \mathbb{C}^{m \times n}$, ${\bf B} \in \mathbb{C}^{n \times m}$, then 	$ \mathrm{Tr} \big( {\bf AB} \big) = \mathrm{Tr} \big( {\bf BA} \big) $.

\vspace{0.1cm}
\noindent{\textbf{Property 5} \textit{(Trace of a Kronecker product of vectors)}:}
	Let ${\bf a} \in \mathbb{C}^{m \times 1}$, ${\bf b} \in \mathbb{C}^{m \times 1}$, then $ \mathrm{Tr} \big( \mathbf{a} \otimes \mathbf{b}^T \big) = \mathrm{Tr} \big( \mathbf{a} \mathbf{b}^T \big) $.
		
\section{Cramer-Rao Bound of \Tool{}} \label{s:crb}
\setcounter{equation}{0}
\renewcommand{\theequation}{B\arabic{equation}}

We analyze the performance of \Tool{} and compare it to the Cramer-Rao bound (CRB) bound. We show that \Tool{} is a near-optimal estimator of the \gls{CSI}, as defined in (\ref{e-B22}). For notation simplicity and without loss of generality, in the sequel, we drop the subcarrier index $ k $ and consider the analysis for a single subcarrier for which $ N $ measurements are available. 

Let $ m_n $ be a measurement (or observation) in a given subcarrier defined as
\begin{equation} \label{e-B1} 
m_n = h e^{j Z_n} + w_n ,
\end{equation}
where $ Z_n $ is a random phase rotation and $ h $ is the complex-valued channel.  Recalling Section \ref{s_algorithm}, $ Z_n $ is introduced by our proposed approach \Alg{} to prevent attackers from acquiring the channel accurately. Thus, let $ p (m_n \mid Z_n; h) $ denote the joint likelihood function of $ Z_n $ and $ h $, given the observation $ m_n $ 
\begin{equation} \label{e-B2} 
	p (m_n \mid Z_n; h) = \frac{1}{\sqrt{\pi \sigma^2}} e^{ - \frac{1}{\sigma^2} \left| m_n - h e^{j Z_n} \right|^2 }.
\end{equation}

For $ N $ uncorrelated measurements, we have the likelihood function 
\begin{equation} \label{e-B3} 
	p (m_1, \cdots, m_N \mid Z_1, \cdots, Z_N; h) = \Pi^N_{n = 1} \frac{1}{\sqrt{\pi \sigma^2}} e^{ - \frac{1}{\sigma^2} \left| m_n - h e^{j Z_n} \right|^2 },
\end{equation}
which can be equivalently recast as
\begin{equation} \label{e-B4} 
	p (\mathbf{m} \mid \mathbf{Z}; h) = \frac{1}{(\pi \sigma^2)^{N/2}} e^{ - \frac{1}{\sigma^2} \left\| \mathbf{m} - h e^{j \mathbf{Z}} \right\|^2_2  },
\end{equation}
where $ \mathbf{m} = \left[m_1, \cdots, m_N \right]^T $ and $ \mathbf{Z} = \left[Z_1, \cdots, Z_N \right]^T $. To compute the CRB of $ h $, we require the likelihood function $ p (\mathbf{m}; h) $. Note that this function can be obtained through averaging $ 	p (\mathbf{m} \mid \mathbf{Z}; h) $ over the random nuisance variables $ \mathbf{Z} $. Thus, the likelihood function $ p \left( \mathbf{m}; h\right) $ is computed as
\begin{equation} \label{e-B5} 
\begin{split} 
	p (\mathbf{m}; h) & = \mathbb{E}_{\mathbf{Z}} \left[ \frac{1}{(\pi \sigma^2)^{N/2}} e^{ - \frac{1}{\sigma^2} \left\| \mathbf{m} - h e^{j \mathbf{Z}} \right\|^2_2  } \right], \\
					  & = \int_{\mathcal{D}_\mathbf{Z}} \frac{1}{(\pi \sigma^2)^{N/2}} e^{ - \frac{1}{\sigma^2} \left\| \mathbf{m} - h e^{j \mathbf{Z}} \right\|^2_2  } p_\mathbf{Z} (\mathbf{z}) d\mathbf{z},
\end{split}
\end{equation} 
where $ \mathbf{z} = \left[z_1, \cdots, z_N \right]^T $ denote the integration variables, and $ \mathcal{D}_\mathbf{Z} $ is the domain of the random variables $ \mathbf{Z} $. In addition, $ \mathbb{E}_{\mathbf{Z}} $ denotes statistical expectation with respect to $ \mathbf{Z} $, which has a priori probability density function $ p_{\mathbf{Z}}(\mathbf{z}) $. Assuming that the random phases are independent, then $ p_{\mathbf{Z}}(\mathbf{z}) = \prod_{n = 1}^{N} p_{Z_n}(z_n) $. Thus, (\ref{e-B5}) can be expressed as
\begin{align} \label{e-B6} 
	p (\mathbf{m}; h) = \int_{\mathcal{D}_{Z_1}} \cdots \int_{\mathcal{D}_{Z_N}} \frac{1}{(\pi \sigma^2)^{N/2}} e^{ - \frac{1}{\sigma^2} \sum^N_{n = 1}  \left| m_n - h e^{j Z_n} \right|^2 } p_{Z_1}(z_1) \cdots p_{Z_N}(z_N) ~ dz_1 \cdots dz_N.
\end{align}

The CRB of any unbiased estimator $ \widehat{h} $ of the channel $ h $ is given by 
\begin{align} \label{e-B7}  
\mathsf{CRB} \left( \widehat{h} \right) = \mathbb{E}_{\mathbf{m}} \left[ \frac{\partial }{\partial h} \ln p (\mathbf{m} ; h) \frac{\partial}{\partial h^*} \ln p (\mathbf{m} ; h) \right] ^{-1},
\end{align} 

where $ \mathbb{E}_{\mathbf{m}} $ denotes statistical expectation with respect to $ \mathbf{m} $ \cite{kay1993:fundamentals-statistical-signal-processing-volume-I}. Nonetheless, the computation of this expression is analytically intractable due to the embedded integration with respect to the random variables $ Z_1, \cdots, Z_N $. As a result, a simpler (but looser) bound called the modified CRB (MCRB) has been derived in \cite{dandrea1994:modified-cramer-rao-synchronization-problems, miller1978:modified-cramer-rao-bound-applications}. Specifically, the MCRB is a lower bound of the CRB, i.e., $ \mathsf{MCRB} \left( \widehat{h} \right)  \leq \mathsf{CRB} \left( \widehat{h} \right) $ is defined as
\begin{align} \label{e-B8}
	\mathsf{MCRB} \left( \widehat{h} \right) = \mathbb{E}_{\mathbf{Z}} \left[ \mathbb{E}_{\mathbf{m} \mid \mathbf{Z}} \left[ \frac{\partial}{\partial h} \ln p (\mathbf{m} \mid \mathbf{Z}; h) \frac{\partial}{\partial h^*} \ln p (\mathbf{m} \mid \mathbf{Z}; h) \right] \right]^{-1}.
\end{align}

From (\ref{e-B6}), we compute the derivatives with respect to $ h $ and $ h^* $, 
\begin{align} \label{e-B9}
	\frac{\partial}{\partial h} \ln p (\mathbf{m} \mid \mathbf{z}; h) = \frac{1}{\sigma^2} \sum^N_{n = 1} \left( e^{j Z_n} m^{*}_n - h^* \right) = \frac{1}{\sigma^2} \sum^N_{n = 1} w^*_n e^{j Z_n},
\end{align}
\begin{align} \label{e-B10}
	\frac{\partial}{\partial h^*} \ln p (\mathbf{m} \mid \mathbf{z}; h) = \frac{1}{\sigma^2} \sum^N_{n = 1} \left( e^{-j Z_n} m_n - h \right) = \frac{1}{\sigma^2} \sum^N_{n = 1} w_n e^{-j Z_n},
\end{align}

Upon replacing (\ref{e-B9}) and (\ref{e-B10}) in (\ref{e-B8}), we obtain that
\begin{equation} \label{e-B11}
\begin{split} 
	\mathsf{MCRB} \left( \widehat{h} \right) 
	& = \mathbb{E}_{\mathbf{Z}} \left[ \mathbb{E}_{\mathbf{m} \mid \mathbf{Z}} \left[ \frac{1}{\sigma^2} \sum^N_{n = 1} w^*_n e^{j Z_n} \frac{1}{\sigma^2} \sum^N_{l = 1} w_l e^{-j Z_l} \right] \right]^{-1},  \\
	& = \mathbb{E}_{\mathbf{Z}} \left[ \mathbb{E}_{\mathbf{w} \mid \mathbf{Z}} \left[ \frac{1}{\sigma^2} \sum^N_{n = 1} w^*_n e^{j Z_n} \frac{1}{\sigma^2} \sum^N_{l = 1} w_l e^{-j Z_l} \right] \right]^{-1}, \\
	& = \mathbb{E}_{\mathbf{Z}} \left[ \mathbb{E}_{\mathbf{w} \mid \mathbf{Z}} \left[ \frac{1}{\sigma^4} \sum^N_{n = 1} \sum^N_{l = 1} w^*_n w_l e^{j Z_n} e^{-j Z_l} \right] \right]^{-1}, \\
	& = \mathbb{E}_{\mathbf{Z}} \left[ \mathbb{E}_{\mathbf{w} \mid \mathbf{Z}} \left[ \frac{1}{\sigma^4} \sum^N_{n = 1} \left| w_n \right|^2 + \frac{1}{\sigma^4} \sum^N_{n = 1} \sum^N_{l \neq n} w^*_n w_l e^{j Z_n} e^{-j Z_l} \right] \right]^{-1},  \\
	& = \mathbb{E}_{\mathbf{Z}} \left[ \mathbb{E}_{\mathbf{w} \mid \mathbf{Z}} \left[ \frac{1}{\sigma^4} \sum^N_{n = 1} \left| w_n \right|^2 \right] + \mathbb{E}_{\mathbf{w} \mid \mathbf{Z}} \left[ \frac{1}{\sigma^4} \sum^N_{n = 1} \sum^N_{l \neq n} w^*_n w_l e^{j Z_n} e^{-j Z_l} \right] \right]^{-1}, \\
	& = \mathbb{E}_{\mathbf{Z}} \left[ \frac{1}{\sigma^4} \sum^N_{n = 1} \mathbb{E}_{w_n \mid \mathbf{Z}} \left[ \left| w_n \right|^2 \right] + \frac{1}{\sigma^4} \sum^N_{n = 1} \sum^N_{l \neq n} \mathbb{E}_{w_n \mid \mathbf{Z}} \left[ w^*_n \right] \mathbb{E}_{w_l \mid \mathbf{Z}} \left[ w_l \right] e^{j Z_n} e^{-j Z_l} \right]^{-1}, \\
	& = \mathbb{E}_{\mathbf{Z}} \left[ \frac{N \sigma^2}{\sigma^4} \right]^{-1}.
\end{split}
\end{equation}

In the second step of (\ref{e-B11}), $ \mathbb{E}_{\mathbf{m} \mid \mathbf{Z}} $ has been changed to $ \mathbb{E}_{\mathbf{w} \mid \mathbf{Z}} $ due to the direct dependence of $ \mathbf{m} $ on $ \mathbf{w} $ (when $ \mathbf{Z} $ is fixed). Note that $ \mathbb{E}_{w_n \mid \mathbf{Z}} \left[ w_n \right] = 0 $ and $ \mathbb{E}_{w_n \mid \mathbf{z}} \left[ \left| w_n \right|^2 \right] = \sigma^2 $ since $ \mathbf{w} \sim \mathcal{CN} \left( 0, \sigma^2 \mathbf{I} \right) $. Thus, $ \sum^N_{n = 1} \mathbb{E}_{w_n \mid \mathbf{z}} \left[ \left| w_n \right|^2 \right] = N \sigma^2 $ and $ \sum^N_{n = 1} \sum^N_{l \neq i} \mathbb{E}_{w_n \mid \mathbf{z}} \left[ w^*_n \right] \mathbb{E}_{w_l \mid \mathbf{z}} \left[ w_l \right] e^{j Z_n} e^{-j Z_l} = 0 $ yielding
\begin{align} \label{e-B12}
	\mathsf{MCRB} \left( \widehat{h} \right) = \frac{\sigma^2}{N}.
\end{align}

From (\ref{e-B12}), we realize that the performance of an optimal estimator $ \hat{h} $ improves with $ N $. Essentially, as more measurements become available, the estimation error decreases. From Section \ref{ss_bison}, the channel estimated by \Tool{} for a single subcarrier was found to be $ u = \frac{1}{N} \sum^N_{n = 1} m_n $. To evaluate the performance of \Tool{} we compute its mean square error (MSE). To this purpose, we assume that the random phase rotations are distributed according to a Gaussian probability density function defined as $ p_{Z_n}(z_n) = \frac{1}{\sqrt{2 \pi \xi^2}} e^{-\frac{(z_n - \mu)^2}{2 \xi^2}} $ with mean $ \mu $ and variance $ \xi^2 $. Note that $ \xi^2 $ and $ \mu $ are the same for all the measurements because these are collected for a single subcarrier. Thus,
\begin{equation} \label{e-B13}
\begin{split}
	\mathrm{MSE} \left( u \right) & = \mathbb{E} \left[ \left( u - h \right)^* \left( u - h \right) \right], \\ 
								  & = \mathbb{E} \left[ \frac{1}{N^2} \sum^N_{n = 1} \sum^N_{i = 1} m^*_n m_i - \frac{h^*}{N} \sum^N_{n = 1} m_n - \frac{h}{N} \sum^N_{n = 1} m^*_i + \left| h \right|^2 \right], \\
								  & = \underbrace{ \mathbb{E} \left[ \frac{1}{N^2} \sum^N_{n = 1} \sum^N_{i = 1} m^*_n m_i \right] }_{S_1} - \underbrace{ \mathbb{E} \left[ \frac{h^*}{N} \sum^N_{n = 1} m_n \right] }_{S_2} - \underbrace{ \mathbb{E} \left[ \frac{h}{N} \sum^N_{n = 1} m^*_i \right] }_{S^*_2} + \mathbb{E} \left[ \left| h \right|^2 \right],
\end{split}
\end{equation}

Now, by using (\ref{e-B1}), we expand $ S_1 $
\begin{equation} \label{e-B14}
\begin{split}
	S_1 & = \mathbb{E} \left[ \frac{1}{N^2} \sum^N_{n = 1} \sum^N_{i = 1} m^*_n m_i \right], \\
	    & = \frac{\left| h \right|^2 }{N^2} \mathbb{E} \left[ \sum^N_{n = 1} \sum^N_{i = 1} e^{-j (Z_n - Z_i)} \right]  + \frac{h}{N^2} \mathbb{E} \left[ \sum^N_{n = 1} \sum^N_{i = 1} w^*_n e^{j Z_i} \right] + 
	    \frac{h^*}{N^2} \mathbb{E} \left[ \sum^N_{n = 1} \sum^N_{i = 1} w_i e^{-j Z_n} \right] + 
	    \frac{1}{N^2} \mathbb{E} \left[ \sum^N_{n = 1} \sum^N_{i = 1} w^*_n w_i \right] \\
	    & = \frac{\left| h \right|^2 }{N^2} \mathbb{E} \left[ \sum^N_{n = 1} \sum^N_{i = 1} e^{-j (Z_n - Z_i)} \right] + 
	    \frac{h}{N^2} \underbrace{ \sum^N_{n = 1} \sum^N_{i = 1} \mathbb{E} \left[ w^*_n \right] \mathbb{E} \left[ e^{j Z_i} \right] }_{0} + \frac{h^*}{N^2} \underbrace{ \sum^N_{n = 1} \sum^N_{i = 1} \mathbb{E} \left[ w_i \right] \mathbb{E} \left[ e^{-j Z_n} \right] }_{0} + \frac{1}{N^2} \underbrace{ \mathbb{E} \left[ \sum^N_{n = 1} \sum^N_{i = 1} w^*_n w_i \right] }_{N \sigma^2}, \\
	    & = \frac{\left| h \right|^2 }{N^2} \mathbb{E} \left[ \sum^N_{n = 1} \sum^N_{i = 1} e^{-j (Z_n - Z_i)} \right] + \frac{\sigma^2}{N} 
\end{split}
\end{equation}

The sum of complex exponentials in (\ref{e-B14}) can be equivalently expressed as, 
\begin{equation} \label{e-B15}
\begin{split} 
	\sum^N_{n = 1} \sum^N_{i = 1} e^{-j (Z_n - Z_i)} = & N + \sum^N_{n = 1} \sum^N_{i \neq n } e^{-j (Z_n - Z_i)} \\
	                                                 = & N + 2 \sum^{N-1}_{i = 1} \sum^N_{n = i+1 } \cos (Z_i - Z_n) \\
	                                                 = & N + 2 \sum^{\frac{N(N-1)}{2}}_{l = 1} \cos (X_l) \\
	                                                 = & N + N(N-1) \cos (X).
\end{split}
\end{equation}

In (\ref{e-B15}), $ X = Z_i - Z_n $, $ \forall i,n $ denotes the difference of two Gaussian random variables. The resulting random variable $ X $ is also Gaussian, which can be obtained by means of the convolution theorem. Specifically, $ X $ has mean zero and twice the variance of $ Z_i $, i.e., the probability density function of $ X $ is given by $ p_X(x) = \frac{1}{\sqrt{4 \pi \xi^2}} e^{-\frac{x^2}{4 \xi^2}} $. Replacing (\ref{e-B15}) in (\ref{e-B14}), $ S_1 $ can be recast as
\begin{equation} \label{e-B16}
\begin{split}
	S_1 & = \frac{\left| h \right|^2 }{N^2} \mathbb{E} \left[ \sum^N_{n = 1} \sum^N_{i = 1} e^{-j (Z_n - Z_i)} \right] + \frac{\sigma^2}{N}, \\
      	& = \frac{\left| h \right|^2 }{N^2} \left( N + N(N-1) \mathbb{E} \left[ \cos (X)  \right] \right) + \frac{\sigma^2}{N}, \\
      	& = \frac{\left| h \right|^2 }{N} + \frac{\left| h \right|^2 (N-1) }{N} \mathbb{E} \left[ \cos (X) \right] + \frac{\sigma^2}{N}, \\
      	& = \frac{\left| h \right|^2 }{N} + \frac{\left| h \right|^2 (N-1) }{N} \int_{-\infty}^{\infty} \cos (x) \frac{1}{\sqrt{4 \pi \xi^2}} e^{\frac{-x^2}{4 \xi^2}} dx + \frac{\sigma^2}{N}, \\
      	& = \frac{\left| h \right|^2 }{N} + \frac{\left| h \right|^2 (N-1) }{N} e^{-\xi^2} + \frac{\sigma^2}{N},
\end{split}
\end{equation}
where $ e^{-\xi^2} = \int_{-\infty}^{\infty} \cos (x) \frac{1}{\sqrt{4 \pi \xi^2}} e^{\frac{-x^2}{4 \xi^2}} dx $. Besides, the term $ S_2 $ collapses to
\begin{equation} \label{e-B17}
\begin{split} 
	S_2 & = \mathbb{E} \left[ \frac{h^*}{N} \sum^N_{n = 1} m_n \right] \\
	    & = \frac{\left| h \right|^2 }{N}  \cdot \mathbb{E} \left[ \sum^N_{n = 1} e^{j Z_n} \right] + \frac{h^*}{N} \mathbb{E} \left[ \sum^N_{n = 1} w_n \right] \\
	    & = \left| h \right|^2 \mathbb{E} \left[ e^{j Z_n} \right] + \underbrace{ \frac{h^*}{N} \mathbb{E} \left[ \sum^N_{n = 1} w_n \right] }_{0} \\
	    & = \left| h \right|^2 \int_{-\infty}^{\infty} e^{j Z_n} \frac{1}{\sqrt{2 \pi \xi^2}} e^{\frac{-(Z_n - \mu)^2}{2 \xi^2}} dZ_n \\
	    & = \left| h \right|^2 e^{-\xi^2/2} e^{j \mu}.
\end{split}
\end{equation}

Collecting the results in (\ref{e-B16}) and (\ref{e-B17}), the MSE collapses to
\begin{equation} \label{e-B18}
\begin{split} 
	\mathrm{MSE} \left( u \right) & = \frac{\left| h \right|^2 }{N} + \left| h \right|^2 e^{-\xi^2} - \frac{\left| h \right|^2 e^{-\xi^2}}{N} + \frac{\sigma^2}{N} - \left| h \right|^2 e^{-\xi^2/2} e^{j \mu}  - \left| h \right|^2 e^{-\xi^2/2} e^{-j \mu}  + \left| h \right|^2, \\
								  & = \left| h \right|^2 + \left| h \right|^2 e^{-\xi^2} - 2 \left| h \right|^2 \cos(\mu) e^{-\xi^2/2} + \frac{\left| h \right|^2 }{N} - \frac{\left| h \right|^2 e^{-\xi^2}}{N} + \frac{\sigma^2}{N}.
\end{split}
\end{equation}

By definition, the MSE of any estimator consists of the bias and the variance as shown in
\begin{align} \label{e-B19}
	\mathrm{MSE} \left( u \right) = \mathrm{bias} \left( u \right)^2  + \mathrm{var} \left( u \right).
\end{align}

The bias of the estimator is computed as
\begin{equation} \label{e-B20}
\begin{split}
	\mathrm{bias} \left( u \right) & = \mathbb{E} \left[ u - h \right], \\
	                               & = \mathbb{E} \left[ u \right] - h,  \\
	                               & = h \mathbb{E} \left[ e^{j Z} \right] - h,  \\
	                               & = h \int_{-\infty}^{\infty} e^{j Z} \frac{1}{\sqrt{2 \pi \xi^2}} e^{-\frac{(z - \mu)^2}{2 \xi^2}} dz - h,  \\
	                               & = h e^{-\xi^2/2} e^{j \mu} - h. 
\end{split}
\end{equation}

Thus, the squared bias is
\begin{equation} \label{e-B21}
\begin{split}
	\mathrm{bias} \left( u \right)^2 & = (h e^{-\xi^2/2} e^{j \mu} - h)^* (h e^{-\xi^2/2} e^{j \mu} - h), \\
	 								 & = \left| h \right|^2 + \left| h \right|^2 e^{-\xi^2} - 2 \left| h \right|^2 \cos(\mu) e^{-\xi^2/2}.
\end{split}
\end{equation}

By comparing (\ref{e-B18}), (\ref{e-B19}) and (\ref{e-B21}), we can extract the variance of the estimator. Therefore, 

\begin{align} \label{e-B22}
	\mathrm{var} \left(  u \right) = \frac{ \left|h \right|^2 }{N} - \frac{ \left|h \right|^2 }{N} e^{- \xi^2} + \frac{\sigma^2}{N}.
\end{align}

\begin{figure}
	\begin{tikzpicture}
	\begin{semilogyaxis}
		[
			xlabel = {$ \xi^2 [\text{deg}^2] $},
			ylabel = {$ \mathrm{bias} \left( u \right)^2 / \left| h \right|^2  $},
			xtick = {0.1, 0.4, 0.7, 1.0},
			xticklabels = {5, 20, 40, 60},
			height = 5.4cm,
			width = 9cm,
			legend columns = 1,
			legend pos = north east,
			legend style = {at = {(1.01,0.0)}, anchor = south west, font = \scriptsize, fill = white,  font=\fontsize{7}{7}\selectfont, /tikz/every even column/.append style={column sep=0.5cm}, legend style={row sep=1pt}, inner sep = 2pt, align = left},
			legend cell align={left},
	  	]
	
	  \addplot[black,thick] expression[domain=0:1] {1 + e^(-x) - 2*e^(-x/2)};
	  
	  \addplot[color = cyan, semithick, densely dashed] expression[domain=0:1] {1 + e^(-x) - 2*cos(1)*e^(-x/2)};
	  
	  \addplot[color = pink, semithick, densely dashed] expression[domain=0:1] {1 + e^(-x) - 2*cos(2)*e^(-x/2)};
	  
	  \addplot[color = green, semithick, densely dashed] expression[domain=0:1] {1 + e^(-x) - 2*cos(5)*e^(-x/2)};
	  
	  \addplot[color = orange, semithick, densely dashed] expression[domain=0:1] {1 + e^(-x) - 2*cos(10)*e^(-x/2)};

	  \addplot[color = blue, semithick, densely dashed] expression[domain=0:1] {1 + e^(-x) - 2*cos(20)*e^(-x/2)};
	  
	  \addplot[color = red, semithick, densely dashed] expression[domain=0:1] {1 + e^(-x) - 2*cos(45)*e^(-x/2)};
	  
	  \addplot[color = brown, semithick, densely dashed] expression[domain=0:1] {1 + e^(-x) - 2*cos(90)*e^(-x/2)};
	  
	  \addplot[color = gray, semithick, densely dashed] expression[domain=0:1] {1 + e^(-x) - 2*cos(180)*e^(-x/2)};
	  
	  \legend{Legitimate user, Attacker | $ \mu = 1^{\circ} $, Attacker | $ \mu = 2^{\circ} $, Attacker | $ \mu = 5^{\circ} $, Attacker | $ \mu = 10^{\circ} $, Attacker | $ \mu = 20^{\circ} $, Attacker | $ \mu = 45^{\circ} $, Attacker | $ \mu = 90^{\circ} $, Attacker | $ \mu = 180^{\circ} $};
	  
	\end{semilogyaxis}
	\end{tikzpicture}
	\caption{Comparison of normalized biases between legitimate users and attackers considering various configurations of $\mu$ and $ \xi^2 $. \textit{In the case of attackers, the bias increases since the additional shift $ \mu $ cannot be removed. Specifically, this occurs due to the impossibility to attackers of generating the sequence of random numbers that renders $ \mu $, which can only obtained by legitimate users.}}
	\label{fig_bias_comparison}
	\vspace{-4mm}
\end{figure}
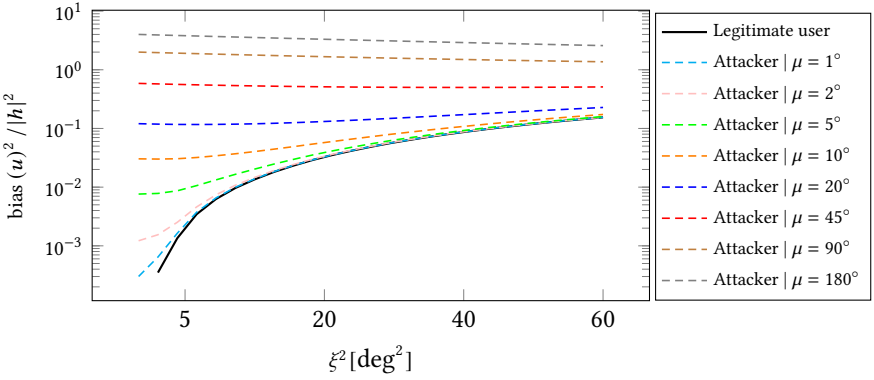

Upon comparing (\ref{e-B12}) and (\ref{e-B22}), we note that a large variance $ \xi^2 $ ($ \xi^2 $ in radians) of the random variables $ Z_n $ leads to a high estimation error according to (\ref{e-B22}). In such a case, $ \mathrm{var} \left(  u \right) \approx \frac{ \left|h \right|^2 }{N} + \frac{\sigma^2}{N} $. However, for small values of $ \xi^2 $, the variance collapses to $ \mathrm{var} \left(  u \right) \approx \frac{\sigma^2}{N} $, thus showing the equivalence between (\ref{e-B12}) and (\ref{e-B22}). While this observation demonstrates that the estimation error of \Tool{} is near-optimal in the variance sense, we also need to consider the bias in (\ref{e-B21}), which is nonzero. Ideally, the estimator needs to be unbiased, i.e., $ \mathrm{bias} \left( u \right)^2 = 0 $. As explained  in Section \ref{ss_obfus_tx}, a legitimate user is aware of the synchronization index and key, and can therefore generate the same sequence of random numbers that yield $ \mu $ (i.e., shifts of the probability density functions). As a result, a legitimate user can remove the additional shift, thus making $ \mu = 0 $. In contrast, for an attacker, $ \mu \neq 0 $. The bias for legitimate users and attackers are respectively defined as
\begin{equation} \label{e-B23}
\begin{split}
	\mathrm{bias}_l \left( u \right)^2 & = \left| h \right|^2 + \left| h \right|^2 e^{-\xi^2} - 2 \left| h \right|^2 e^{-\xi^2/2},
\end{split}
\end{equation}
\begin{equation} \label{e-B24}
\begin{split}
	\mathrm{bias}_a \left( u \right)^2 & = \left| h \right|^2 + \left| h \right|^2 e^{-\xi^2} - 2 \left| h \right|^2 \cos(\mu) e^{-\xi^2/2}, ~ \mu \neq 0,
\end{split}
\end{equation}
showing that $ \mathrm{bias}_l \left( u \right)^2 \leq \mathrm{bias}_a \left( u \right)^2 $. 

To illustrate the difference between (\ref{e-B23}) and (\ref{e-B24}), we show in Fig. \ref{fig_bias_comparison} the biases for several configurations of $ \mu $ and $ \xi^2 $. We observe that for only small $ \mu = \left\lbrace 1^\circ, 5^\circ \right\rbrace $ the biases of the attacker and the legitimate users are similar. However, for sufficiently large $ \mu $ the difference between the two biases becomes noticeable. In our approach, \Alg{}, $ \mu $ is not fixed but is instead randomly generated for every subcarrier using the randomization index and the key. Therefore, for potential attackers---not aware of this information---the bias for each subcarrier varies within the range of values shown in Fig. \ref{fig_bias_comparison}, hindering accurate \gls{CSI} acquisition. Further, for small $ \xi^2 $ we observe that $ \mathrm{bias}_l \left( u \right)^2 \approx 0 $, thus indicating that \Tool{} can be seen as an unbiased estimator in the case of legitimate users when the variance of the phase rotations is low. To clarify this aspect, in Fig. \ref{fig_mse_comparison_a} we show the MCRB bound and the variance of the estimator \Tool{} when \Alg{} is used to conceal the \gls{CSI}. In Fig. \ref{fig_mse_comparison_a} we have neglected the effect of bias and assumed that $ \left| h \right|^2 = \sigma^2 = 1 $. We realize that even for large values of $ \xi^2 $, \Tool{} is capable of performance similarly to the MCRB bound in terms of its variance. In Fig. \ref{fig_mse_comparison_b}, we consider the overall effect of bias and variance in channel estimation for both legitimate users and attackers. We observe that for legitimate users, \Tool{} performs near-optimally when $ \xi^2 $ is relatively small (i.e., $ \xi^2 = 5 $) whereas the error increases for larger $ \xi^2 $. However, as demonstrated in Section \ref{s_eval}, even a small value of $ \xi^2 $ is effective in hindering the \gls{CSI} acquisition by attackers. Thus, $ \xi^2 = 5 $ can be used to successfully protect the \gls{CSI} and the radiometric fingerprint while assuring near-optimality. For potential attackers, the errors between $ 10 $ and $ 100 $ times higher for the shown setting.

\definecolor{ref}{HTML}{252525}
\definecolor{v01}{HTML}{4d4d4d}
\definecolor{v04}{HTML}{a50f15}
\definecolor{v07}{HTML}{006d2c}
\definecolor{v10}{HTML}{396AB1}
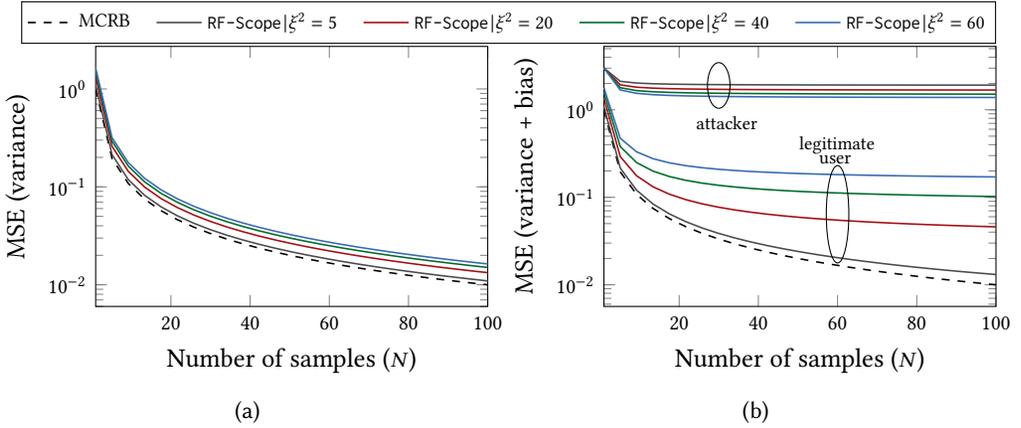
\begin{figure*}[!t]
	\centering
	\begin{subfigure}[t]{0.48\columnwidth}
		\begin{tikzpicture}
		\begin{semilogyaxis}
			[
				xlabel = {Number of samples ($ N $)},
				ylabel = {MSE (variance)},
				xmin=1,
				xmax=100,
				height=5cm,
				width=6.8cm,
				legend columns = 5,
				legend pos = north east,
				legend style = {at = {(-0.19,1.01)}, anchor = south west, font = \scriptsize, fill = white,  font=\fontsize{7}{7}\selectfont, /tikz/every even column/.append style={column sep=0.27cm}, legend style={row sep=1pt}, inner sep = 2pt, align = left},
				legend cell align={left},
		  	]
		
		 \addplot[black,semithick, dashed] expression[domain=1:100] {1/x};
		 
		 \addplot[color = v01, semithick] expression[domain=1:100] {1/x -e^(-0.1)/x + 1/x}; 
		 
		 \addplot[color = v04, semithick] expression[domain=1:100] {1/x - e^(-0.4)/x + 1/x};
		 
		 \addplot[color = v07, semithick] expression[domain=1:100] {1/x - e^(-0.7)/x + 1/x};
		 	  
		 \addplot[color = v10, semithick] expression[domain=1:100] {1/x - e^(-1.0)/x + 1/x};	  	  
		 
		 \legend{$ \mathrm{MCRB} $, $ \Tool{} | \xi^2 = 5 $, $ \Tool{} | \xi^2 = 20 $, $ \Tool{} | \xi^2 = 40 $, $ \Tool{} | \xi^2 = 60 $};
	 
		\end{semilogyaxis}
		\end{tikzpicture}
		\caption{}
		\label{fig_mse_comparison_a}
	\end{subfigure}
	\centering
	\begin{subfigure}[t]{0.48\columnwidth}
		\begin{tikzpicture}
		\begin{semilogyaxis}
			[
				xlabel = {Number of samples ($ N $)},
				ylabel = {MSE (variance + bias)},
				xmin=1,
				xmax=100,
				height=5cm,
				width=6.8cm
		  	]
		  	
		\addplot[black,semithick, dashed] expression[domain=1:100] {1/x};	  	
		  	
		 
		 
	 	  
		
		 \addplot[color = v01, semithick] expression[domain=1:100] {1/x -e^(-0.1)/x + 1/x + 1 + e^(-0.1) - 2*e^(-0.1/2)}; 
		 
		 \addplot[color = v04, semithick] expression[domain=1:100] {1/x - e^(-0.4)/x + 1/x + 1 + e^(-0.4) - 2*e^(-0.4/2)};
		 
		 \addplot[color = v07, semithick] expression[domain=1:100] {1/x - e^(-0.7)/x + 1/x + 1 + e^(-0.7) - 2*e^(-0.7/2)};
		 	  
		 \addplot[color = v10, semithick] expression[domain=1:100] {1/x - e^(-1.0)/x + 1/x + 1 + e^(-1.0) - 2*e^(-1.0/2)};	  	
		 
		 \addplot[color = v01, semithick] expression[domain=1:100] {1/x -e^(-0.1)/x + 1/x + 1 + e^(-0.1) - 2*cos(90)*e^(-0.1/2)}; 
		 
		 \addplot[color = v04, semithick] expression[domain=1:100] {1/x - e^(-0.4)/x + 1/x + 1 + e^(-0.4) - 2*cos(90)*e^(-0.4/2)};
		 
		 \addplot[color = v07, semithick] expression[domain=1:100] {1/x - e^(-0.7)/x + 1/x + 1 + e^(-0.7) - 2*cos(90)*e^(-0.7/2)};
		 	  
		 \addplot[color = v10, semithick] expression[domain=1:100] {1/x - e^(-1.0)/x + 1/x + 1 + e^(-1.0) - 2*cos(90)*e^(-1.0/2)};	 	
		 
		 \draw (30, 1.9) ellipse(0.15cm and 0.3cm);	
		 \node[text width=1.2cm, anchor=west, right] at (22, 0.7) {\scriptsize attacker};   
		 
		 \draw (60, 0.064) ellipse(0.15cm and 0.64cm);	
		 \node[text width=1.2cm, anchor=west] at (48, 0.40) {\scriptsize legitimate};  
		 \node[text width=1.2cm, anchor=west] at (53, 0.27) {\scriptsize user};  
	 
		\end{semilogyaxis}
		\end{tikzpicture}
		\caption{}
		\label{fig_mse_comparison_b}
	\end{subfigure}
	\caption{Comparison of MCRB and \Tool{} when using \Alg{} as a tool to conceal the \gls{CSI} from potential attackers.}
	\label{fig_mse_comparison}
	\vspace{-5mm}
\end{figure*}

As complementary discernment, we show non-tight MCRBs bounds for the channel magnitude, channel phase, and variance of the disturbance that reveal relations that can be used to guide the design of alternative estimators.
\begin{align} \label{e-B25} 
	\mathsf{MCRB} \left( \widehat{\left| h \right|} \right) = \frac{\sigma^2}{2 N}
\end{align}
\begin{align} \label{e-B26} 
	\mathsf{MCRB} \left( \widehat{\phi} \right) = \frac{\sigma^2}{2 N \left| h \right|^2} = \frac{ \mathsf{MCRB} \left( \widehat{\left| h \right|} \right) }{ \left| h \right|^2 }
\end{align}
\begin{align} \label{e-B27} 
	\mathsf{MCRB} \left( \widehat{\mathbf{z}} \right) = \frac{1}{\frac{2 N \left| h \right|^2}{\sigma^2} + \frac{N}{\xi^2}} = \frac{1}{\mathsf{MCRB} \left( \widehat{\phi} \right)^{-1} + \frac{N}{\xi^2}}
\end{align}

\section{Experimental results for 802.11a}
\label{appendix_experimental_results_80211a}

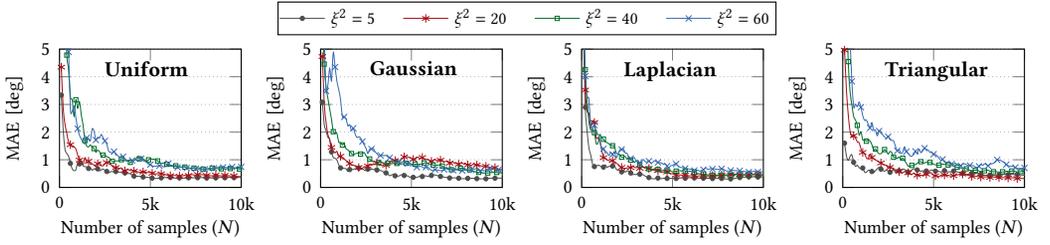
\begin{figure*}[!t]
	\centering
	\vspace{-5mm}
\definecolor{ref}{HTML}{252525}
\definecolor{v01}{HTML}{4d4d4d}
\definecolor{v04}{HTML}{a50f15}
\definecolor{v07}{HTML}{006d2c}
\definecolor{v10}{HTML}{396AB1}

\begin{subfigure}[t]{0.23\columnwidth}
\centering
\pgfplotsset{scaled y ticks=false}
\begin{tikzpicture}
		\begin{axis}
		[
		xlabel = {Number of samples ($ N $)},
		ylabel = {MAE [deg]},
		xmin = 0,
		xmax = 10000, 
		ymin = 0, 
		ymax = 5,
		width = 4cm,
		height = 3.4cm,
		ymajorgrids,
    	grid style = {line width=0.3pt, draw=gray!70},
		ticklabel style = {font=\fontsize{7}{7}\selectfont},
		ytick = {0, 1, 2, 3, 4, 5, 6},
		xtick = {0, 5000, 10000},
		xticklabels = {0, 5k, 10k},
		ylabel style = {text width = 5cm, align = center, font=\fontsize{7}{7}\selectfont},
		xticklabel style={
        /pgf/number format/fixed,
        /pgf/number format/precision=5000
		},
		scaled x ticks=false,
		xlabel style = {align = center, font=\fontsize{7}{7}\selectfont, yshift=+3pt},
		legend columns = 4,
		legend pos = north east,
		legend style = {at = {(1.2, 1.1)}, anchor = south west, font = \scriptsize, fill = white,  font=\fontsize{6}{6}\selectfont, /tikz/every even column/.append style={column sep=3mm}, legend style={row sep=-1pt}, inner sep = 0.5pt, align = left},
		legend cell align={left},
		]

		\addplot[color = v01, mark = *, mark options = {scale = 0.4, fill = v01, solid}, line width = 0.3pt, smooth, mark repeat=5, mark phase=1] table {figs_data/Fig11/Fig11-ErrorUniformMean0Var01BISON_80211a.txt}; 
		\addlegendentry[]{$\xi^2 = 5$} 
		
		\addplot[color = v04, mark = asterisk, mark options = {scale = 0.8, fill = v04, solid}, line width = 0.3pt, smooth, mark repeat=5, mark phase=1] table {figs_data/Fig11/Fig11-ErrorUniformMean0Var04BISON_80211a.txt}; 
		\addlegendentry[]{$\xi^2 = 20$} 
		
		\addplot[color = v07, mark = square, mark options = {scale = 0.4, fill = v07, solid}, line width = 0.3pt, smooth, mark repeat=5, mark phase=1] table {figs_data/Fig11/Fig11-ErrorUniformMean0Var07BISON_80211a.txt}; 
		\addlegendentry[]{$\xi^2 = 40$} 
		
		\addplot[color = v10, mark = x, mark options = {scale = 0.8, fill = v10, solid}, line width = 0.3pt, smooth, mark repeat=5, mark phase=1] table {figs_data/Fig11/Fig11-ErrorUniformMean0Var1BISON_80211a.txt}; 
		\addlegendentry[]{$\xi^2 = 60$} 
		
		\node at (axis cs:2000,4.9) [anchor=north west] {\footnotesize{\textbf{Uniform}}};
		
		\end{axis}
		\end{tikzpicture}
\end{subfigure}	
\hspace{1mm}
\begin{subfigure}[t]{0.23\columnwidth}
\centering
\pgfplotsset{scaled y ticks=false}
\begin{tikzpicture}
		\begin{axis}
		[
		xlabel = {Number of samples ($ N $)},
		ylabel = {MAE [deg]},
		xmin = 0,
		xmax = 10000, 
		ymin = 0, 
		ymax = 5,
		width = 4cm,
		height = 3.4cm, 
		ymajorgrids,
    	grid style={line width=0.3pt, draw=gray!70},
		ticklabel style = {font=\fontsize{7}{7}\selectfont},
		ytick = {0, 1, 2, 3, 4, 5, 6},
		xtick = {0, 5000, 10000},
		xticklabels = {0, 5k, 10k},
		ylabel style = {text width = 5cm, align = center, font=\fontsize{7}{7}\selectfont},
		xlabel style = {align = center, font=\fontsize{7}{7}\selectfont, yshift=+3pt},
		xticklabel style={
        /pgf/number format/fixed,
        /pgf/number format/precision=5000},
		scaled x ticks=false,
		legend columns = 1,
		legend pos = north east,
		legend style = {at = {(0.25,0.67)}, anchor = south west, font = \scriptsize, fill = white,  font=\fontsize{6}{6}\selectfont, /tikz/every even column/.append style={column sep=0.03cm}, legend style={row sep=-1pt}, inner sep = 0.5pt, align = left},
		legend cell align={left},
		]

		\addplot[color = v01, mark = *, mark options = {scale = 0.4, solid}, line width = 0.3pt, smooth, mark repeat=5, mark phase=1] table {figs_data/Fig11/Fig11-ErrorGaussianMean0Var01BISON_80211a.txt}; 

		\addplot[color = v04, mark = asterisk, mark options = {scale = 0.8, fill = v04, solid}, line width = 0.3pt, smooth, mark repeat=5, mark phase=1] table {figs_data/Fig11/Fig11-ErrorGaussianMean0Var04BISON_80211a.txt}; 

		\addplot[color = v07, mark = square, mark options = {scale = 0.4, fill = v07, solid}, line width = 0.3pt, smooth, mark repeat=5, mark phase=1] table {figs_data/Fig11/Fig11-ErrorGaussianMean0Var07BISON_80211a.txt}; 

		\addplot[color = v10, mark = x, mark options = {scale = 0.8, fill = v10, solid}, line width = 0.3pt, smooth, mark repeat=5, mark phase=1] table {figs_data/Fig11/Fig11-ErrorGaussianMean0Var1BISON_80211a.txt}; 
		
		\node at (axis cs:2200,4.9) [anchor=north west] {\footnotesize{\textbf{Gaussian}}};
		\end{axis}
		\end{tikzpicture}
\end{subfigure}
\hspace{1mm}
\begin{subfigure}[t]{0.23\columnwidth}
\centering
\pgfplotsset{scaled y ticks=false}
\begin{tikzpicture}
		\begin{axis}
		[
		xlabel = {Number of samples ($ N $)},
		ylabel = {MAE [deg]},
		xmin = 0,
		xmax = 10000, 
		ymin = 0, 
		ymax = 5,
		width = 4cm,
		height = 3.4cm, 
		ymajorgrids,
    	grid style={line width=0.3pt, draw=gray!70},
		ticklabel style = {font=\fontsize{7}{7}\selectfont},
		ytick = {0, 1, 2, 3, 4, 5, 6},		
		xtick = {0, 5000, 10000},
		xticklabels = {0, 5k, 10k},
		ylabel style = {text width = 5cm, align = center, font=\fontsize{7}{7}\selectfont},
		xlabel style = {align = center, font=\fontsize{7}{7}\selectfont, yshift=+3pt},
		xticklabel style={
        /pgf/number format/fixed,
        /pgf/number format/precision=5000},
		scaled x ticks=false,
		legend columns = 1,
		legend pos = north east,
		legend style = {at = {(0.25,0.67)}, anchor = south west, font = \scriptsize, fill = white,  font=\fontsize{6}{6}\selectfont, /tikz/every even column/.append style={column sep=0.03cm}, legend style={row sep=-1pt}, inner sep = 0.5pt, align = left},
		legend cell align={left},
		]

		\addplot[color = v01, mark = *, mark options = {scale = 0.4, solid}, line width = 0.3pt, smooth, mark repeat=5, mark phase=1] table {figs_data/Fig11/Fig11-ErrorLaplacianMean0Var01BISON_80211a.txt}; 

		\addplot[color = v04, mark = asterisk, mark options = {scale = 0.8, fill = v04, solid}, line width = 0.3pt, smooth, mark repeat=5, mark phase=1] table {figs_data/Fig11/Fig11-ErrorLaplacianMean0Var04BISON_80211a.txt}; 

		\addplot[color = v07, mark = square, mark options = {scale = 0.4, fill = v07, solid}, line width = 0.3pt, smooth, mark repeat=5, mark phase=1] table {figs_data/Fig11/Fig11-ErrorLaplacianMean0Var07BISON_80211a.txt}; 

		\addplot[color = v10, mark = x, mark options = {scale = 0.8, fill = v10, solid}, line width = 0.3pt, smooth, mark repeat=5, mark phase=1] table {figs_data/Fig11/Fig11-ErrorLaplacianMean0Var1BISON_80211a.txt}; 
		
		\node at (axis cs:1800,4.9) [anchor=north west] {\footnotesize{\textbf{Laplacian}}};
		\end{axis}
		\end{tikzpicture}
\end{subfigure}	
\hspace{1mm}
\begin{subfigure}[t]{0.23\columnwidth}
\centering
\pgfplotsset{scaled y ticks=false}
\begin{tikzpicture}
		\begin{axis}
		[
		xlabel = {Number of samples ($ N $)},
		ylabel = {MAE [deg]},
		xmin = 0,
		xmax = 10000, 
		ymin = 0, 
		ymax = 5,
		width = 4cm,
		height = 3.4cm, 
		ymajorgrids,
    	grid style={line width=0.3pt, draw=gray!70},
		ticklabel style = {font=\fontsize{7}{7}\selectfont},
		ytick = {0, 1, 2, 3, 4, 5, 6},
		xtick = {0, 5000, 10000},
		xticklabels = {0, 5k, 10k},
		ylabel style = {text width = 5cm, align = center, font=\fontsize{7}{7}\selectfont},
		xlabel style = {align = center, font=\fontsize{7}{7}\selectfont, yshift=+3pt},
		xticklabel style={
        /pgf/number format/fixed,
        /pgf/number format/precision=5000},
		scaled x ticks=false,
		legend columns = 1,
		legend pos = north east,
		legend style = {at = {(0.2,0.67)}, anchor = south west, font = \scriptsize, fill = white,  font=\fontsize{6}{6}\selectfont, /tikz/every even column/.append style={column sep=0.03cm}, legend style={row sep=-1pt}, inner sep = 0.5pt, align = left},
		legend cell align={left},
		]

		\addplot[color = v01, mark = *, mark options = {scale = 0.4, solid}, line width = 0.3pt, smooth, mark repeat=5, mark phase=1] table {figs_data/Fig11/Fig11-ErrorTriangularMean0Var01BISON_80211a.txt}; 

		\addplot[color = v04, mark = asterisk, mark options = {scale = 0.8, fill = v04, solid}, line width = 0.3pt, smooth, mark repeat=5, mark phase=1] table {figs_data/Fig11/Fig11-ErrorTriangularMean0Var04BISON_80211a.txt}; 

		\addplot[color = v07, mark = square, mark options = {scale = 0.4, fill = v07, solid}, line width = 0.3pt, smooth, mark repeat=5, mark phase=1] table {figs_data/Fig11/Fig11-ErrorTriangularMean0Var07BISON_80211a.txt}; 

		\addplot[color = v10, mark = x, mark options = {scale = 0.8, fill = v10, solid}, line width = 0.3pt, smooth, mark repeat=5, mark phase=1] table {figs_data/Fig11/Fig11-ErrorTriangularMean0Var1BISON_80211a.txt}; 
		
		\node at (axis cs:1800,4.9) [anchor=north west] {\footnotesize{\textbf{Triangular}}};
		\end{axis}
		\end{tikzpicture}
\end{subfigure}
\vspace{-10mm}	
\caption{Mean absolute error (MAE) for different symmetric zero-mean distributions using \Tool{} in 802.11a. \textit{The MAE values with $10000$ samples are below $1^\circ$ for all distributions.}}
\vspace{-1mm}	
	\vspace{-8mm}
	\label{fig:zero_mean_mae_80211a}
\end{figure*}

We analyze the performance of naive randomization in 802.11a\footnote{The main difference between these two technologies is the number of subcarriers, which is $ 52 $ and $ 56 $ for 802.11a and 802.11ac, respectively.} and show the results in \fref{fig:zero_mean_mae_80211a}.
The results suggest that naive randomization performs similarly in 802.11a and 802.11ac, i.e., through \Tool{} the  MAE error diminishes with larger $ N $, which allows an adversary to restore the original fingerprint.

The results of \Alg{} over 802.11a are shown in \fref{fig:x_mean_mae_80211a}, which are reminiscent of the behavior observed with 802.11ac in \fref{fig:non_zero_mean_mae}. In particular, the MAE is 10-fold the error achieved with naive randomization. This experiment corroborates that \Alg{} is not only feasible in 802.11ac but also in other technologies with a similar underlying structure.

\section{NI 802.11 Application Framework}
\label{appendix_application_framework}
\begin{figure*}[]
	\begin{subfigure}{0.47\textwidth}
		\centering
		\includegraphics[width=\textwidth]{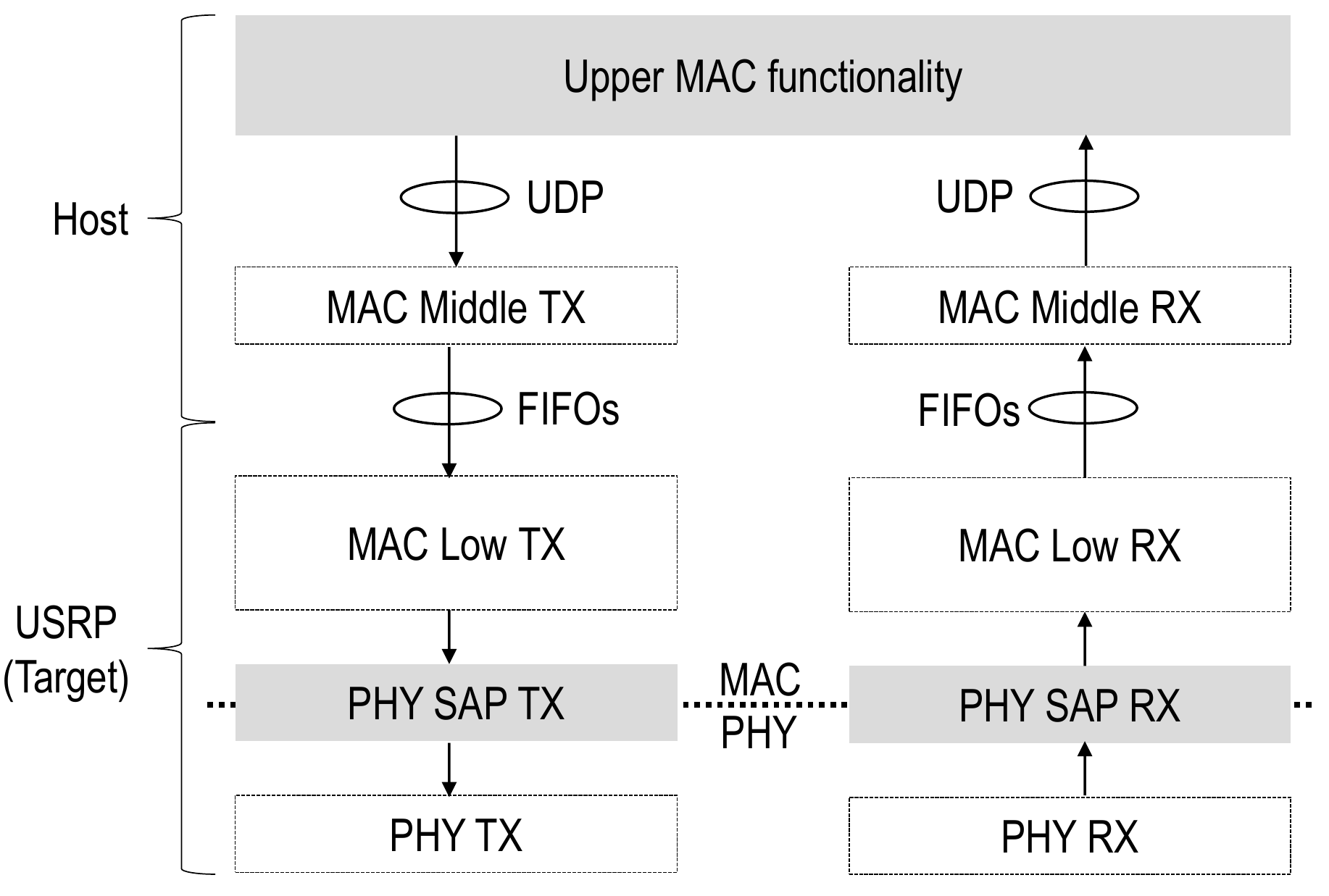}
		\caption{}
		\label{fig:afw_overview}
	\end{subfigure}
	\begin{subfigure}{0.47\textwidth}
		\centering
		\vspace{1mm}
		\includegraphics[width=\textwidth]{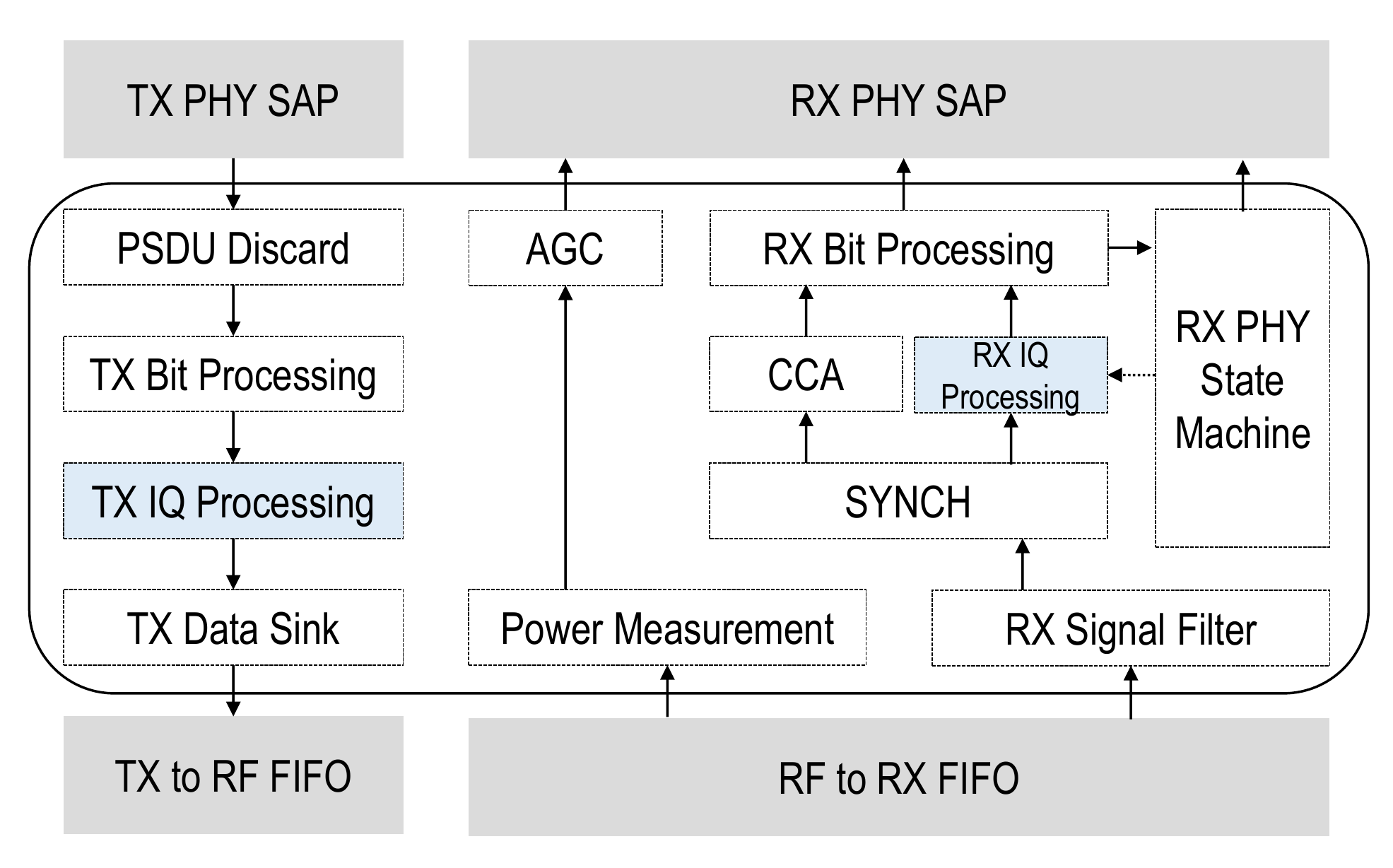}
		\caption{}
		\label{fig:fpga_block}
	\end{subfigure}
	\vspace{-3mm}
	\caption{NI 802.11 implementation details. \fref{fig:afw_overview} shows the interfaces used to connect the different layers of the WiFi implementation. \fref{fig:fpga_block} provides an overview of the physical layer implementation on the FPGA. We mainly adjusted the IQ processing blocks (marked in blue) in both, RX and TX directions.}
\end{figure*}

The implementation of the NI 802.11 AFW is separated into a host and an FPGA module, as depicted in \fref{fig:afw_overview}.
The host module of the NI 802.11 AFW mainly implements middle MAC layer functionalities as well as a MAC high abstraction layer.
The latter allows third party 802.11 higher MAC applications, such as the ns-3 network simulator, to connect to the stack.
The MAC high abstraction layer is connected to MAC middle layer via UDP, which is responsible for duplicate detection in RX and synchronization index assignment in TX direction.

Note that the higher MAC abstraction layer does not implement association or authentication procedures, so the NI 802.11 AFW cannot complete the connection setup to \gls{COTS} hardware. 
To still allow data streams for demo  and measurement purposes between two instances of the 802.11 AFW, simplistic data sinks and sources (random data / \gls{UDP}) are available.

The FPGA module includes the implementation of the lower MAC layer as well as the whole physical layer.
Even though both layers are running on the FPGA, they are executed in different clock domains, as the timing requirements are different.
This results in clock rates of 100 MHz (10 nanoseconds) and 250 MHz (4 nanoseconds) for MAC and physical layer, respectively. 

\begin{figure*}[!t]
	\centering
	\input{figs/performance_ideal_pdfs_80211a}
	\vspace{-3mm}
	\label{fig:x_mean_mae_80211a}
\end{figure*}

The main building blocks of the physical layer, as implemented on the FPGA, are depicted in \fref{fig:fpga_block}.
In TX direction, the lower MAC layer passes the digital data as bits through the TX PHY \gls{SAP} to the physical layer.
The first block (PSDU Discard) removes the PHY service data unit (PSDU) and verifies the consistency of the packet before passing it to the TX Bit Processing block.
This following block takes care of serialization, scrambling, convolutional encoding, puncturing, and interleaving. 
The subsequent TX IQ Processing block modulates the data and creates the OFDM symbols, including all training fields.
While the STF is pre-calculated in time-domain, all other training fields, such as the LTFs are modulated in frequency-domain.
After that, the Inverse Fast Fourier Transform (IFFT) transforms the data form the frequency domain into the time-domain.
The time domain samples are written to the \textit{TX to RF FIFO} in the last block (TX Data Sink).
The final steps, consisting of up-conversion to carrier frequency and sending out the up-converted samples over the air, are left out of \fref{fig:fpga_block} for brevity.

The PHY code in RX direction involves more steps to accurately and correctly identify and retrieve a packet.
The RX Signal Filter reads the already down-converted samples in the baseband and filters for 40 or 20 MHz channels.
Upon that, the synchronization detects the packet start using the Schmidl and Cox algorithm \cite{schmidl:1997}, which also estimates and compensates the CFO. 
The Clear Channel Assessment (CCA) in the subsequent step calculates the received signal power and compares it against a given CCA threshold.
The RX IQ Processing module then transforms the samples from time-domain into frequency-domain using FFT.
Furthermore, it equalizes the channel and detects the format of the frame (non-HT, HT, VHT).
The last step consists of transforming the IQ samples into bits, creating field assignments, bit deinterleaving, decoding, and descrambling.
The RX Bit Processing writes the finished frame into the RX PHY SAP for the lower MAC layer to further process it.
The RX PHY State Machine generates control information for the IQ and bit processing modules based on the bitstream. 
It keeps track of meta-information of the packet, such as the number of OFDM symbols inside a packet, bandwidth, and PSDU length, and signals the lower MAC layer when a packet is completely received. 
Parallel to this whole process, power measurements on the baseband are performed (Power Measurement) and provided to the Automatic Gain Control (AGC), which dynamically determines the gain of the amplifiers.

\end{document}